\documentclass{elsart}

\usepackage{graphicx}

\usepackage{amssymb}
\usepackage{amsmath}
\usepackage{verbatim}
\RequirePackage{fancyvrb} \DefineVerbatimEnvironment{boxedverbatim} {Verbatim} {frame=single,
fontsize=\footnotesize}

\newcommand{\beq}{\begin{equation}}
\newcommand{\eeq}{\end{equation}}
\newcommand{\bdm}{\begin{displaymath}}
\newcommand{\edm}{\end{displaymath}}

\newcommand{\bF}{{\mbox{\boldmath $F$}}}

\newcommand{\bB}{{\mbox{\boldmath $B$}}}
\newcommand{\bu}{{\mbox{\boldmath $u$}}}
\newcommand{\bv}{{\mbox{\boldmath $v$}}}
\newcommand{\bw}{{\mbox{\boldmath $w$}}}

\newcommand{\Dt}{\Delta t}
\newcommand{\Dx}{\Delta x}

\def\simless{\mathbin{\lower 3pt\hbox
      {$\rlap{\raise 5pt\hbox{$\char'074$}}\mathchar"7218$}}}
\def\simgreat{\mathbin{\lower 3pt\hbox
      {$\rlap{\raise 5pt\hbox{$\char'076$}}\mathchar"7218$}}} 

\newcommand{\grad}{{\mbox{\boldmath $\nabla$}}}

\usepackage{algorithm}
\usepackage{algorithmic}

\journal{Computer Physics Communications}

\begin{document}

\begin{frontmatter}

\title{Efficient magnetohydrodynamic simulations on graphics processing units with CUDA}

\author[label1,label2]{Hon-Cheng Wong}
\author[label1]{\& Un-Hong Wong}
\author[label3]{\& Xueshang Feng}
\author[label1]{\& Zesheng Tang}

\address[label1]{Institute of Space Science,\\
Macau University of Science and Technology, Macao, China}
\address[label2]{Faculty of Information Technology,\\
Macau University of Science and Technology, Macao, China}
\address[label3]{SIGMA Weather Group, State Key Laboratory for Space Weather,\\
Center for Space Science and Applied Research,\\
Chinese Academy of Sciences, Beijing 100190, China}

\begin{abstract}
Magnetohydrodynamic (MHD) simulations based on the ideal MHD equations have become a powerful tool
for modeling phenomena in a wide range of applications including laboratory, astrophysical, and
space plasmas. In general, high-resolution methods for solving the ideal MHD equations are
computationally expensive and Beowulf clusters or even supercomputers are often used to run the
codes that implemented these methods. With the advent of the Compute Unified Device Architecture
(CUDA), modern graphics processing units (GPUs) provide an alternative approach to parallel
computing for scientific simulations. In this paper we present, to the author's knowledge, the
first implementation of MHD simulations entirely on GPUs with CUDA, named {\em GPU-MHD}, to
accelerate the simulation process. {\em GPU-MHD} supports both single and double precision
computation. A series of numerical tests have been performed to validate the correctness of our
code. Accuracy evaluation by comparing single and double precision computation results is also
given. Performance measurements of both single and double precision are conducted on both the
NVIDIA GeForce GTX 295 (GT200 architecture) and GTX 480 (Fermi architecture) graphics cards. These
measurements show that our GPU-based implementation achieves between one and two orders of
magnitude depending on the used graphics card, problem size, and precision when comparing to the
original serial CPU MHD implementation. In addition, we extend {\em GPU-MHD} to support the
visualization of the simulation results and thus the whole MHD simulation and visualization process
can be performed entirely on GPUs.
\end{abstract}

\begin{keyword}
MHD simulations \sep GPUs \sep CUDA \sep Parallel computing

\end{keyword}
\end{frontmatter}

\section{Introduction}
\label{sec:intro}
\par
Magnetohydrodynamic (MHD) equations can be used in modeling phenomena in a wide range of
applications including laboratory~\cite{Ciardi2007}, astrophysical~\cite{Stone2007}, and space
plasmas~\cite{Feng2007}. For example, 3D MHD simulations have been widely adopted in space weather
simulations. The historical review and current status of the existing popular 3D MHD models can be
found in~\cite{Dryer1998} and~\cite{Dryer2007}, respectively. However, MHD equations form a
nonlinear system of hyperbolic conservation laws, which is so complex that high-resolution methods
are necessary to solve them in order to capture shock waves and other discontinuities. These
high-resolution methods are in general computationally expensive and parallel computational
resources such as Beowulf clusters or even supercomputers are often utilized to run the codes that
implemented these
methods~\cite{Meijer1996}~\cite{Hayes2006}~\cite{Lee2006}~\cite{Huang2008}~\cite{Ziegler2008}.

In the last few years, the rapid development of graphics processing units (GPUs) makes them more
powerful in performance and more programmable in functionality. By comparing the computational
power of GPUs and CPUs, GPUs exceed CPUs by orders of magnitude. The theoretical peak performance
of the current consumer graphics card NVIDIA GeForce GTX 295 (with two GPUs) is 1788.48G
floating-point operations per second (FLOPS) per GPU in single precision while a CPU (Core 2 Quad
Q9650 --- 3.0 GHz) gives a peak performance of around 96GFLOPS in single precision. The release of
the {\em Compute Unified Device Architecture (CUDA)}~\cite{NVIDIA2009} hardware and software
architecture is the culmination of such development. With CUDA, one can directly exploit a GPU as a
data-parallel computing device by programming with the standard C language and avoid working with a
high-level shading language such as Cg~\cite{Mark2003}, which requires a significant amount of
graphics specific knowledge and was previously used for performing computation on GPUs. Detailed
performance studies on GPUs with CUDA can be found in~\cite{Che2008} and~\cite{Schenk2008}.

CUDA is a general purpose parallel computing architecture developed by NVIDIA. It includes the CUDA
Instruction Set Architecture (ISA) and the parallel compute engine. An extension to C programming
language and its compiler are provided, making the parallelism and high computational power of GPUs
can be used not only for rendering and shading, but also for solving many computationally intensive
problems in a fraction of the time required on a CPU. CUDA also provides basic linear algebra
subroutines (CUBLAS) and fast Fourier transform (CUFFT) libraries to leverage GPUs' capabilities.
These libraries release developers from rebuilding the frequently used basic operations such as
matrix multiplication. Graphics cards from G8x series support the CUDA programming mode; and the
latest generation of NVIDIA GPUs (GT2x0 series or later) unifies vertex and fragment processors and
provides shared memory for interprocessor communication.

A increasing number of new GPU implementations with CUDA in different astrophysical simulations
have been proposed. Belleman {\em et al.}~\cite{Belleman2008} re-implemented the direct
gravitational $N$-body simulations on GPUs using CUDA. For $N\gtrsim 10^5$, they reported a speedup
of about 100 compared to the host CPU and about the same speed as the GRAPE-6Af. A library
\verb=Sapporo= for performing high precision gravitational $N$-body simulations was developed on
GPUs by Gaburov {\em et al.}~\cite{Gaburov2009}. This library achieved twice as fast as commonly
used GRAPE6A/GRAPE6-BLX cards. Stantchev {\em et al.}~\cite{Stantchev2008}~\cite{Stantchev2009}
implemented a Particle-In Cell (PIC) code on GPUs for plasmas simulations and visualizations and
demonstrated a speedup of 11-22 for different grid sizes. Sainio~\cite{Sainio2010} presented an
accelerated GPU cosmological lattice program for solving the evolution of interacting scalar fields
in an expanding universe, achieving speedups between one and two orders of magnitude in single
precision. In the above works, no discussion on using double precision on GPUs was reported. In MHD
simulations, the support of double precision is important, especially for nonlinear problems. We
will evaluate the performance and accuracy of double precision on GPUs in this work.

In this paper, we present an efficient implementation to accelerate computation of MHD simulations
on GPUs, called {\it GPU-MHD}. To our knowledge, this is the first work describing MHD simulations
on GPUs in detail. The goal of our work is to perform a pilot study on numerically solving the
ideal MHD equations on GPUs. In addition, the trend of today's chip design is moving to streaming
and massively parallel processor models, developing new MHD codes to exploit such architecture is
essential. {\it GPU-MHD} can be easily ported to other many-core platforms such as Intel's upcoming
Larrabee~\cite{Seiler2008}, making it more flexible for the user's choice of hardware. This paper
is organized as follows: A brief description of the CUDA programming model is given in Section 2.
The numerical scheme in which {\it GPU-MHD} adopted is presented in Section 3. In Section 4, we
present the GPU implementation in detail. Numerical tests are given in Section 5. Accuracy
evaluation by comparing single and double precision computation results is given in Section 6.
Performance measurements are reported in Section 7 and visualization of the simulation results is
described in Section 8. We conclude our work and point out the future work in Section 9.

\section{A brief description of the CUDA}
\par

The Compute Unified Device Architecture (CUDA) was introduced by NVIDIA as a general purpose
parallel computing architecture, which includes GPU hardware architecture as well as software
components (CUDA compiler and the system drivers and libraries). The CUDA programming
model~\cite{Kirk2010}~\cite{NVIDIA2009}~\cite{Sanders2011} consists of functions, called {\it
kernels}, which can be executed simultaneously by a large number of lightweight {\it threads} on
the GPU. These threads are grouped into one-, two-, or three-dimensional {\it thread blocks}, which
are further organized into one- or two-dimensional {\it grids}. Only threads in the same block can
share data and synchronize with each other during execution. Thread blocks are independent of each
other and can be executed in any other. A graphics card that supports CUDA, for example, the GT200
GPU~\cite{Lindholm2008}, consisting of 30 streaming multiprocessors (SMs). Each multiprocessor
consists of 8 streaming processors (SPs), providing a total of 240 SPs. Threads are grouped into
batches of 32 called {\it warps} which are executed in single instruction multiple data (SIMD)
fashion independently. Threads within a warp execute a common instruction at a time.

For memory access and usage, there are four types of memory, namely, {\it global memory}, {\it
constant memory}, {\it texture memory} as well as {\it shared memory}. Global memory has a separate
address space for obtaining data from the host CPU's main memory through the PCIE bus, which is
about 8 GB/sec in the GT200 GPU. Any valued stored in global memory can be accessed by all SMs via
load and store instructions. Constant memory and texture memory are cached, read-only and shared
between SPs. Constants that are kept unchanged during kernel execution may be stored in constant
memory. Built-in linear interpolation is available in texture memory. Shared memory is limited (16
KB for GT200 GPU) and shared between all SPs in a MP. For detailed information concerning memory
optimizations, we refer the reader to ``CUDA Best Practice Guide"~\cite{NVIDIA2009c}.

Double precision is one important concern in many computational physics applications, however,
support of double precision is limited to the NIVDIA cards having Compute Capability 1.3 (See
Appendix A in~\cite{NVIDIA2009}) such as the GTX 260, GTX 280, Quadro FX 5800 (contains one GT200
GPU), and Tesla C1060 (contains one GT200 GPU) and S1070 (contains four GT200 GPUs). In GT200 GPU,
there are eight single precision floating point (FP32) arithmetic logic units (ALUs) (one per SP)
in SM, but only one double precision floating point (FP64) ALU (shared by eight SPs). The
theoretical peak performance of GT200 GPU is 936 GFLOPS in single precision and 78 GFLOPS in double
precision. In CUDA, double precision is disabled by default, ensuring that all double numbers are
silently converted into float numbers inside kernels and any double precision calculations computed
are incorrect. In order to use double precision floating point numbers, we need to call
\verb=nvcc=: ``\verb=-arch= \verb=== \verb=sm_13=". The flag ``\verb=-arch= \verb=== \verb=sm_13="
in the command tells ``\verb=nvcc=" to use the Compute Capability 1.3 which means enabling the
double precision support. The recent Fermi architecture~\cite{NVIDIA2009c} (GTX 480, for example)
significantly improves the performance of double precision calculations by introducing better
memory access mechanisms.

In Sections 6 and 7 we compare the accuracy and actual performance of {\it GPU-MHD} in single and
double precision on both GT200 and Fermi architectures.

\section{Numerical scheme}\label{MHD}
\par

The ideal MHD equations with the assumption of the magnetic permeability $\mu=1$ can be represented
as hyperbolic system of conservation laws as follows~\cite{Goedbloed2004}
\begin{align}
 \frac{\partial \rho}{\partial t} + {\grad\cdot} \left(\rho{\bv}\right) & = 0\\
%
\frac{\partial \rho {\bv}}{\partial t} + {\grad\cdot} \left(\rho{\bv\bv} - {\bB\bB}\right) +
{\grad} P^* & = 0 \\
\frac{\partial {\bB}}{\partial t} -
{\grad\times} ({\bv\times\bB}) & = 0 \\
%
\frac{\partial E}{\partial t} + \grad\cdot((E + P^*) {\bv} - {\bB} ({\bB \cdot \bv})) & = 0
\end{align}

Here, $\rho$ is the mass density, $\rho\bv$ the momentum density, $\bB$ the magnetic field, and $E$
the total energy density. The total pressure $P^* \equiv P + \frac{B^2}{2}$ where $P$ is the gas
pressure that satisfies the equation of state, $P \equiv (\gamma - 1)(E - \rho\frac{ v^2}{2} -
\frac{B^2}{2})$. In addition, the MHD equations should obey the divergence-free constraint
${\grad\cdot\bB}=0$.

Over the last few decades, there has been a dramatic increase in the number of publications on the
numerical solution of ideal MHD equations. In particular the development of shock-capturing
numerical methods for ideal MHD equations. We do not provide an exhaustive review of the literature
here. A comprehensive treatment of numerical solution of MHD equations can be found
in~\cite{Kulikovskii2001}, for example. Pen {\em et al.}~\cite{Pen2003} proposed a free, fast,
simple, and efficient total variation diminishing (TVD) MHD code featuring modern high-resolution
shock capturing on a regular Cartesian grid. This code is second-order accuracy in space and time
and enforces the ${\grad\cdot \bB}=0$ constraint to machine precision and it was successfully used
for studying nonradiative accretion onto the supermassive black hole~\cite{Pen2003b} and fast
magnetic reconnection~\cite{Pang2010}. Due to these advantages and convenience for GPU verse CPU
comparison, the underlying numerical scheme in {\it GPU-MHD} is based on this work. A detailed
comparison of shock capturing MHD codes can be found in~\cite{Toth1996}, for example. We plan to
explore other recent high-order Godunov schemes such as~\cite{Lee2009} and~\cite{Stone2009} for
{\it GPU-MHD} as our future work.

We briefly review the numerical scheme~\cite{Pen2003} we adopted in {\it GPU-MHD} here. In this
numerical scheme, the magnetic field is held fixed first and then the fluid variables are updated.
A reverse procedure is then performed to complete a one time step. Three dimensional problem is
split into one-dimensional sub-problems by using a Strang-type directional
splitting~\cite{Strang1968}.

Firstly, we describe the fluid update step in which the fluid variables are updated while holding
the magnetic field fixed. The magnetic field is interpolated to cell centers for second-order
accuracy. By considering the advection along the $x$ direction, the ideal MHD equations can be
written in flux-conservative vector form as
\begin{equation}\label{eq:euler}
\frac{\partial{\bu}}{\partial t} + \frac{\partial{\bF}(u)}{\partial x} = 0
\end{equation}

where the flux vector is given by
\begin{equation}\label{eq:flux}
{\bF} = \left( \begin{array}{c}
\rho v_x \\
\rho v_x^2 + P^* - B_x^2 \\
\rho v_x v_y - B_x B_y \\
\rho v_x v_z - B_x B_z \\
(E + P^*) v_x - B_x {\bf B} \cdot {\bf v}
\end{array} \right)
\end{equation}

Equation (\ref{eq:euler}) is then solved by Jin \& Xin's relaxing TVD method~\cite{Jin1995}. With
this method, a new variable ${\bw} = {\bF}({\bu})/c$ is defined, where $c(x, t)$ is a free positive
function called the {\em flux freezing speed}. For ideal MHD equations, we have
${\bu}=(u_1,u_2,u_3,u_4,u_5)=(\rho,\rho v_x,\rho v_y,\rho v_z,E)$ and equations
\begin{align}
\frac{\partial {\bu}}{\partial t} + \frac{\partial}{\partial x}(c{\bw}) & = 0\\[8pt]
\frac{\partial {\bw}}{\partial t} + \frac{\partial}{\partial x}(c{\bu}) & = 0
\end{align}

These equations can be decoupled through a change of left- and right-moving variables
${\bu}^R=({\bu} + {\bw})/2$ and ${\bu}^L=({\bu} - {\bw})/2$
\begin{align}\label{eq:lr}
\frac{\partial {\bu}^R}{\partial t} + \frac{\partial}{\partial x}(c{\bu}^R) & = 0\\[8pt]
\frac{\partial {\bu}^L}{\partial t} - \frac{\partial}{\partial x}(c{\bu}^L) & = 0
\end{align}

The above pair of equations is then solved by an upwind scheme, separately for right- and
left-moving waves, using cell-centered fluxes. Second-order spatial accuracy is achieved by
interpolating of fluxes onto cell boundaries using a monotone upwind schemes for conservation laws
(MUSCL)~\cite{vanLeer1979} with the help of a flux limiter. Runge-Kutta scheme is used to achieve
second-order accuracy of time integration.

We denote ${\bu}_n^t$ as the cell-centered values of the cell $n$ at time $t$, ${\bF}_n^t$ as the
cell-centered flux in cell $n$,  As an example, we consider the positive advection velocity,
negative direction can be obtained in a similar way. We obtain the first-order upwind flux
$F_{n+1/2}^{(1),t}$ from the averaged flux $F_n^t$ in cell $n$. Two second-order flux corrections
can be defined using three local cell-centered fluxes as follows
\begin{align}\label{eq:Delta_F}
\Delta\bF_{n + 1/2}^{L, t} & = \frac{\bF_n^t - \bF_{n - 1}^t}{2}\\[8pt]
\Delta\bF_{n + 1/2}^{R, t} & = \frac{\bF_{n + 1}^t - \bF_n^t}{2}
\end{align}

When the corrections have opposite signs,there is no second-order correction in the case of near
extrema. With the aid of a flux limiter $\phi$ we then get the second-order correction
\begin{equation}\label{eq:Delta_F_n}
\Delta\bF_{n+1/2}^t=\phi(\Delta\bF_{n+1/2}^{L,t},\Delta\bF_{n+1/2}^{R,t})
\end{equation}
The van Leer limiter~\cite{vanLeer1974}
\begin{equation}\label{eq:limiter}
{\rm vanleer} (a, b) = \frac{2ab}{a + b}
\end{equation}
is used in {\it GPU-MHD}. By adding the second-order correction to the first-order fluxes we obtain
second-order fluxes. For example, the second-order accurate right-moving flux ${\bF}_{n + 1/2}^{R,
t}$ can be calculated
\begin{equation}\label{eq:F_n}
{\bF}_{n + 1/2}^{R, t} = {\bF}_n^t + \Delta\bF_{n+1/2}^t
\end{equation}

The time integration is performed by calculating the fluxes $\bF(u_n^{t})$ and the freezing speed
$c_n^t$ in the first half time step is given as follows
\begin{equation}\label{eq:Delta_u_half}
{\bu}_n^{t+\Dt/2}={\bu}_n^t-\left(\frac{{\bF}_{n+1/2}^t-{\bF}_{n-1/2}^t}{\Dx}\right)\frac{\Dt}{2}
\end{equation}
where${\bF}_{n+1/2}^t={\bF}_{n+1/2}^{R,t}-{\bF}_{n+1/2}^{L,t}$ is computed by the first-order
upwind scheme. By using the second-order TVD scheme on ${\bu}_n^{t+\Dt/2}$, we obtain the full time
step ${\bu}_n^{t+\Dt}$
\begin{equation}\label{eq:Delta_u_full}
{\bu}_n^{t + \Dt}={\bu}_n^t - \left(\frac{{\bF}_{n + 1/2}^{t + \Dt/2} - {\bF}_{n - 1/2}^{t +
\Dt/2}}{\Dx}\right)\Dt\ ,
\end{equation}

To keep the TVD condition, the flux freezing speed $c$ is the maximum speed information can travel
and should be set to $|v_x|+(\gamma p/\rho + B^2/\rho)^{1/2}$ as the maximum speed of the fast MHD
wave over all directions is chosen. As the time integration is implemented using a second-order
Runge-Kutta scheme, the time step is determined by satisfying the CFL condition
\begin{equation}\label{equat_CFL}
\begin{aligned}
c_{max} & = [max(|v_{x}|, |v_{y}|, |v_{z}|) + (\gamma p/\rho + B^2/\rho)^{1/2}]\\
\Delta t & = cfl / c_{max}
\end{aligned}
\end{equation}
where $cfl$ is the Courant-Number and $cfl \simless 1$ is generally set to $cfl \simeq 0.7$ for
stability, and $B$ is the magnitude of the magnetic field. Constrained transport
(CT)~\cite{Evans1988} is used to keep the ${\nabla\cdot\bB}=0$ to machine precision. Therefore, the
magnetic field is defined on cell faces and it is represented in arrays~\cite{Pen2003}
\begin{equation}\label{equat_b}
\begin{aligned}
Bx(i, j, k) & = (B_x)_{i - 1/2, j, k}\\
By(i,j,k) & = (B_y)_{i, j - 1/2, k}\\
Bz(i,j,k) & = (B_z)_{i, j, k - 1/2}
\end{aligned}
\end{equation}
where the cell centers are denoted by $(i, j, k) \equiv (x_i, y_j, z_k)$, and faces by $(i\pm 1/2,
j, k)$, $(i, j\pm 1/2, k)$, and $(i,j,k\pm 1/2)$, etc. The cells have unit width for convenience.

Secondly, we describe the update of the magnetic field in separate two-dimensional
advection-constraint steps along $x$-direction while holding the fluid variables fixed. The
magnetic field updates along $y$ and $z$-directions can be handled in a similar matter. We follow
the expressions used in~\cite{Kappeli2009}. For example, we can calculate the averaging of $v$
along $x$ direction as follows
\begin{equation}\label{eq:averaging_x}
  (v_x)_{i, j + 1/2, k} = \frac{1}{4}
  \left[    \left(v_x\right)_{i + 1, j + 1/2, k}
        + 2 \left(v_x\right)_{i, j + 1/2, k}
        +   \left(v_x\right)_{i - 1, j + 1/2, k} \right]
\end{equation}
A first-order accurate flux is then obtained by
\begin{equation}\label{eq:first_flux}
  \left(v_x B_y \right)_{i + 1/2, j + 1/2, k}
  = \left\{
    \begin{array}{l l l}
      \left(v_x B_y \right)_{i, j + 1/2, k}   & , \; & (v_x)_{i + 1/2, j + 1/2, k} > 0 \\
      \\
      \left(v_x B_y \right)_{i + 1, j + 1/2, k} & , \; & (v_x)_{i + 1/2, j + 1/2, k} \leq 0 \\
    \end{array}
  \right.
\end{equation}
where the velocity average is
\begin{equation}
  (v_x)_{i + 1/2, j + 1/2, k} = \frac{1}{2}
  \left[  \left(v_x\right)_{i, j + 1/2, k}
        + \left(v_x\right)_{i + 1, j + 1/2, k}\right]
\end{equation}

$B_x$ is updated by constructing a second-order-accurate upwind electromotive force (EMF) $v_yB_x$
using Jin \& Xin's relaxing TVD method~\cite{Jin1995} in the advection step. Then this same EMF is
immediately used to update $B_y$ in the constraint step.

Extension to three dimensions can be done through a Strang-type directional
splitting~\cite{Strang1968}. Equation (\ref{eq:euler}) is dimensionally split into three separate
one-dimensional equations. For a time step $\Delta t$, let ${fluid}_x$ be the fluid update along
$x$, $B_{x \rightarrow y}$ be the update of $B_x$ along $y$, and $L_i$ be the update operator of
$\bu^t$ to $\bu^{t + \Delta t}$ by including the flux along $i$ direction. Each $L_i$ includes
three update operations in sequence, for example, $L_x$ includes ${fluid}_x$, $B_{y \rightarrow
x}$, and $B_{z \rightarrow x}$. A forward sweep and a reverse sweep are defined as $\bu^{t + \Delta
t} = L_zL_yL_x\bu^t$ and $\bu^{t + 2\Delta t} = L_xL_yL_z\bu^{t + \Delta t}$, respectively. A
complete update combines a forward sweep and reverse sweep. The dimensional splitting of the
relaxing TVD can be expressed as follows~\cite{Trac2003}
\begin{align}\label{eq:L_sequence}
\bu^{t_2}& = \bu^{t_1 + 2\Delta t_1} = L_xL_yL_zL_zL_yL_x\bu^{t_1}\\[8pt]
\bu^{t_3}& = \bu^{t_2 + 2\Delta t_2} = L_zL_xL_yL_yL_xL_z\bu^{t_2}\\[8pt]
\bu^{t_4}& = \bu^{t_3 + 2\Delta t_3} = L_yL_zL_xL_xL_zL_y\bu^{t_3}
\end{align}
where $\Delta t_1$, $\Delta t_2$, and $\Delta t_3$ are sequential time steps after each double
sweep. For Cartesian coordinate system, it is easy to apply Strang-type directional
splitting~\cite{Strang1968} on a high-dimensional problem and split it into one-dimensional
sub-problems in Cartesian coordinate system~\cite{LeVeque2002}. In principle, we can also apply
directional splitting for cylindrical or spherical coordinate systems. We may need to split the
edges of grid in any direction into equal-distance pieces and determine the positions of the cell
centers and face centers. Similar techniques from Li and Li~\cite{Li2004} can be utilized to extend
the usage of directional splitting for cylindrical or spherical coordinate systems. This extension
will be left as our future work.

\section{GPU implementation}
\par

In this section, we provide the implementation details of {\it GPU-MHD}. With {\it GPU-MHD}, all
computations are performed entirely on GPUs and all data is stored in the GRAM of the graphics
card. Currently, {\it GPU-MHD} works on a regular Cartesian grid and supports both single and
double precision modes. Considering the rapid development of graphics hardware, our GPU
implementation was design general enough for GT200 architecture (GTX 295 in our study) and Fermi
architecture (GTX 480 in our study). Therefore, {\it GPU-MHD} can be used on newer architectures
without significant modification.

Before we explain our GPU implementation in detail, the consideration and strategy of our design is
presented first. During the computational process, the TVD numerical scheme for solving the MHD
equations will generate many intermediate results such as the ``flux" and some interpolated values
of each grid point. These intermediate results will then be used in the next calculation step. One
important thing is not only these intermediate results of the current grid point but also those of
the neighboring grid points are needed to be stored. This means the intermediate results of the
neighboring grid points have to be calculated before going to the next calculation step. As a
result, each calculation step in the algorithm was designed with one or several kernels and huge
amount of data should be stored. In order to avoid the data transmission between CPU and GPU during
the computation, {\it GPU-MHD} was designed to be run entirely on GPUs. To reduce the memory usage,
the storage for the intermediate results will be reused to store the intermediate results generated
by the next step. The eight components ($u_1, u_2, u_3, u_4, u_5, B_x, B_y, B_z$) for solving the
MHD equations are stored in the corresponding eight arrays. Each component of a grid point is
stored close to the same component of the neighboring gird points. In any calculation step, only
the necessary component of a calculation (kernel) will be accessed, thus providing more effective
I/O. The strategy of our design is summarized as follows:
\begin{itemize}
\item
Each step of the numerical scheme is handled with one or several kernels to exploit the parallelism of GPUs;\\
\item
Storage of the intermediate results are reused to reduce memory usage;\\
\item
Components of the MHD equations are stored in separate arrays to provide effective memory access.
\end{itemize}

\subsection{Memory arrangement}

Although shared memory provides much faster access rate than global memory, its size is very
limited (16 kB in GTX 295 and 48 kB in GTX 480). As we have to process many intermediate results in
each calculation step, shared memory is too small to fit in our GPU implementation. Of course there
are some techniques of using shared memory, the basic idea is to copy the data from global memory
to the shared memory first, and then use the data in shared memory to do calculations. After the
calculations have been completed, write these results back to global memory. This will benefit
those computations that need many data accesses during the calculation period. However, as we
mentioned in the beginning of this section, due to the nature of the algorithm, {\it GPU-MHD} was
designed with many separated CUDA kernels. Calculation of each kernel actually is simple and
variables of grid points in each kernel are mostly accessed only once (read) or twice (read and
then write the result). In order to provide fast access speed, parameters and temporary results
(generated and used only within kernel) in each kernel are stored with registers. The parameters
for the whole simulation such as the data size and size of dimensions are stored using constant
memory. Thus in our case, shared memory does not show its advantages . On the other hand, the size
of shared memory is too small for our problem, especially when double precision is used in the
calculations. We did try to use shared memory in {\it GPU-MHD} by coping the capable amount of data
to shared memory for the calculations, but there is no speedup compared to our current approach.
Therefore, our code mainly uses global memory. There are three phases in our code: transform of
data from the host memory into the global memory, execution of the kernels, and transfer of data
from the GPU into the host memory.

\begin{figure}[hbt]
\begin{center}
\includegraphics*[width=5.0in]{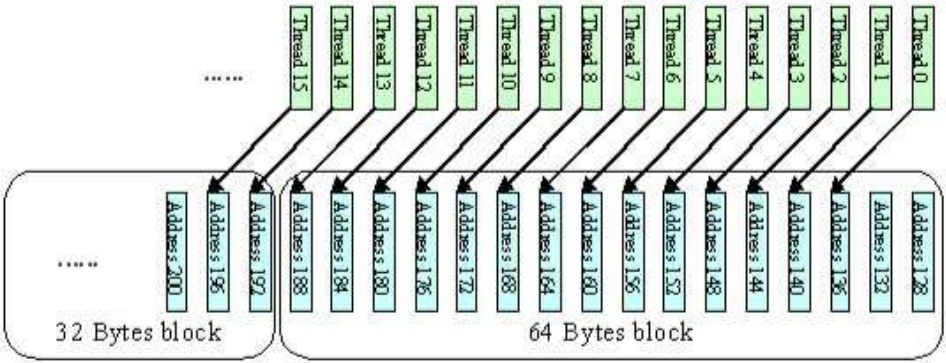}
\hfill \caption{An example demonstrating the automatic coalescing in GT200
architecture.}\label{fig:GTX200_coalescing}
\end{center}
\end{figure}

For global memory, if the data is well organized in global memory with the form that a load
statement in all threads in a warp accesses data in the same aligned 128-byte block, then the
threads can efficiently access data from the global memory. The process of organizing the data in
such a form is so called coalescing~\cite{NVIDIA2009b}~\cite{Kirk2010}. Actually, GT200
architecture (with Compute Capability 1.2 or 1.3) has more flexible in handling data in global
memory than those cards with Compute Capability 1.1 or lower. Coalescing of loading and storing
data that are not aligned perfectly to 128-byte boundaries is handled automatically on this
architecture (see Appendix G.3.2.2 in~\cite{NVIDIA2009b}). We illustrate this new feature in
Figure~\ref{fig:GTX200_coalescing}. GT200 architecture supports 32 bytes memory block and has less
limitation to memory address, which is accessed by the header (first) thread. Even without
``shifting" the address to aligned 64 bytes or 128 bytes, the GPU kernels can still keep good
performance, especially when we only process with $2^{n}$ data.

\begin{figure}[h]
\begin{center}
\includegraphics*[width=4 in]{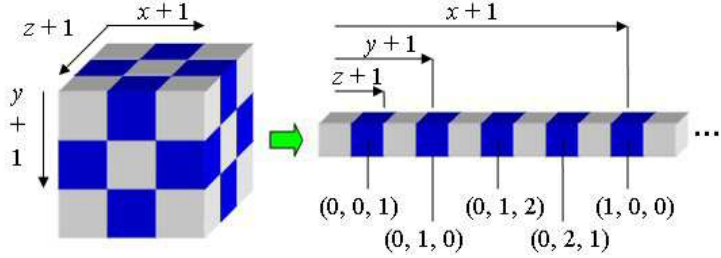}
\hfill \caption{Mapping from 3D array to 1D array in column major.}\label{fig:matrix}
\end{center}
\end{figure}

The memory arrangement of {\it GPU-MHD} is presented here. The most intuitive way to write a
parallel program to solve a multidimensional problem is to use multidimensional arrays for data
storage and multidimensional threads for computation. However, the ability of the current CUDA is
limited in supporting multidimensional threads, therefore, we could not implement our code in such
a straightforward way. Especially in three dimensions or higher dimensions, there are still some
limitations in handling multidimensional arrays and multidimensional threads. As a result, the most
primitive way is to store data in one-dimension and perform parallel computation with one-dimension
threads. By using an indexing technique, our storage and threading method can be extended to to
solve multidimensional problems. Our data storage arrangement is expressed in Fig.~\ref{fig:matrix}
and in Equations (\ref{index1}) to (\ref{index2}).
\begin{equation}
\label{index1} \left\{
\begin{array}{c} INDEX_{x} \\
INDEX_{y} \\ INDEX_{z}
\end{array}
\begin{array}{ccl} = index &/& (SIZE_{y} \times SIZE_{z}) \\
\mbox{ = [}index & \mathbf{mod} & (SIZE_{y} \times SIZE_{z})\mbox{]} / SIZE_{z} \\
= index & \mathbf{mod} & SIZE_{z}
\end{array}\right .
\end{equation}

\begin{align}
INDEX_{x} \pm 1 & = index \pm (SIZE_{y} \times SIZE_{z})\\[8pt]\label{index2}
INDEX_{y} \pm 1 & = index \pm SIZE_{z}\\[8pt]
INDEX_{z} \pm 1 & = index \pm 1\label{index3}
\end{align}

Here $INDEX_{x}$, $INDEX_{y}$, and $INDEX_{z}$ are the indexes of a 3D matrix. $index$ is the 1D
index used in {\it GPU-MHD}, $SIZE_{y}$, and $SIZE_{z}$ are the matrix size (number of grid points
in our study) of a 3D matrix.

Equation (\ref{index1}) expresses the mapping of three-dimensional (3D) indexes to one-dimensional
(1D) indexes. Equations (\ref{index2}) to (\ref{index3}) express the shift operations. Shift
operations are very important in numerical solution of conservation laws because some calculations
are based on the neighboring grid points. The above indexing technique is used to prepare suitable
values (vectors) as input values for the calculation kernels we implemented in CUDA. As an example,
we give a conceptual calculation kernel for a calculation in $x$-dimension to show how the indexing
technique works for this task in the following. This kernel calculates the result with the grid
point itself and neighboring gird points in $x$-dimension. The calculations in $y$- or
$z$-dimension have the similar form.

\begin{boxedverbatim}
Calculate_X(data, result) {
   index = getID(); //self-increment index for multi-threading
   grid_point_index = index; //(x, y, z)
   neighbor_1 = grid_point_index + (SIZEy * SIZEz); //(x + 1, y, z)
   neighbor_2 = grid_point_index - (SIZEy * SIZEz); //(x - 1, y, z)

   calculate_kernel(data, result, grid_point_index,
                    neighbor_1, neighbor_2, ...);
   ......
}
\end{boxedverbatim}
The indexing technique is a common way to represent multidimensional arrays using 1D arrays by
mapping a 3D index $(x, y, z)$ to an 1D index $(x \times Y_{size} \times Z_{size} + y \times
Z_{size} + z)$. The GPU kernels of TVD were designed such that each kernel calculates using the
actual index of a particular grid point and its neighbors. For example, if the calculation needs
the information in a particular gird point and its neighboring grid points in $z$-dimension, then
the indexing operation will retrieve $[x \times Y_{size} \times Z_{size} + y \times Z_{size} + (z -
1)]$, $[x \times Y_{size} \times Z_{size} + y \times Z_{size} + z]$ and $[z \times Y_{size} \times
Z_{size} + y \times Z_{size} + (z + 1)]$ and proceed the calculation. If the calculation needs the
information in a particular grid point and its neighboring grid points in $y$-dimension, then the
indexing operation will retrieve $[x \times Y_{size} \times Z_{size} + (y - 1) \times Z_{size} +
z]$, $[x \times Y_{size} \times Z_{size} + y \times Z_{size} + z]$ and $[x \times Y_{size} \times
Z_{size} + (y + 1) \times Z_{size} + z]$. Then these resulting indexes from indexing operation will
pass to the GPU kernels of TVD for performing the calculation. As a result, for $N$-dimension
problem, what we need are $N$-dimension indexing operation kernels while only one TVD kernel is
needed at all the time.

For efficiency, {\it GPU-MHD} only processes the problem with the number of grid points satisfying
$2^{n}$ condition. One reason is that the size of a warp of GPU contains 32 threads, problems with
grid point number of $2^{n}$ are easier to determine the number of threads and blocks to fit in
multiple of a warp before the GPU kernel is called. That means we do not need to check if the ID of
the grid point being processed (calculated by the block ID and thread ID) is out of the range. It
is very helpful in making the GPU code run more efficient. On the other hand, it is also effective
to reduce logical operations in a GPU kernel, which is known to be a little bit slow in the current
GPU architecture. As a result, warp divergence caused by the number of the data is avoided (there
is still a little bit warp divergence caused by ``if" operation in the calculation of our
algorithm). Similar method is used in the CUDA SDK code sample ``reduction".

The actual memory pattern used in {\it GPU-MHD} will be presented at the end of next subsection
after introducing our algorithm.
\subsection{Program flow}
A ``CUDA kernel" is a function running on GPU~\cite{Kirk2010}~\cite{NVIDIA2009}~\cite{Sanders2011}.
Noted that the CUDA kernel will process all grid points in parallel, therefore, a \verb=For=
instruction is not needed for going through all grid points. {\it GPU-MHD} includes the following
steps:
\begin{enumerate}
\item CUDA initialization
\item Setup the initialize condition for the specified MHD problem:\\
${\bu}=(u_1, u_2, u_3, u_4, u_5)$ of all grid points, ${\bB}=(B_{x}, B_{y}, B_{z})$ of cell faces,
and set parameters such as time $t$, etc.
\item Copy the the initialize condition ${\bu}$, ${\bB}$ to device memory (CUDA global memory)
\item For all grid points, calculate the $c_{max}$ by Equation (\ref{equat_CFL}) (implemented with a CUDA kernel)
\item Use \verb=cublasIsamax= (in single precision mode) function or
\verb=cublasIdamax= (in double precision mode) function of the CUBLAS library to find out the
maximum value of all $c_{max}$, and then determine the $\Delta t$
\item Since the value of $\Delta t$ is stored in device memory, read it back to host memory (RAM)
\item Sweeping operations of the relaxing TVD (Calculation of the $L_{i}, i = {x, y, z}$, implemented
with several CUDA kernels, will be explained in the next subsection)
\item $t = t + 2\Delta t$
\item \verb=If= $t$ reaches the target time, go to next step\\
\verb=else= repeats the procedure from step (4)
\item Read back data ${\bu}$, ${\bB}$ to host memory
\item Output the result
\end{enumerate}

\begin{figure}[h]
\begin{center}
\includegraphics*[width=4.0in]{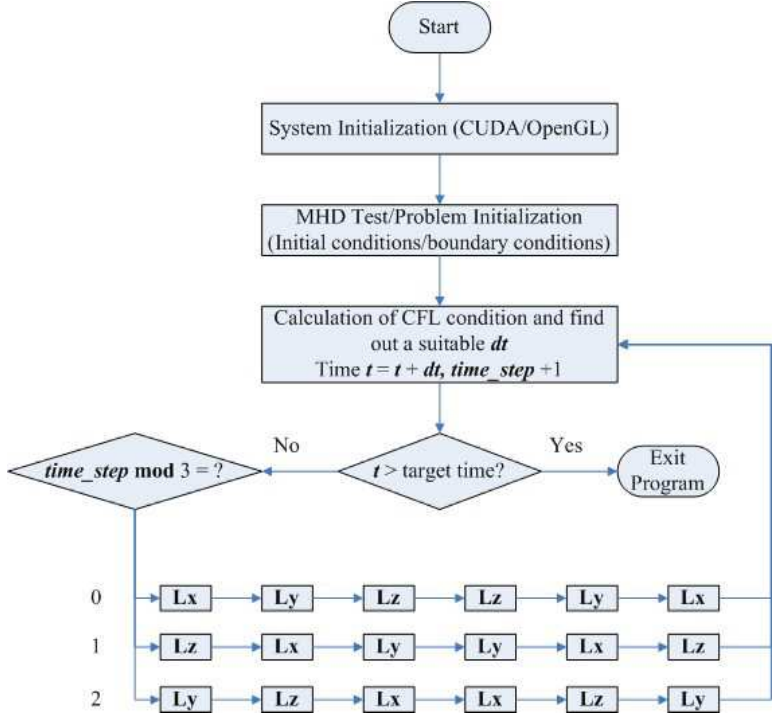}
\hfill \caption{The flow chart of {\it GPU-MHD}.} \label{fig:FlowChart}
\end{center}
\end{figure}

The program flow of {\it GPU-MHD} is shown in Fig.~\ref{fig:FlowChart}. After the calculation of
the CFL condition, the sweeping operations will be performed. The sweeping operation $L_{i}$ will
update both the fluid variables and orthogonal magnetic fields along $i$ dimension. This is a core
computation operation in the relaxing TVD scheme described in Section~\ref{MHD}.

The CFL condition for the three-dimensional relaxing TVD scheme is obtained by Equation
(\ref{equat_CFL}). The procedure is to calculate all the $c_{max}$ of each grid point and find out
the maximum value. In {\it GPU-MHD}, parallel computation power of CUDA is exploited to calculate
the $c_{max}$ of each grid point in parallel and all the $c_{max}$ values are stored in a matrix.
Then the \verb=cublasIsamax= function is used (in double precision mode, the \verb=cublasIdamax=
function is used) to find out the maximum $c_{max}$ of the matrix in parallel (called the reduction
operation). The \verb=cublasIsamax= function is provided in the CUBLAS library
--- a set of basic operations for vector and matrix provided by NVIDIA with the CUDA
toolkit~\cite{NVIDIA2009}. The reason we read the $\Delta t$ back and store both $\Delta t$ and $t$
in host memory is due to the data in device memory cannot be printed out directly in the current
CUDA version. This information is useful for checking if there is any problem during the simulation
processing. The implementation of sweeping operations will be explained in the next subsection.

\subsection{Sweeping operations}\label{sec:sweeping}

Before we start to describe the sweeping operations, consideration of memory arrangement is
presented first in the following.

Implementing parallel computation using CUDA kernels is somewhat similar to parallel implementation
on a CPU-cluster, but it is not the same. The major concern is the memory constrain in GPUs. CUDA
makes parallel computation process on GPUs which can only access their graphics memory (GRAM).
Therefore, data must be stored in GRAM in order to be accessed by GPUs. There are several kinds of
memory on graphics hardware including registers, local memory, shared memory, and global memory,
etc., and they have different characteristics and usages~\cite{NVIDIA2009}, making memory
management of CUDA quite different compared to parallel computation on a CPU-cluster. In addition,
even the size of GRAM in a graphics card increases rapidly in newer models (for example, the latest
NVIDIA graphics card --- GeForece GTX 295 has 1.75G GRAM), but not all the capacity of GRAM can be
used to store data arbitrarily. Shared memory and local memory are flexible to use, however, their
sizes are very limited in a block and thus they cannot be used for storing data with large size. In
general, numerical solution of conservation laws will generate many intermediate results (for
example, ${\bu}^{t + \Delta t/2}$, ${\bF}$, $c$, $w$, etc.) during the computation process, these
results should be stored for subsequent steps in the process. Therefore, global memory was mainly
used in {\it GPU-MHD}.

\begin{figure}[hbt]
\begin{center}
\includegraphics*[width=4in]{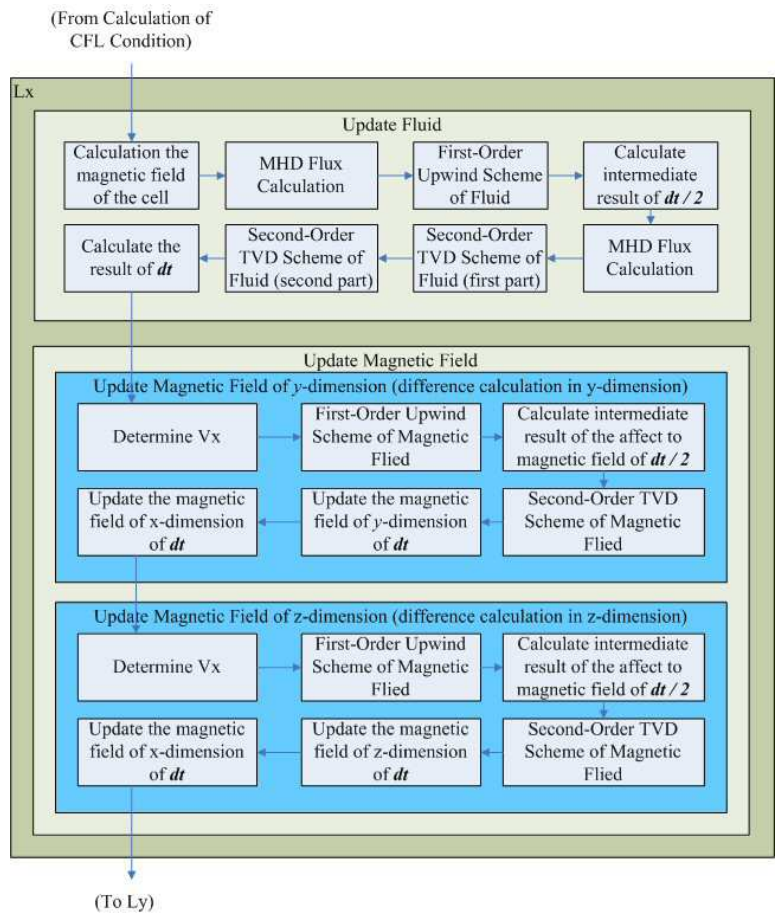}
\hfill \caption{Calculation process of $L_x$.} \label{fig:Lx} \end{center}
\end{figure}

After the maximum value of $c_{max}$ in Equation (\ref{equat_CFL}) is found, we can get the $\Delta
t$ by determining the Courant-Number ($cfl$). The sequential step is the calculation of $L_{i}$ ($i
= x, y, z$). The implementation of $L_{i}$ includes two parts: update the fluid variables and
update the orthogonal magnetic fields. As an example, the process for calculating $L_x$ is shown in
Fig.~\ref{fig:Lx} where each block was implemented with one or several CUDA kernels. The process
for calculating $L_y$ or $L_z$ is almost the same as $L_x$ except that the dimensional indexes are
different.

The first part of the $L_x$ calculation process is ${fluid}_x$. The fluid variables will be updated
along $x$. {\bf Algorithm 1} shows the steps and GPU kernels of this process (the data of ${\bu}$
and ${\bB}$ are already copied to device memory), all the steps are processed on all grid points
with CUDA kernels in parallel.

\begin{algorithm*}
\caption{Algorithm of ${fluid}_x$, all equations and difference calculations are processed using
CUDA kernels}\label{alg:fluid}
\begin{algorithmic}[1]
\STATE load ${\bu}$, ${\bB}$ and $\Delta t$

\STATE memory allocation for the storage of the intermediate results: ${\bB_{temp}}$,
${\bu_{temp}}$, ${\bf flux_{temp}}$, ${\bf other_{temp}}$, (${\bf other_{temp}}$ includes the
storage of ${\bF}$, $c$, $w$, etc)

\STATE ${\bB_{temp}} \leftarrow$ results obtained by Equation (\ref{equat_b}) with ${\bB}$,
(${\bB}$ stored the magnetic field of the cell faces)

\STATE ${\bf other_{temp}} \leftarrow$ results obtained by Equations (\ref{eq:flux}) and
(\ref{eq:lr}) with ${\bu}$

\STATE ${\bf flux_{temp}} \leftarrow$ the flux of a half time step: difference calculation
(``First-Order Upwind Scheme of Fluid" CUDA kernels in Fig.~\ref{fig:Lx}) obtained by Equation
(\ref{eq:Delta_u_half}) using ${\bf other_{temp}}$

\STATE ${\bu_{temp}} \leftarrow$ calculate the intermediate result (${\bu}^{t + \Delta t / 2}$)
using Equation (\ref{eq:Delta_u_half}) with ${\bu}$ and ${\bf flux_{temp}}$

\STATE ${\bf other_{temp}} \leftarrow$ results obtained by Equations (\ref{eq:flux}) and
(\ref{eq:lr}) with ${\bu_{temp}}$ (the same algorithm and same CUDA kernels in Step 4)

\STATE ${\bf flux_{temp}} \leftarrow$ the flux of another half time step: difference calculation
(``Second-Order TVD Scheme of Fluid" CUDA kernels in Fig.~\ref{fig:Lx}) obtained by Equation
(\ref{eq:Delta_u_full}) and the limiter (Equation (\ref{eq:limiter})) using ${\bf other_{temp}}$

\STATE calculate the result of ${\bu}^{t + \Delta t}$ with ${\bf flux_{temp}}$ using Equation
(\ref{eq:Delta_u_full}) and save it back to ${\bu}$

\STATE free the storage of the intermediate results

\STATE (continue to the second part of $L_{x}$, update the orthogonal magnetic fields)
\end{algorithmic}
\end{algorithm*}

In this process, we have to calculate the magnetic fields of the grid point (Equation
\ref{equat_b}) first because all the magnetic fields are defined on the faces of the grid
cell~\cite{Pen2003}. To update the fluid variables of $L_x$, the main process, which includes one
or even several CUDA kernels, is to calculate the affect of the orthogonal magnetic fields to the
fluid variables of Equations (\ref{eq:flux}), (\ref{eq:lr}) and (10). One such main process gives
the flux of the $\Delta t / 2$ step. After two main processes of flux calculation and the other
difference calculations, the value of fluid
--- ${\bu}$ is updated from ${\bu^t}$ to ${\bu^{t+\Delta t}}$ in one $L_{x}$ process.

The second part of the $L_x$ calculation process is to update the orthogonal magnetic fields in
$y$-dimension ($B_{y \rightarrow x}$), and $z$-dimension ($B_{z \rightarrow x}$) with the fluid
along $x$-dimension. The strategy and implementation are similar to those in the first part but
with a different algorithm for the orthogonal magnetic fields.

\begin{algorithm*}
\caption{Algorithm of ($B_{y \rightarrow x}$) and ($B_{z \rightarrow x}$), all equations and
difference calculations are processed using CUDA kernels}\label{alg:magnetic}
\begin{algorithmic}[1]
\STATE (after the processes of fluid, we obtain an updated ${\bu}$)

\STATE load $u_{1}$ (density $\rho$), $u_{2}$ ($\rho v_{x}$), ${\bB}$ and $\Delta t$

\STATE memory allocation for the intermediate results: ${\bB_{temp}}$, ${\bf flux_{temp}}$, ${\bf
vx_{temp}}$ and ${\bf vx_{face}}$

\STATE ${\bf vx_{temp}} \leftarrow$ determine the fluid speed with the updated $u_{1}$ and $u_{2}$
in ${fluid}_x$, with the difference calculated in $y$-dimension

\STATE ${\bf vx_{face}} \leftarrow$ Results obtained by Equation (\ref{eq:averaging_x})

\STATE ${\bf flux_{temp}} \leftarrow$ the flux of a half time step: difference calculation of
``flux of magnetic field in $y$-dimension" (``First-Order Upwind Scheme of Magnetic Field" CUDA
kernels in Fig.~\ref{fig:Lx}) obtained by Equations (\ref{eq:Delta_u_half}) and
(\ref{eq:first_flux}))

\STATE ${\bB_{temp}} \leftarrow$ calculate the intermediate result (${\bu}^{t + \Delta t / 2}$) by
applying Equation (\ref{eq:Delta_u_half}) to $B_{y}$ (not by applying Equation
(\ref{eq:Delta_u_half}) to ${\bu}$) with $B_y$ and ${\bf flux_{temp}}$

\STATE ${\bf flux_{temp}} \leftarrow$ the flux of another half time step: difference calculation
(``Second-Order TVD Scheme of Magnetic Field" CUDA kernels in Fig.~\ref{fig:Lx}) obtained by
Equation (\ref{eq:Delta_u_half}), the limiter of Equation (\ref{eq:limiter}) and Equation
(\ref{eq:first_flux})

\STATE calculate the result of $B_{x}^{t + \Delta t}$ and $B_{z}^{t + \Delta t}$ with ${\bf
flux_{temp}}$ by applying Equation (\ref{eq:Delta_u_full})) to $B_{y}$, and save it back to ${\bB}$

\STATE (the following steps is similar to above steps but the affected orthogonal magnetic field is
changed from $y$ to $z$)

\STATE ${\bf vx_{temp}} \leftarrow$ determine the fluid speed with the updated $u_{1}$ and $u_{2}$
in ${fluid}_x$, with the difference calculated in $z$-dimension

\STATE ${\bf vx_{face}} \leftarrow$ Results obtained with Equation (\ref{eq:averaging_x}) using
index of $i$, $j$, $k + 1/2$

\STATE ${\bf flux_{temp}} \leftarrow$ the flux of a half time step: difference calculation of
``flux of magnetic field in $z$-dimension" (``First-Order Upwind Scheme of Magnetic Flied" CUDA
kernels in Fig.~\ref{fig:Lx}) obtained by Equations (\ref{eq:Delta_u_half}) and
(\ref{eq:first_flux})

\STATE ${\bf b_{temp}} \leftarrow$ calculate the intermediate result (${\bf u}^{t+\Delta t / 2}$)
by applying Equation (\ref{eq:Delta_u_half}) to $B_{z}$ (not by applying Equation
(\ref{eq:Delta_u_half}) to ${\bu}$) with $B_z$ and ${\bf flux_{temp}}$

\STATE ${\bf flux_{temp}} \leftarrow$ the flux of another half time step: difference calculation
(``Second-Order TVD Scheme of Magnetic Flied" CUDA kernels in Fig.~\ref{fig:Lx}) obtained by
Equation (\ref{eq:Delta_u_half}), the limiter of Equation (\ref{eq:limiter}) and Equation
(\ref{eq:first_flux})

\STATE calculate the results of $B_{x}^{t+\Delta t}$ and $B_{z}^{t+\Delta t}$ with ${\bf
flux_{temp}}$ by applying Equation (\ref{eq:Delta_u_full})) to $B_{z}$, and save it back to ${\bB}$

\STATE free the storage of the intermediate results
\end{algorithmic}
\end{algorithm*}

In {\bf Algorithm 1}, the calculations in steps (4) to (9) are the steps for $B_{y \rightarrow x}$,
and steps (11) to (16) are the steps for $B_{z \rightarrow x}$. The steps for $B_{y \rightarrow x}$
and $B_{z \rightarrow x}$ are almost the same, and the only different parts are the dimensional
indexes of the difference calculations, and the affected magnetic fields: $B_{y}$ and $B_{z}$.
After the first part of $L_{x}$ the fluid ${\bu^t}$ is updated to ${\bu^{t+\Delta t}}$. This change
of the fluid affects to the orthogonal magnetic fields. Therefore, the corresponding change (flux)
of orthogonal magnetic fields can be calculated with the density and velocity of the updated fluid
${\bu^{t + \Delta t}}$. Then the orthogonal magnetic fields are also updated to ${\bB_y^{t + \Delta
t}}$ and ${\bB_z^{t + \Delta t}}$, and also, these changes give effects to ${\bB_x}$.

After one process of $L_{x}$, both fluid and magnetic fields are updated to $t + \Delta t$ with the
affect of the flow in $x$-dimension. And a sweeping operation sequence includes two $L_{x}$,
$L_{y}$, and $L_{z}$ (see Equations (\ref{eq:L_sequence})). So we actually get the updated fluid
and magnetic fields of $t + 2\Delta t$ after one sweeping operation sequence. Note that the second
$L_{x}$ in the sequence is a reverse sweeping operation, the order of ${fluid}_x$, $B_{y
\rightarrow x}$ and $B_{z \rightarrow x}$ has to be reversed: $B_{y \rightarrow x}$ and $B_{z
\rightarrow x}$ first, and ${fluid}_x$ second.

As we mentioned before, numerical solution of conservation laws needs lots of memory because there
are many intermediate results generated during the computation process. These intermediate results
should be stored for the next calculation steps which need the information of the neighboring grid
points obtained in the previous calculation steps. Otherwise, in order to avoid the asynchronous
problem in parallel computation, we have to do many redundant processes. This is due to the
processors on GPUs will not automatically start or stop working synchronously. Without storing the
intermediate results, it will be hard to guarantee the values of the neighboring grid points
updated synchronously. With the purpose to minimizing the memory usage, not only the calculation
process of $L_x$ is divided into several steps (CUDA kernels), but also the intermediate results
are stored as little as possible. The processes dealing with the difference calculations are also
divided into several steps to minimize the storage of the intermediate results and to guarantee
there is no wrong result caused by asynchronous problem.

\begin{figure}[h]
\begin{center}
\includegraphics*[width=4in]{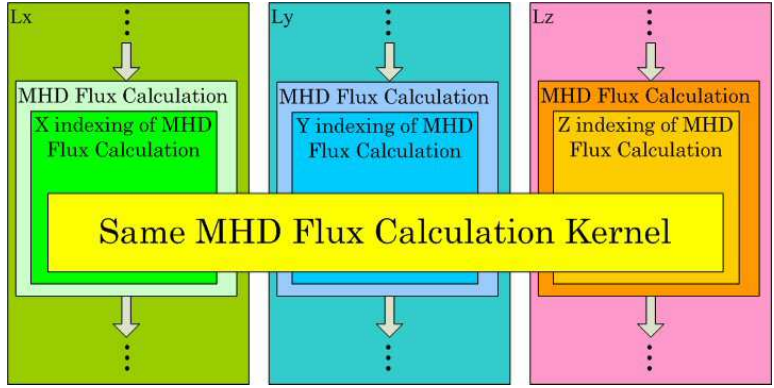}
\hfill \caption{Flux computation in {\it GPU-MHD}.} \label{fig:IndexingKernels}
\end{center}
\end{figure}

It should be realized that most of the processes in the three-dimensional relaxing TVD scheme with
the dimensional splitting technique is similar. Pen {\em et al.}~\cite{Pen2003} swapped the data of
$x$, $y$, and $z$-dimensions while {\it GPU-MHD} used one-dimensional arrays. But the similar
swapping technique can be applied in our case with some indexing operations. Instead of transposing
or swapping the data, we implemented each calculation part of the flux computation with two sets of
CUDA kernels: one set is the CUDA kernels for calculating the relaxing TVD scheme (we call it TVD
kernel here) and the other set is the CUDA kernels actually called by $L_{i}$ operations (we call
them $L_{i}$ kernels here). Indexing operations are contained in all $L_{i}$ kernels. After the
index is calculated, TVD kernels are called and the indexes are passed to the TVD kernels, letting
the TVD kernels calculate the flux of corresponding dimension. Therefore, the difference among
$L_{x}$, $L_{y}$, and $L_{z}$ is the dimensional index. The flux computation of {\it GPU-MHD} is
shown in Fig.~\ref{fig:IndexingKernels}.

The indexing operation swaps the target that will be updated and the neighboring relationship will
also be changed accordingly. For example, the calculation that uses $x + 1$ as the neighboring
element in $L_x$ will be changed to $y + 1$ in $L_y$. As transposing the data in a matrix needs
more processing time, it is efficient and flexible to extend the code to multidimensional by
dividing the indexing operation and flux calculation.

\begin{figure}[hbt]
\begin{center}
\includegraphics*[width=5.0in]{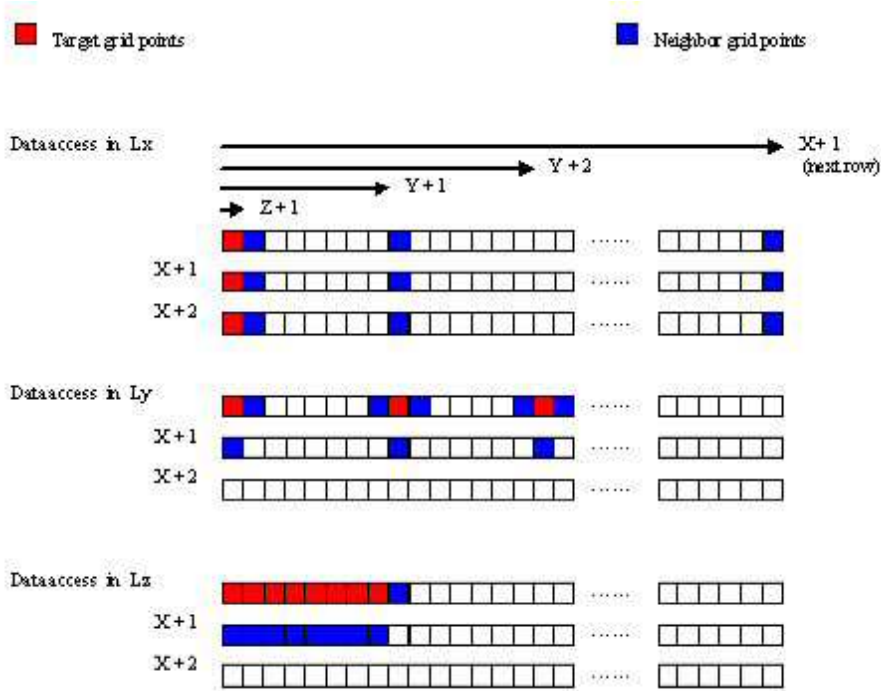}
\hfill \caption{The data accesses of $L_x$, $L_y$ and $L_z$}\label{fig:DataAccess}
\end{center}
\end{figure}

As we mentioned in Section 4.1, the data is stored in 1D array, the data accesses of $L_x$, $L_y$,
and $L_z$ are depicted in Fig.~\ref{fig:DataAccess}. In $L_x$, the data of $(x, y, z), (x, y + 1,
z), (x, y - 1, z), (x, y, z + 1), (x, y, z - 1)$ are used to calculate and update the data of $(x,
y, z)$. The data of $(x + 1, y, z), (x + 1, y + 1, z), (x + 1, y - 1, z), (x + 1, y, z + 1), (x +
1, y, z - 1)$ are used to calculate and update the data of $(x + 1, y, z)$, and so on. Similarly,
in $L_y$, the data of $(x, y, z), (x + 1, y, z), (x - 1, y, z), (x, y, z + 1), (x, y, z - 1)$ are
used to calculate and update the data of $(x, y, z)$. The data of $(x, y + 1, z), (x + 1, y + 1,
z), (x - 1, y + 1, z), (x, y + 1, z + 1), (x, y + 1, z - 1)$ are used to calculate and update the
data of $(x, y + 1, z)$, and so on. In $L_z$, the data of $(x, y, z), (x + 1, y, z), (x - 1, y, z),
(x, y + 1, z), (x, y - 1, z)$ are used to calculate and update the data of $(x, y, z)$. The data of
$(x, y, z + 1), (x + 1, y, z + 1), (x - 1, y, z + 1), (x, y + 1, z + 1), (x, y - 1, z + 1)$ are
used to calculate and update the data of $(x, y, z + 1)$, and so on. It seems that the data
accesses of $L_x$ and $L_y$ will slow down the performance since these accesses are not in so
called ``coalescing" pattern. However, experimental results show that the computational times
spending on calculating each dimensional component such as ${fluid}_x$ and $B_{y \rightarrow x}$ in
$L_x$, $L_y$, and $L_z$ are very close in our current arrangement (see
Tables~\ref{Table_1D_part}~\ref{Table_2D_part}, and~\ref{Table_3D_part} in Section 7). This is due
to the fact that GT200 and Fermi GPU are more flexible to handle the data access that is not
perfectly coalesced (see Section 4.1). Thus we did not further perform the coalescing to make these
data accesses in optimal coalescing pattern.

After the whole pipeline of Fig.~\ref{fig:FlowChart} is completed, the MHD simulation results will
be stored in GRAM and these results are readily to be further processed by the GPU for
visualization or read back to the CPU for other usage. Due to the data-parallel nature of the
algorithm and its high arithmetic intensity, we can expect our GPU implementation will exhibit a
relatively good performance on GPUs.

\section{Numerical tests}
\par

In this section, several numerical tests in one-dimensional (1D), two-dimensional (2D), and
three-dimensional (3D) for validation of {\it GPU-MHD} are given. Two graphics cards NVIDIA GeForce
GTX 295 and GTX 480 were used. GTX 295 has two GPUs inside but only one was used in these numerical
tests. The results shown in this section are computed with single precision mode in {\it GPU-MHD}
on GTX 295. The difference between single precision and double precision computation results will
be discussed in Section~\ref{sec:accuracy}.

\subsection{One-dimensional problems}

\subsubsection{Brio-Wu shock tube}\label{sub:Brio-Wu}

1D Brio-Wu shock tube problem~\cite{Brio1988} which is a MHD version of the Sod
problem~\cite{Sod1978}, consists of a shock tube with two initial equilibrium states as follows

Left side $(x < 0.5)$
\begin{equation}
\label{equat_BW_init1} \left\{\begin{array}{c} v_{x} \\
v_{y} \\ v_{z} \end{array} \right\} \mbox{=} \left\{\begin{array}{c} 0 \\
0 \\ 0 \end{array} \right\}
\end{equation}
\begin{equation}
\label{equat_BW_init2} \left\{\begin{array}{c} B_{x} \\
B_{y} \\ B_{z} \end{array} \right\} \mbox{=} \left\{\begin{array}{c} 0.75 \\
1 \\ 0 \end{array} \right\}
\end{equation}
\begin{equation}
\label{equat_BW_init3} \rho = 1, \,\,\,\,\, p = 1
\end{equation}

Right side $(x \geq 0.5)$
\begin{equation}
\label{equat_BW_init4} \left\{\begin{array}{c} v_{x} \\
v_{y} \\ v_{z} \end{array} \right\} \mbox{=} \left\{\begin{array}{c} 0 \\
0 \\ 0 \end{array} \right\}
\end{equation}
\begin{equation}
\label{equat_BW_init5} \left\{\begin{array}{c} B_{x} \\
B_{y} \\ B_{z} \end{array} \right\} \mbox{=} \left\{\begin{array}{c} 0.75 \\
-1 \\ 0 \end{array} \right\}
\end{equation}
\begin{equation}
\label{equat_BW_init6} \rho = 0.125, \,\,\,\,\, p = 0.1
\end{equation}

\begin{figure}[h]
\begin{center}
\includegraphics*[width=2.5in]{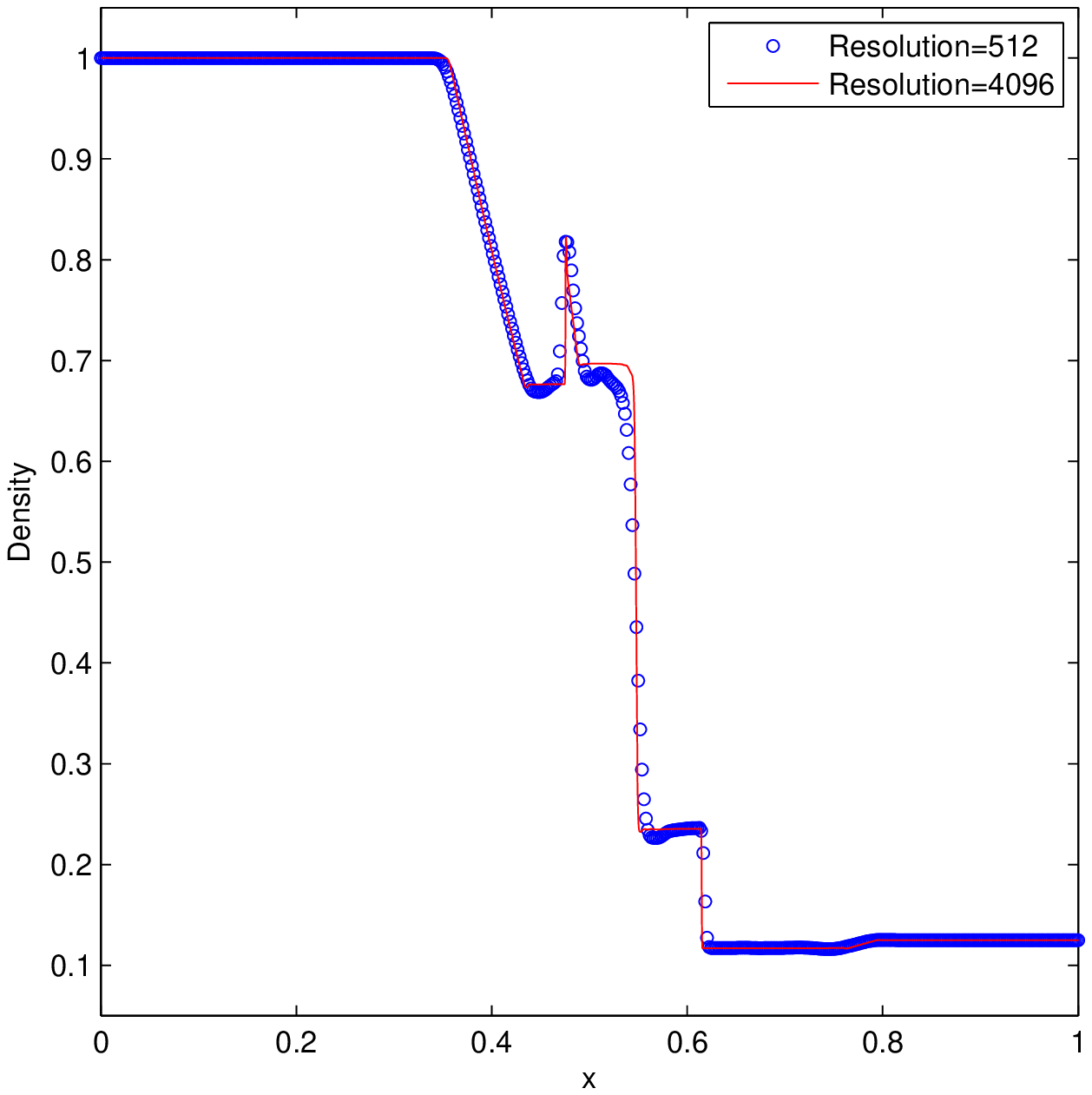}
\includegraphics*[width=2.5in]{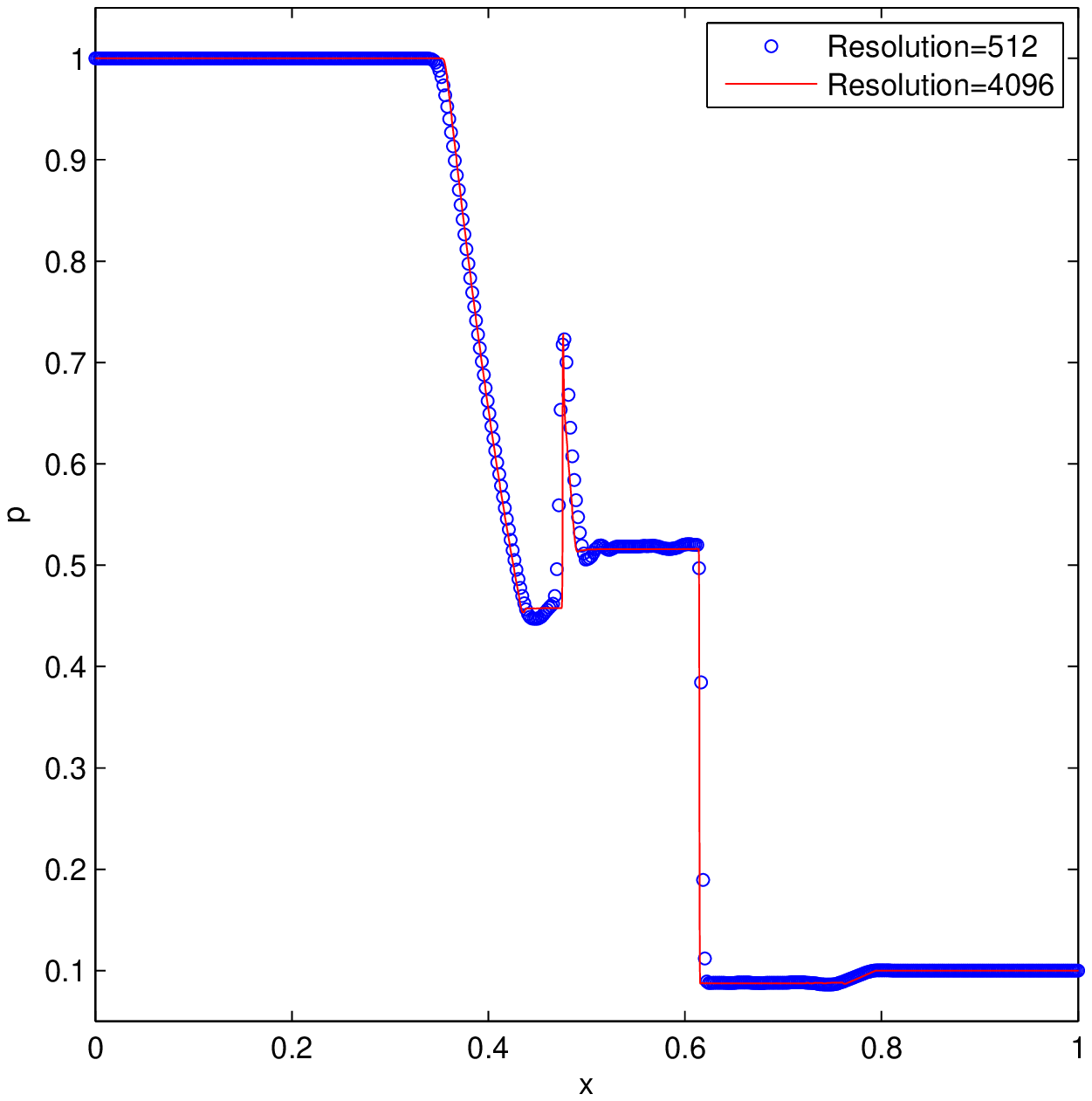}
\includegraphics*[width=2.5in]{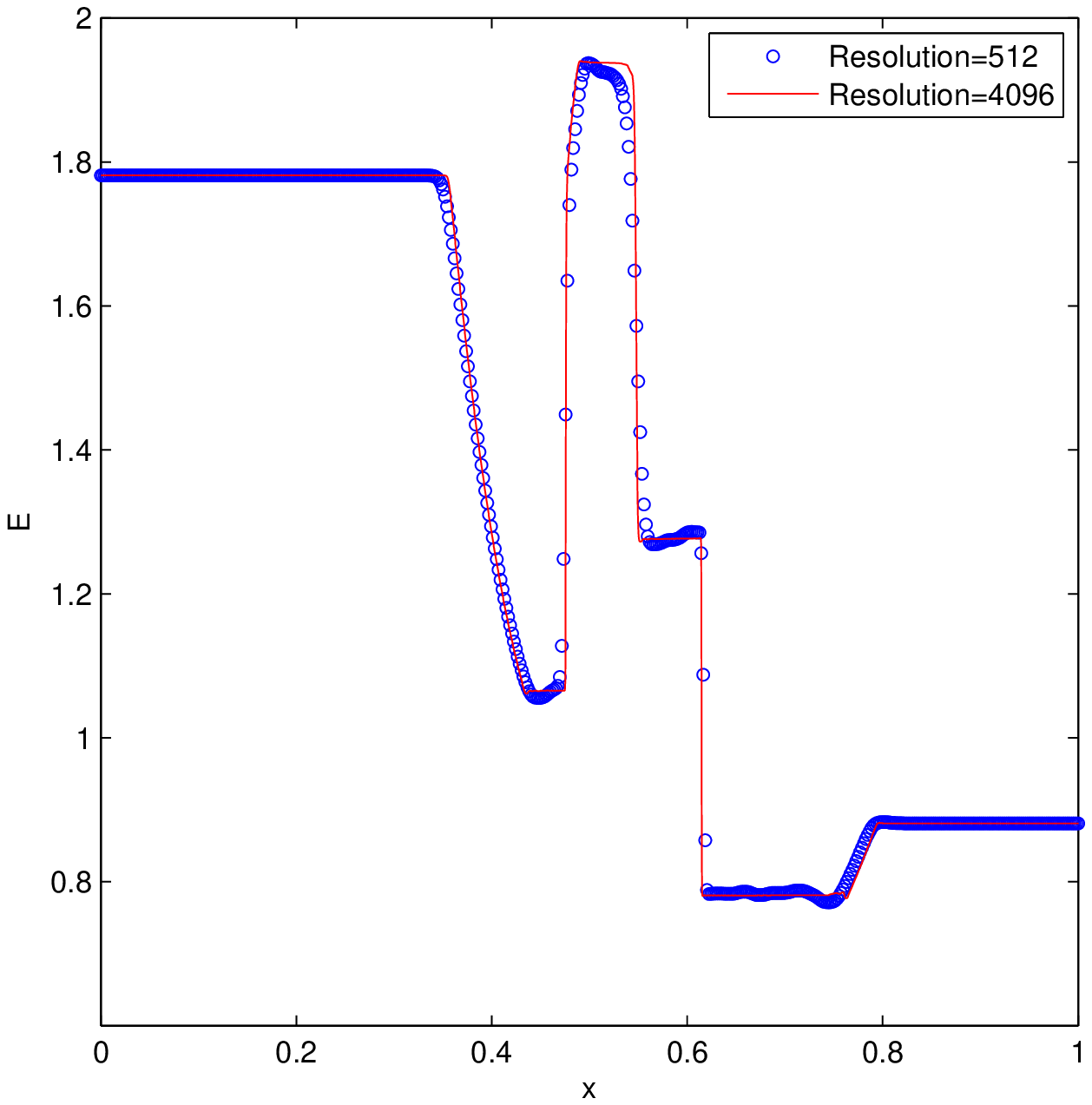}
\includegraphics*[width=2.5in]{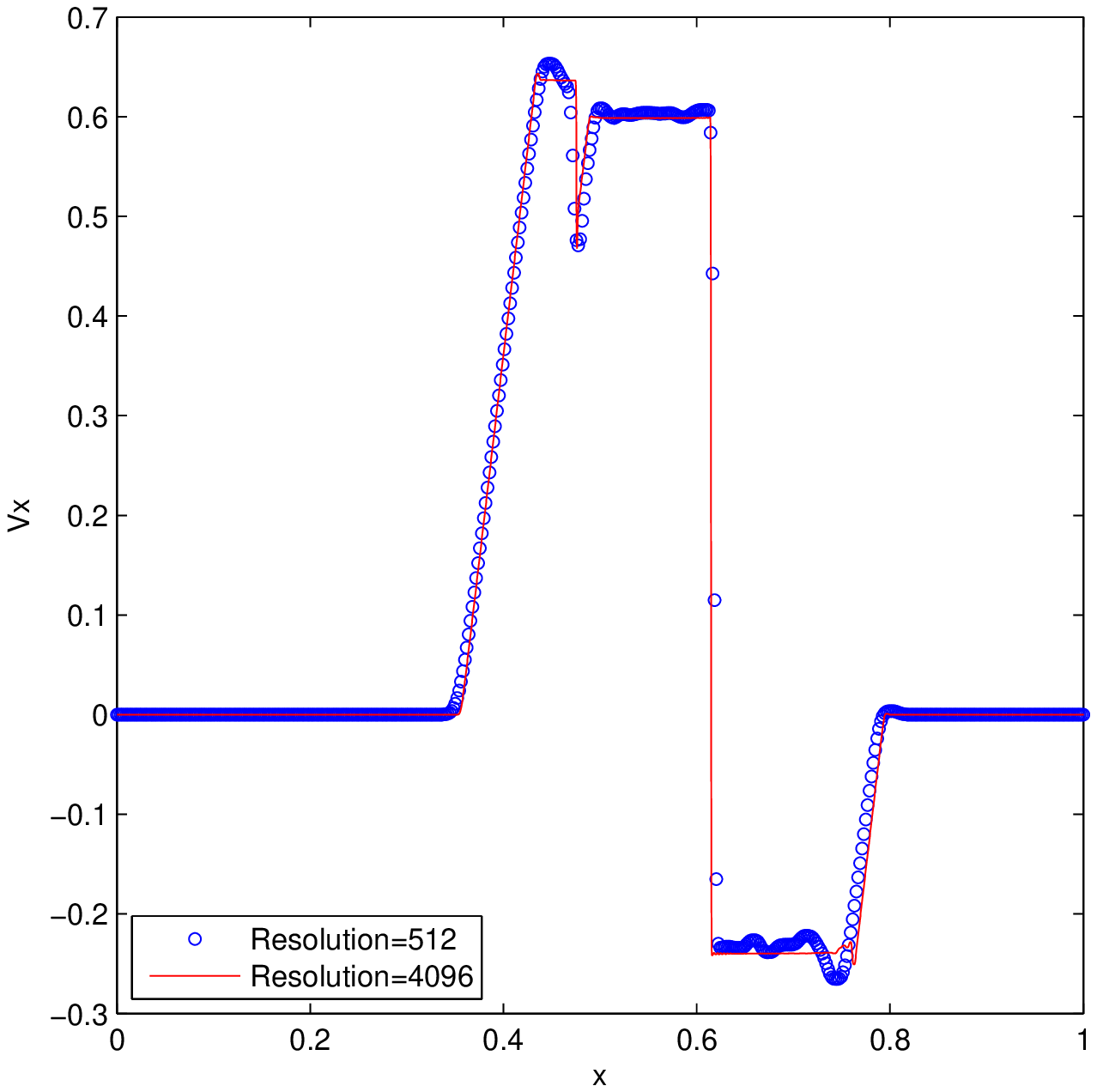}
\hfill \caption{Results (Part I) of Brio-Wu shock tube problem at $t = 0.08L$. The result computed
with 512 grid points is shown with circles and solid line shows reference high resolution result of
4096 grid points.}\label{fig:BrioWu}
\end{center}
\end{figure}

\begin{figure}[h]
\begin{center}
\includegraphics*[width=2.5in]{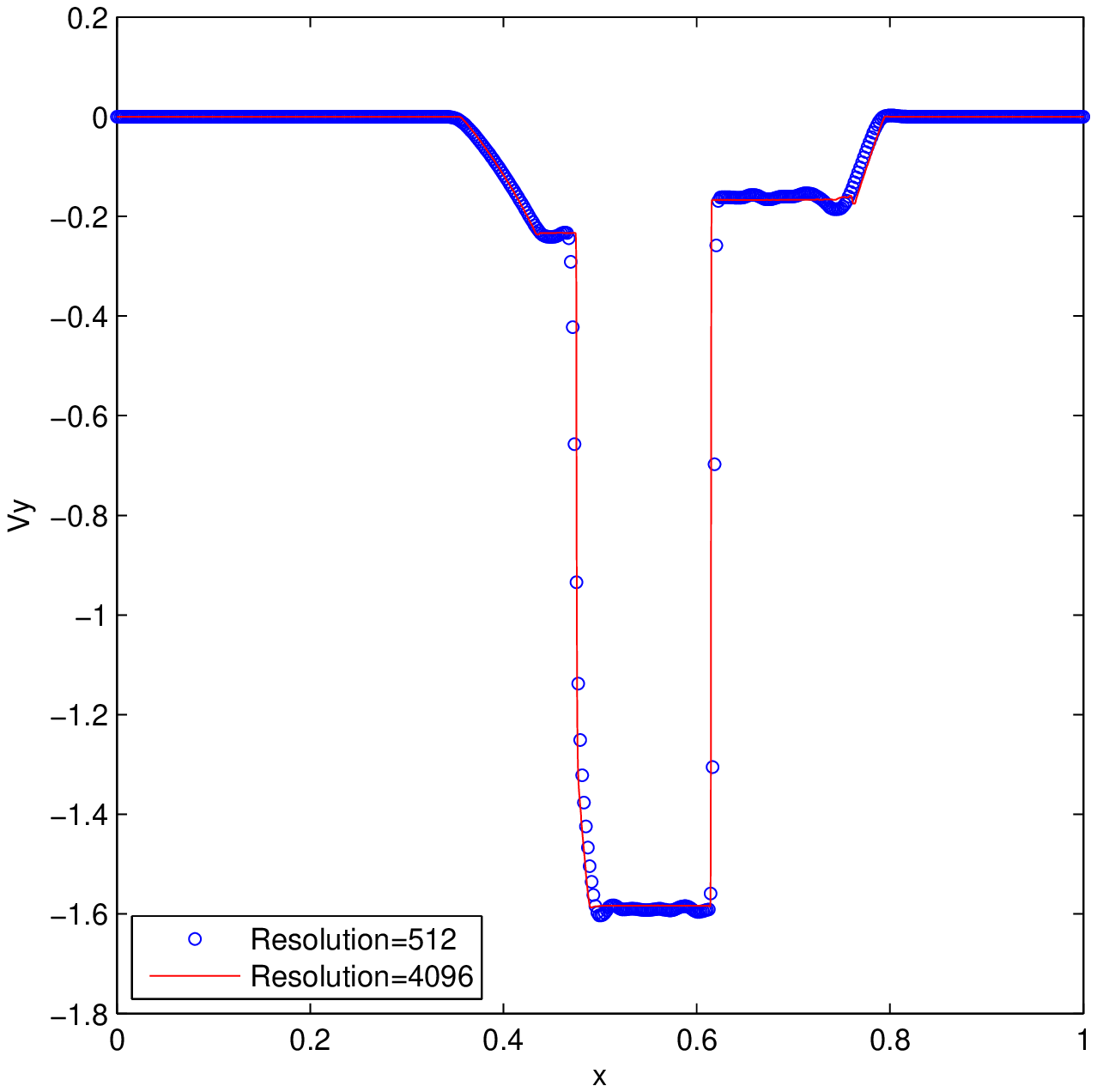}
\includegraphics*[width=2.5in]{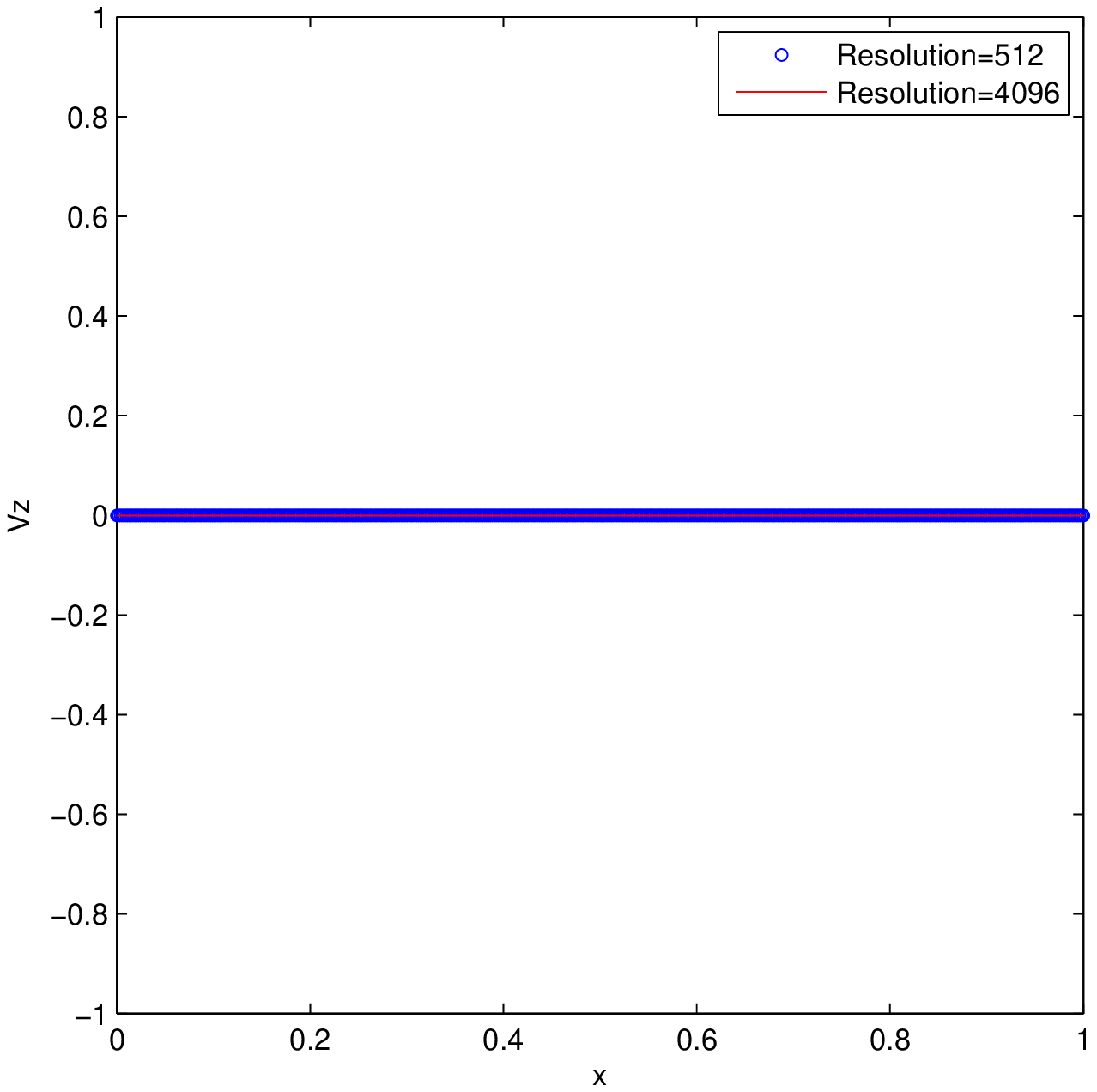}
\includegraphics*[width=2.5in]{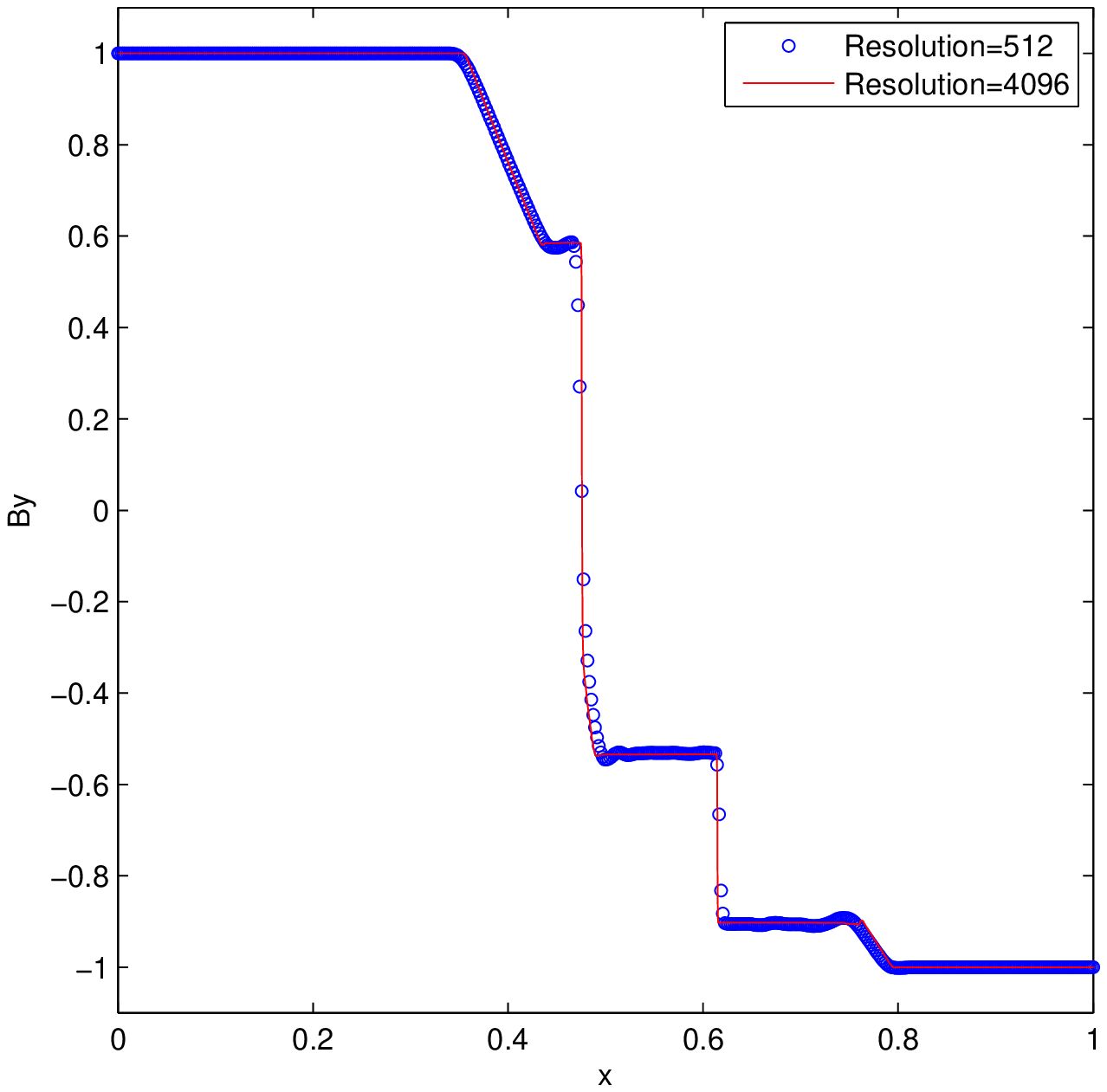}
\includegraphics*[width=2.5in]{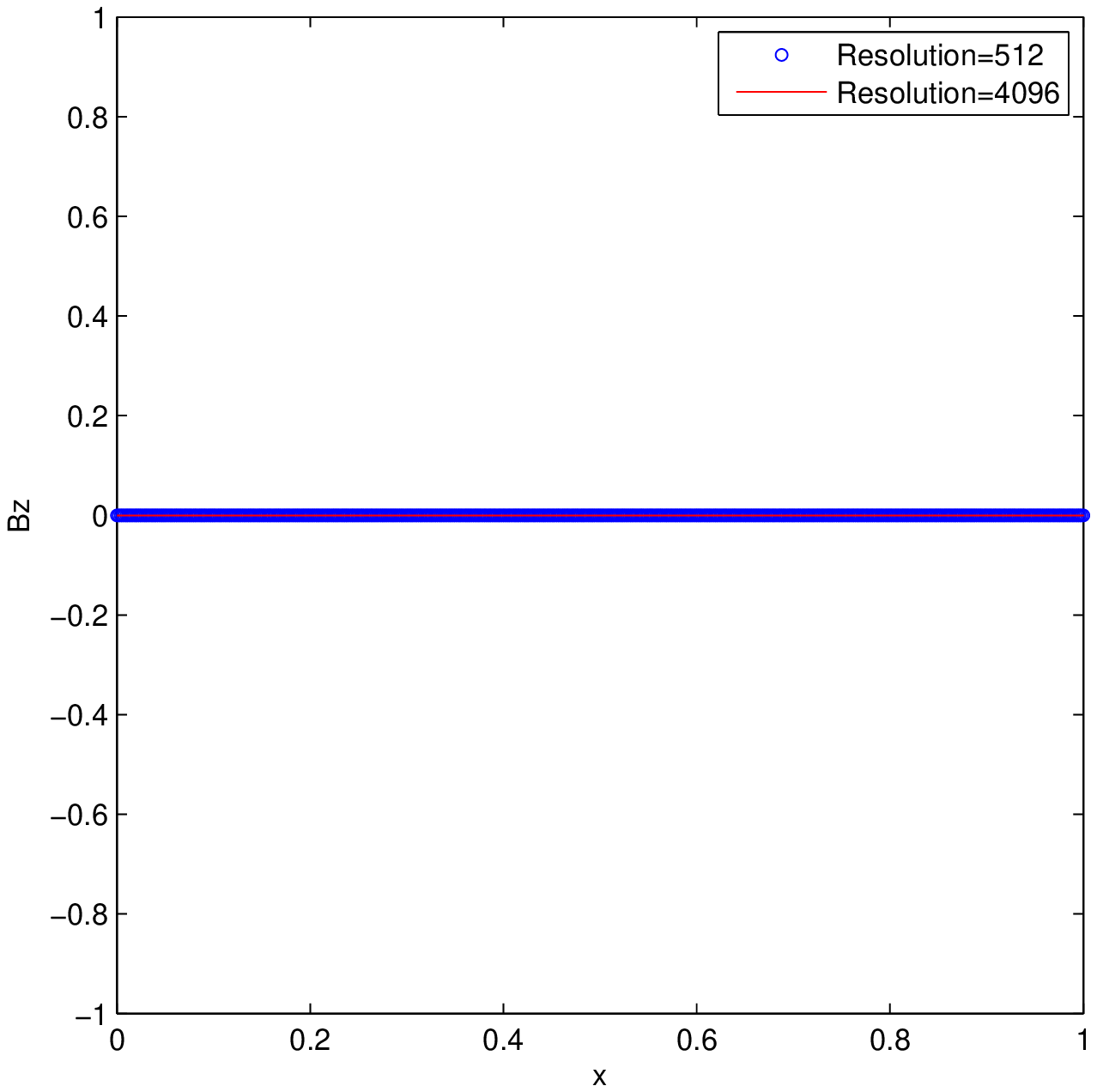}
\hfill \caption{Results (Part II) of Brio-Wu shock tube problem at $t = 0.08L$. The result computed
with 512 grid points is shown with circles and solid line shows reference high resolution result of
4096 grid points.} \label{fig:BrioWu2}
\end{center}
\end{figure}

Constant value of $\gamma = 2$ was used and the problem was solved for $x\in[0,1]$ with 512 grid
points. Numerical results are presented at $t = 0.08L$ in Fig.~\ref{fig:BrioWu} and
Fig.~\ref{fig:BrioWu2}, which include the density, the pressure, the energy, the $y$- and
$z$-magnetic field components, and the $x$-, $y$- and $z$-velocity components. The results are in
agreement with those obtained by Brio and Wu~\cite{Brio1988} and Zachary {\it et
al.}~\cite{Zachary1994}.

\subsubsection{MHD shock tube}

The second 1D test is the MHD shock tube problem considered in~\cite{Dai1994}.

Left side $(x < 0.5)$
\begin{equation}
\label{equat_DW_init1} \left\{\begin{array}{c} v_{x} \\
v_{y} \\ v_{z} \end{array} \right\} \mbox{=} \left\{\begin{array}{c} 1.2 \\
0.01 \\ 0.5 \end{array} \right\}
\end{equation}
\begin{equation}
\label{equat_DW_init2} \left\{\begin{array}{c} B_{x} \\
B_{y} \\ B_{z} \end{array} \right\} \mbox{=} \left\{\begin{array}{c} 2\sqrt{(4\pi)} \\
3.6/\sqrt{(4\pi)} \\ 2/\sqrt{(4\pi)} \end{array} \right\}
\end{equation}
\begin{equation}
\label{equat_DW_init3} \rho = 1.08, \,\,\,\,\, p = 0.95
\end{equation}

Right side $(x \geq 0.5)$
\begin{equation}
\label{equat_DW_init4} \left\{\begin{array}{c} v_{x} \\
v_{y} \\ v_{z} \end{array} \right\} \mbox{=} \left\{\begin{array}{c} 0 \\
0 \\ 0 \end{array} \right\}
\end{equation}
\begin{equation}
\label{equat_DW_init5} \left\{\begin{array}{c} B_{x} \\
B_{y} \\ B_{z} \end{array} \right\} \mbox{=} \left\{\begin{array}{c} 2\sqrt{(4\pi)} \\
4/\sqrt{(4\pi)} \\ 2/\sqrt{(4\pi)} \end{array} \right\}
\end{equation}
\begin{equation}
\label{equat_DW_init6} \rho = 1, \,\,\,\,\, p = 1
\end{equation}

\begin{figure}[h]
\begin{center}
\includegraphics*[width=2.5in]{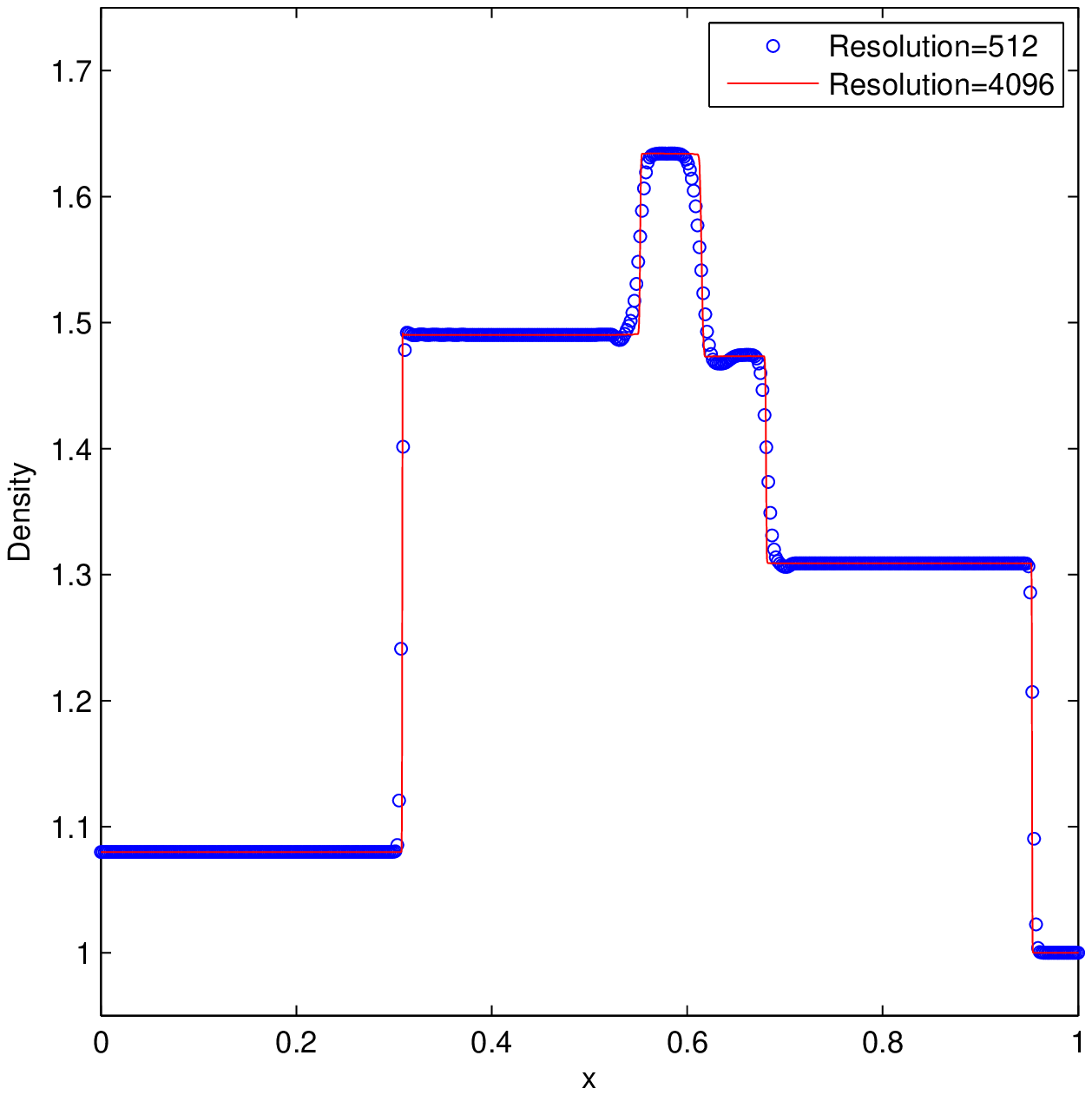}
\includegraphics*[width=2.5in]{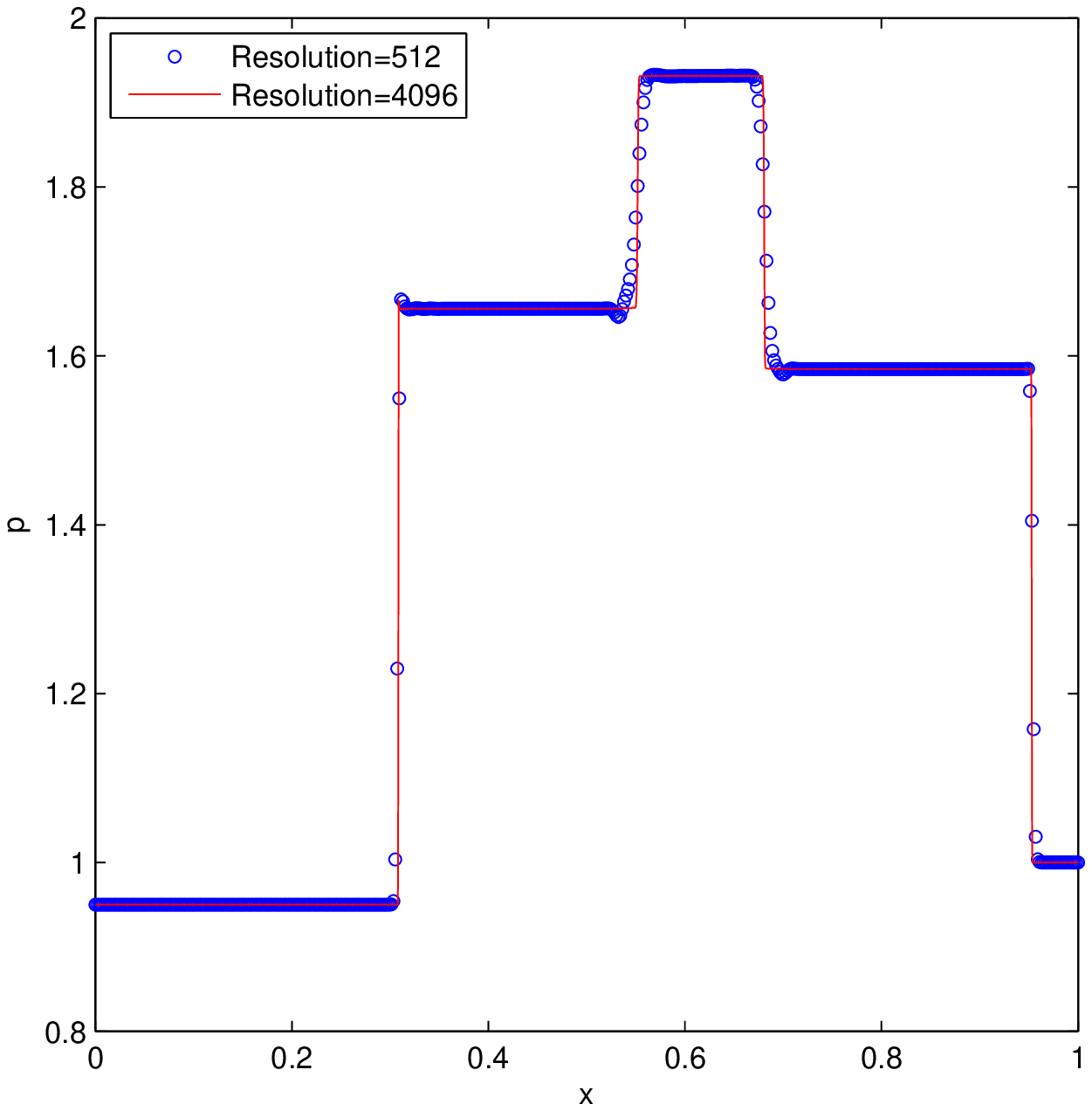}
\includegraphics*[width=2.5in]{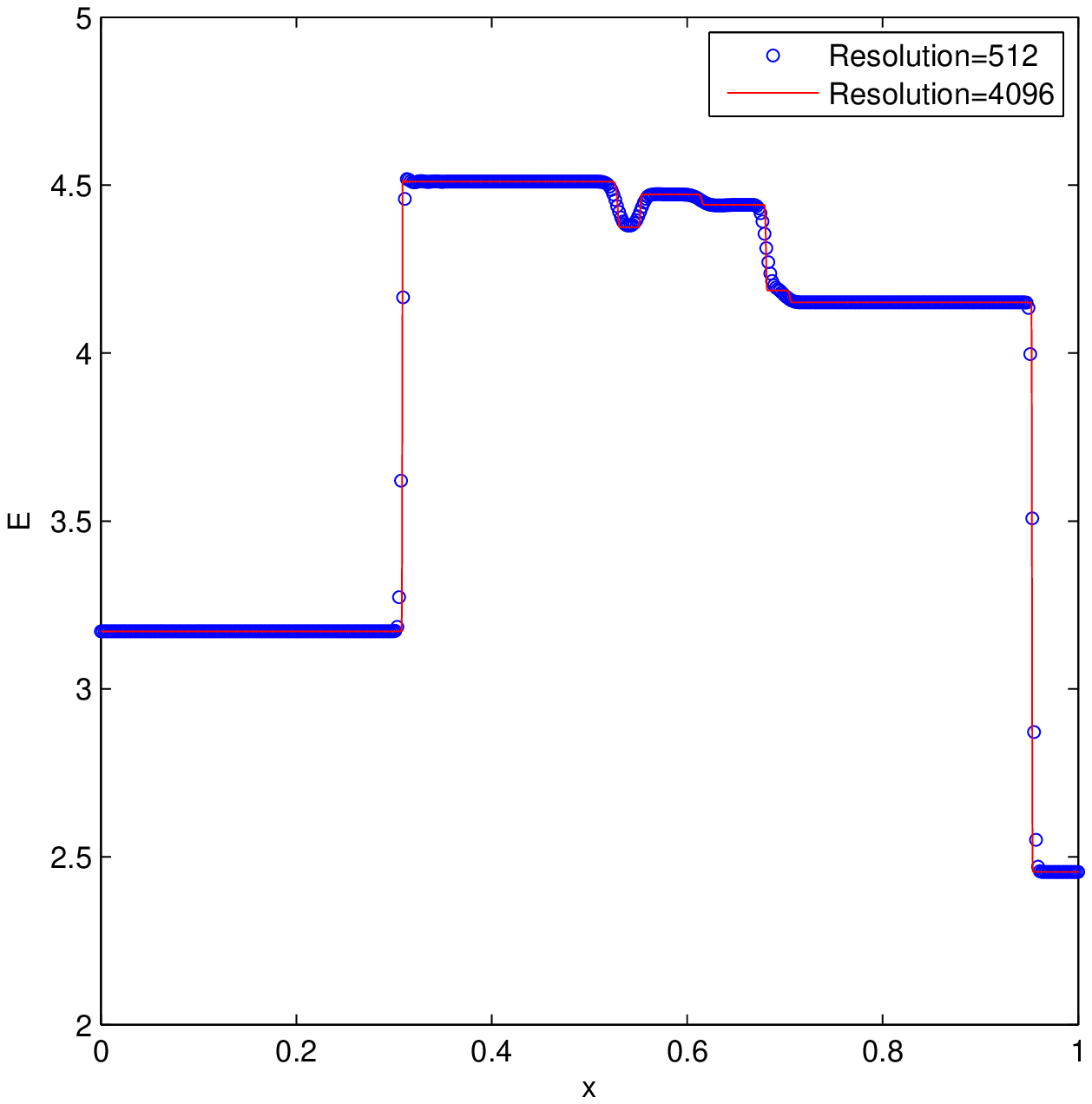}
\includegraphics*[width=2.5in]{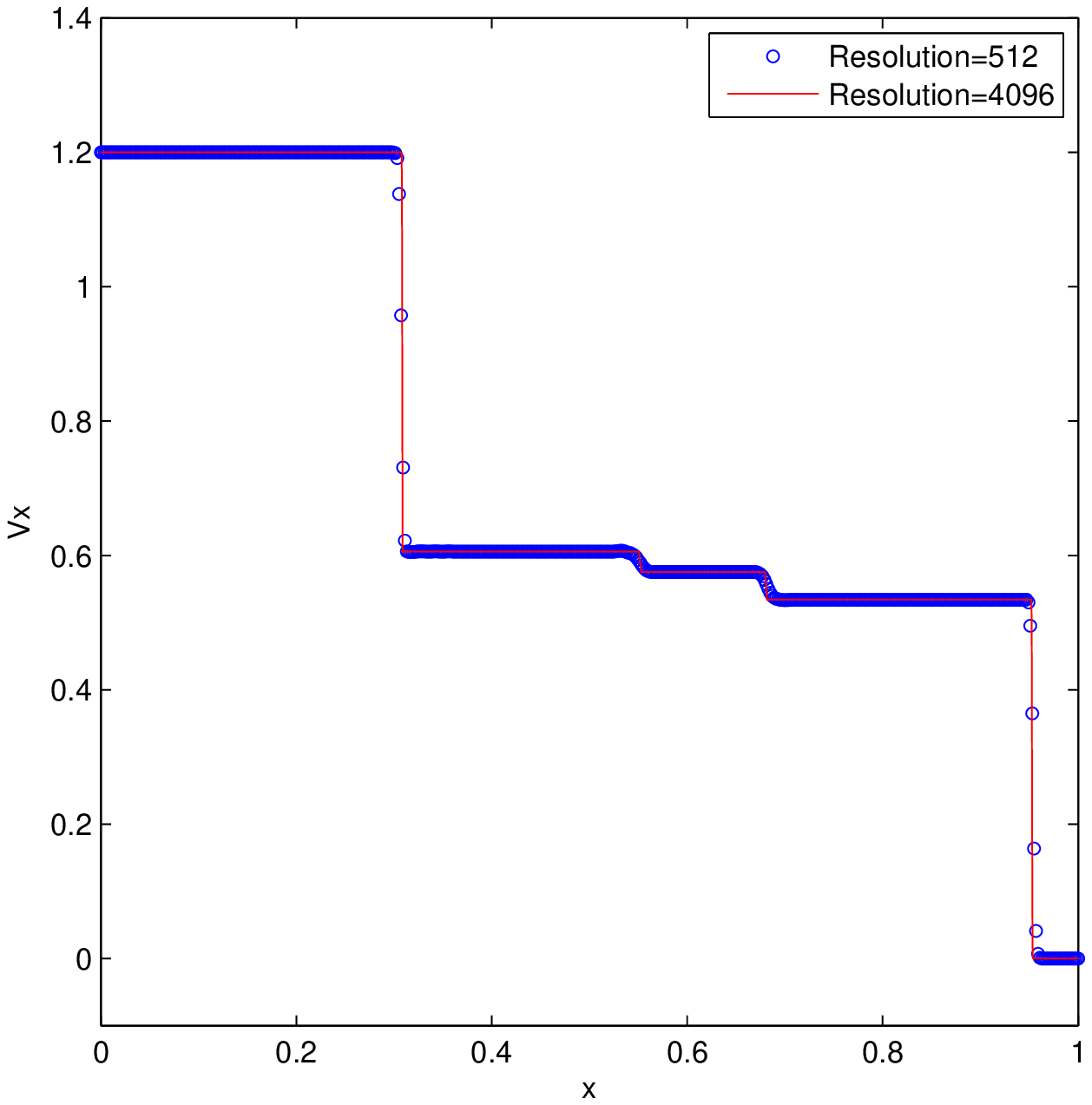}
hfill \caption{Results (Part I) of MHD shock tube test at $t = 0.2L$. The result computed with 512
grid points is shown with circles and solid line shows reference high resolution result of 4096
grid points.}\label{fig:RJ952a}
\end{center}
\end{figure}

\begin{figure}[h]
\begin{center}
\includegraphics*[width=2.5in]{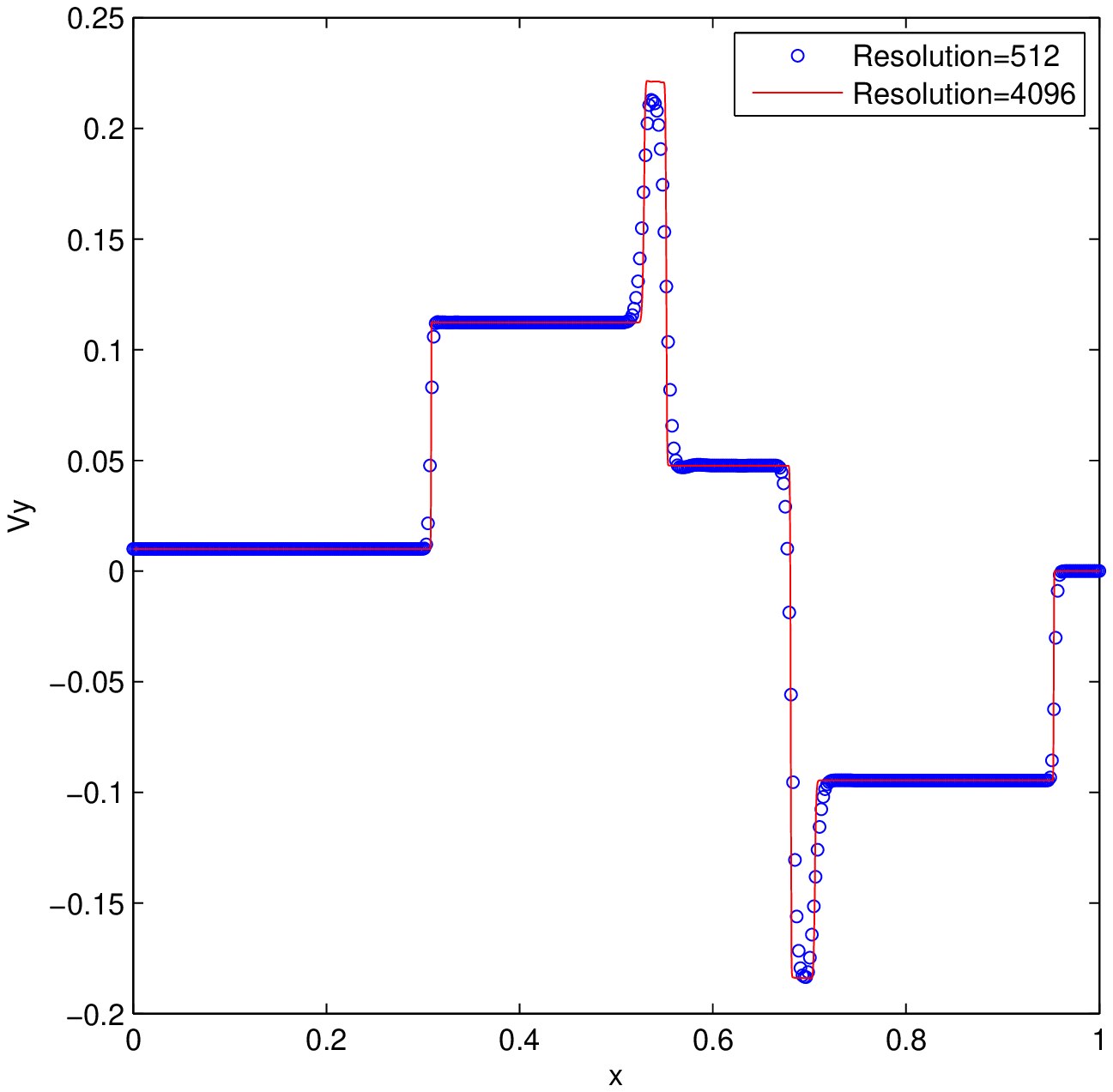}
\includegraphics*[width=2.5in]{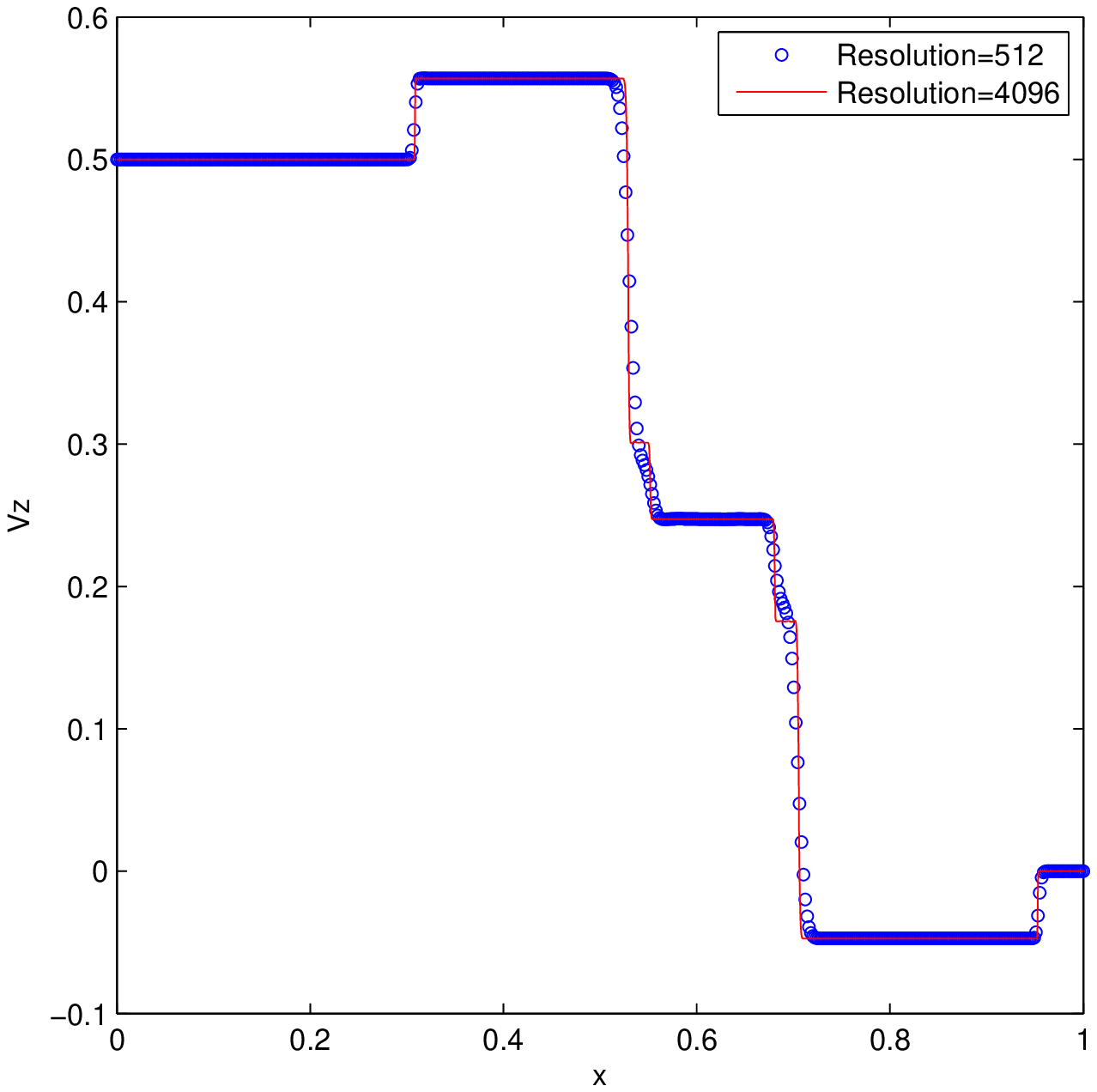}
\includegraphics*[width=2.5in]{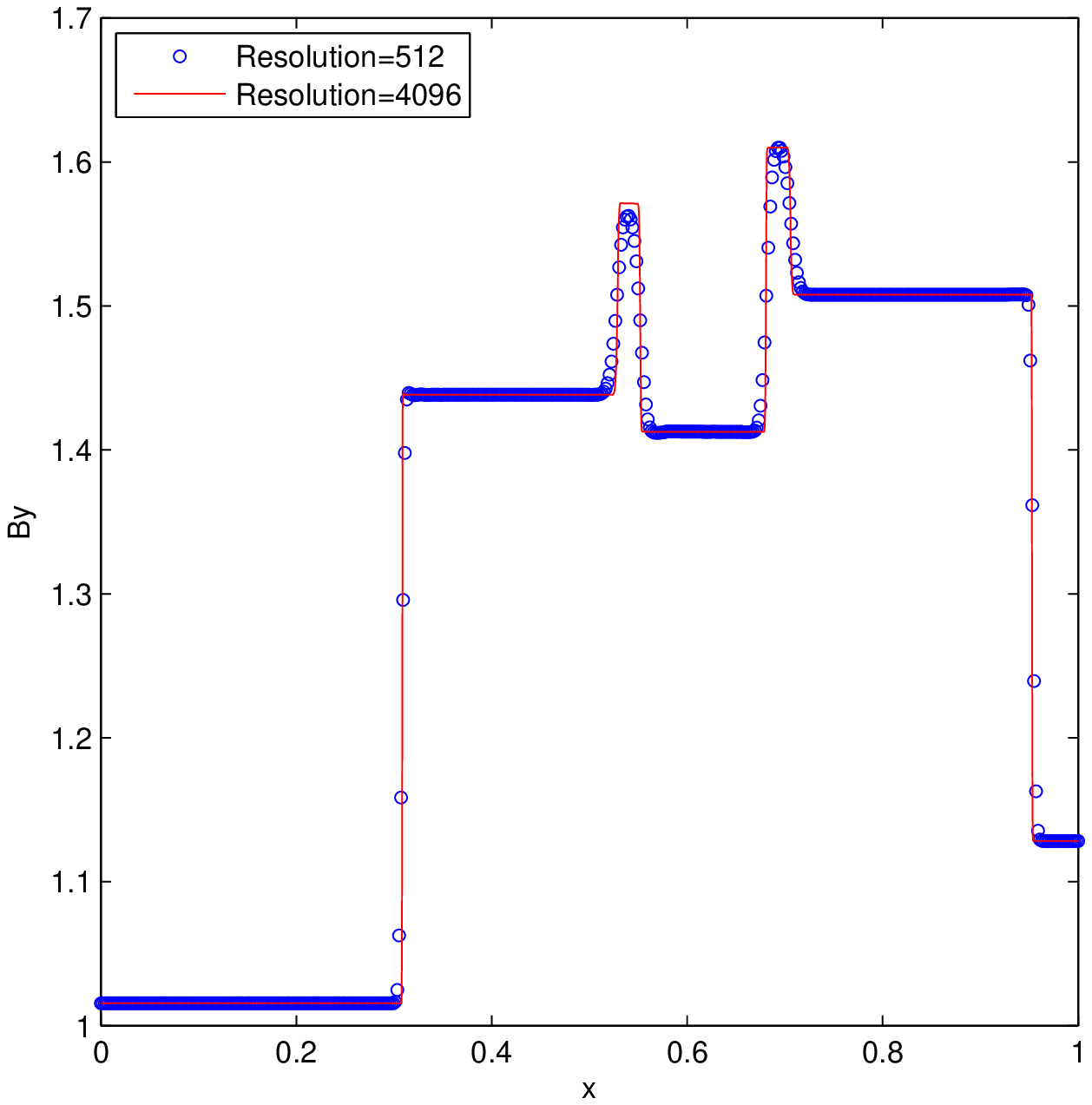}
\includegraphics*[width=2.5in]{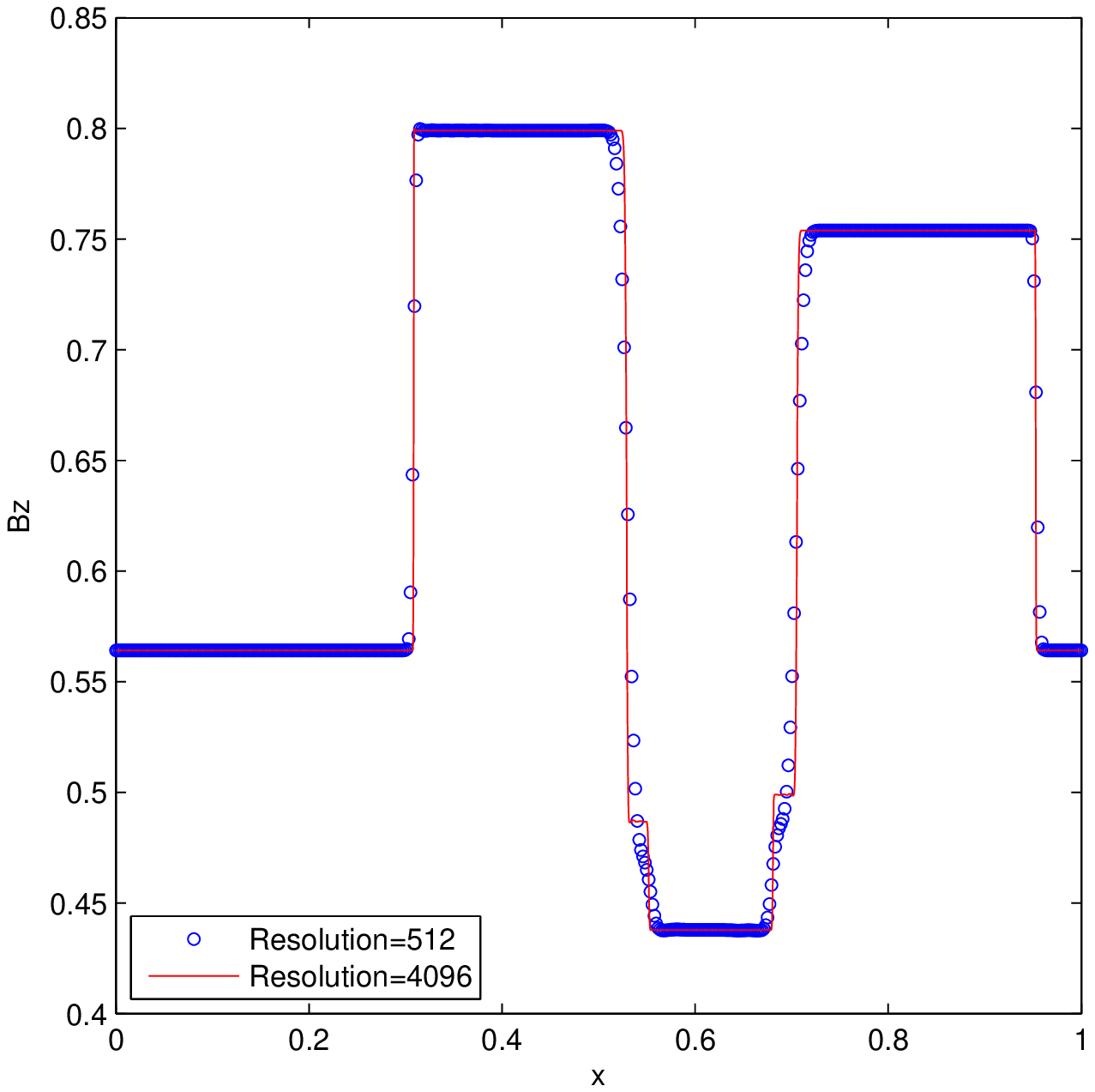}
\hfill \caption{Results (Part II) of MHD shock tube test at $t = 0.2L$. The result computed with
512 grid points is shown with circles and solid line shows reference high resolution result of 4096
grid points.} \label{fig:RJ952a2}
\end{center}
\end{figure}

Constant value of $\gamma = 5/3$ was used and the problem was solved for $x\in[0,1]$ with 512 grid
points. Numerical results are presented at $t = 0.2L$ in Fig.~\ref{fig:RJ952a} and
Fig.~\ref{fig:RJ952a2}, which include the density, the pressure, the energy, the $y$- and
$z$-magnetic field components, and the $x$-, $y$- and $z$-velocity components. The results are in
agreement with those obtained by~\cite{Dai1994} and~\cite{Ryu1995}.


\subsection{Two-dimensional problems}

\subsubsection{Orszag-Tang problem}\label{sub:O-T}

The first 2D test is Orszag-Tang problem~\cite{Orszag1979}, which is used to study incompressible
MHD turbulence. In our test, the boundary conditions are periodic everywhere. The density $\rho$,
pressure $p$, initial velocities $(v_{x}, v_{y}, v_{z})$, and magnetic field $(B_{x}, B_{y},
B_{z})$ are given by
\begin{equation}
\label{equat_OT_init1} \left\{\begin{array}{c} v_{x} \\
v_{y} \\ v_{z} \end{array} \right\} \mbox{=} \left\{\begin{array}{c} -{\rm sin}(2\pi y) \\
{\rm sin}(2\pi x) \\ 0 \end{array} \right\}
\end{equation}
\begin{equation}
\label{equat_OT_init2} \left\{\begin{array}{c} B_{x} \\
B_{y} \\ B_{z} \end{array} \right\} \mbox{=} \left\{\begin{array}{c} -B_{0}{\rm sin}(2\pi y) \\
B_{0}{\rm sin}(4\pi x) \\ 0 \end{array} \right\} \,\,\,\,\, {\rm where}\,\, B_{0} = 1 / \sqrt{4\pi}
\end{equation}
\begin{equation}
\label{equat_OT_init3} \rho = 25/(36\pi), \,\,\,\,\, p = 5/(12\pi), \,\,\,\,\, \gamma =
5/3,\,\,\,\,\, (0 \leq x \leq 1)\,\,(0 \leq y \leq 1)
\end{equation}

\begin{figure}[h]
\begin{center}
\includegraphics*[width=2.5in]{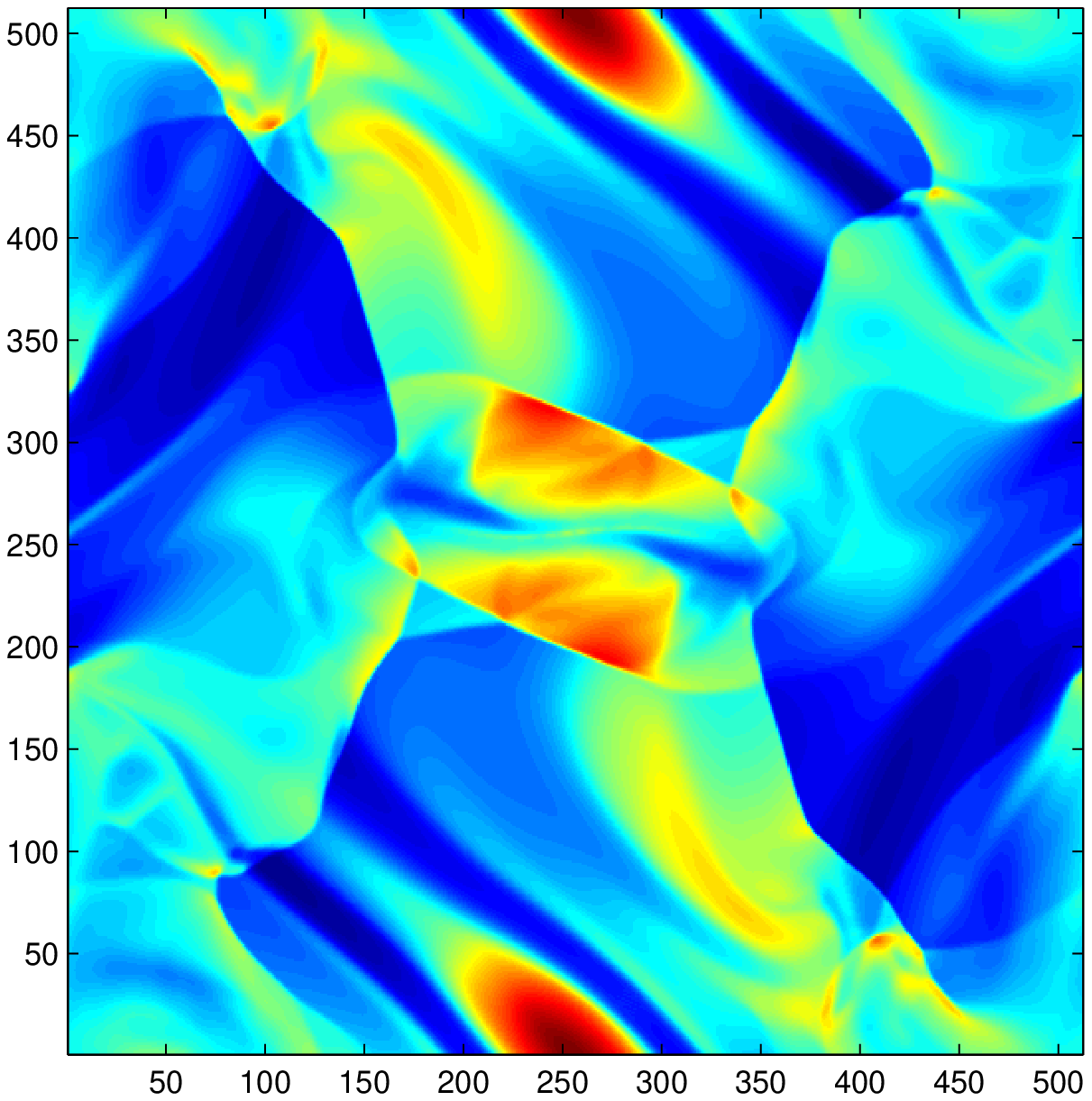}
\includegraphics*[width=2.5in]{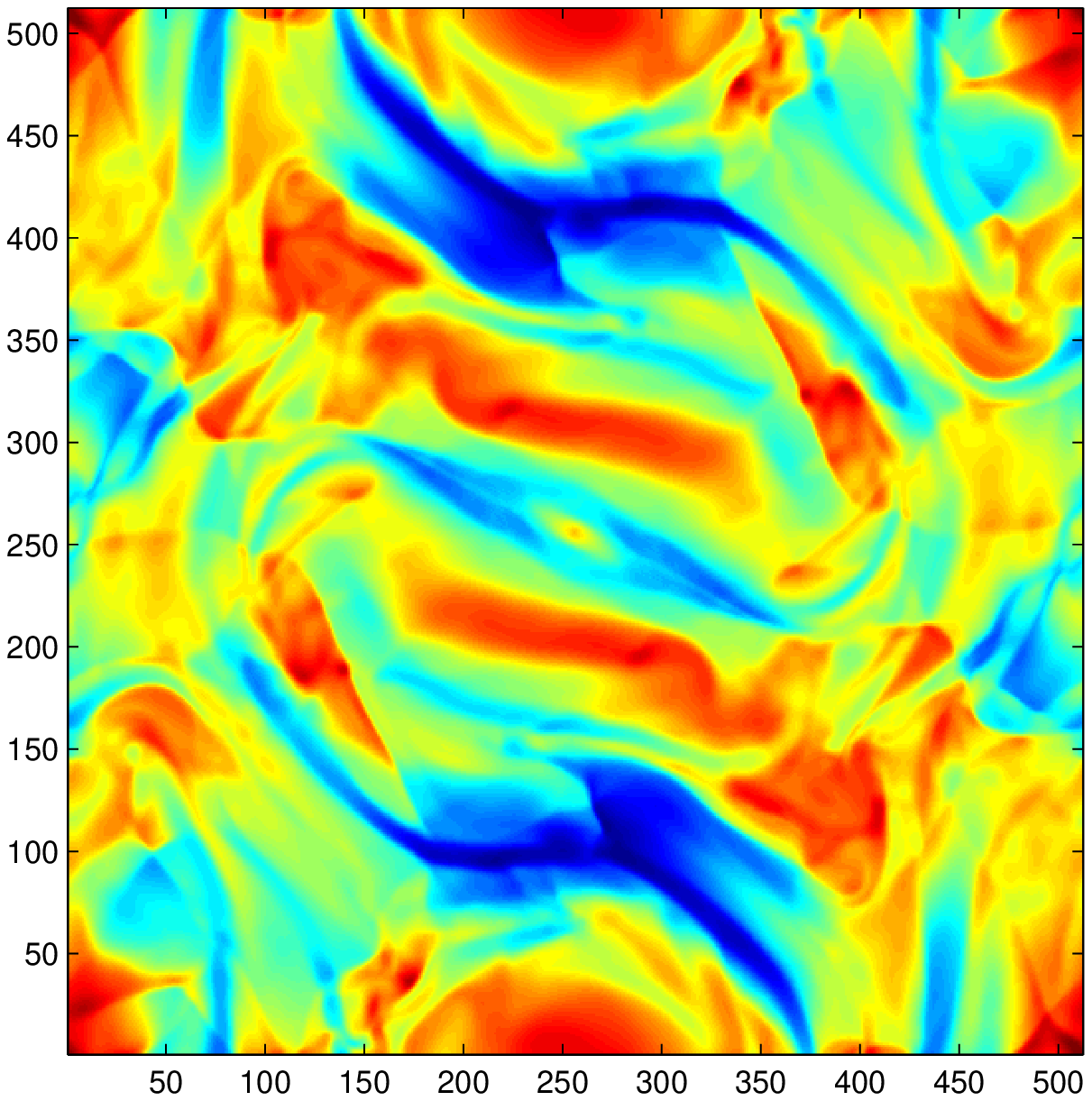}
\includegraphics*[width=2.5in]{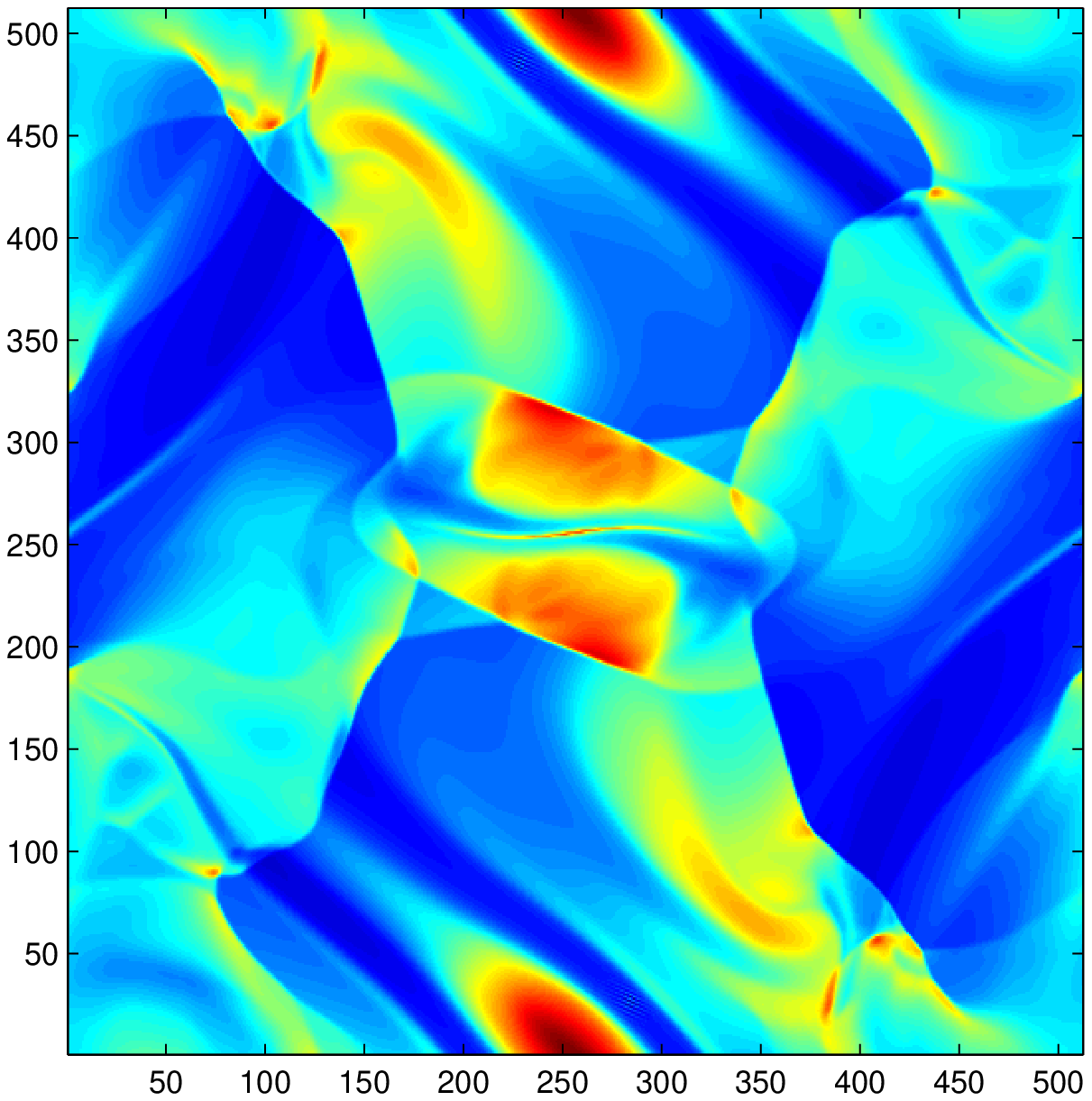}
\includegraphics*[width=2.5in]{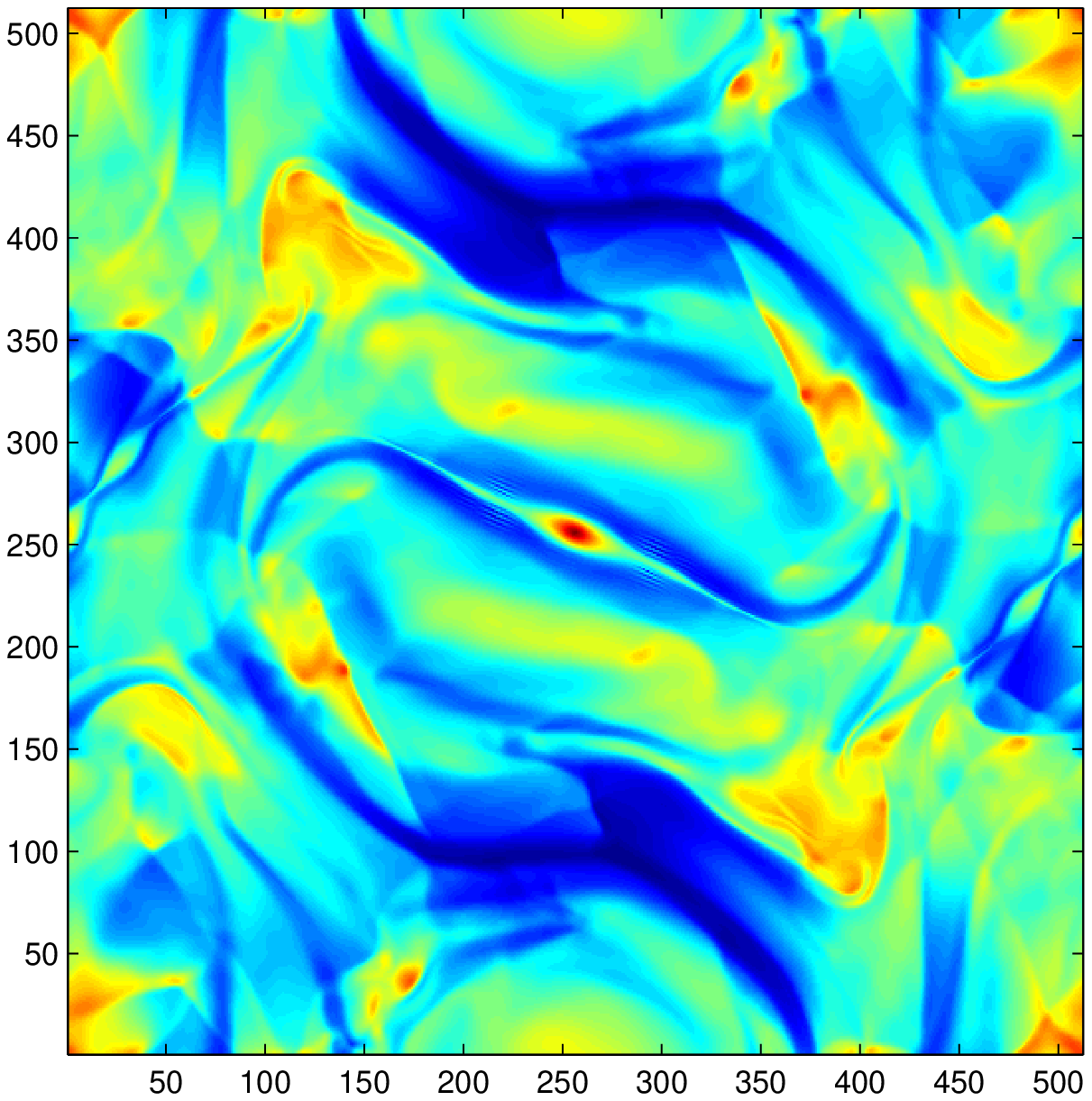}
\hfill \caption{Results of the density (top) and gas pressure (bottom) of Orszag-Tang vortex test
at $t = 0.5L$ (left) and $t = 1.0L$ (right) computed with 512 $\times$ 512 grid
points.}\label{fig:OT}
\end{center}
\end{figure}

The Orszag-Tang vertex test was performed in a two-dimensional periodic box with 512 $\times$ 512
grid points. The results of the density and gas pressure evolution of the Orszag-Tang problem at $t
= 0.5 L$ and $t = 1.0L$ are shown in Fig.~\ref{fig:OT}, where the complex pattern of interacting
waves is perfectly recovered. The results agree well with those in Lee {\it et al.}~\cite{Lee2009}.

\subsubsection{Two-dimensional blast wave problem}

\begin{figure}[h]
\begin{center}
\includegraphics*[width=2.5in]{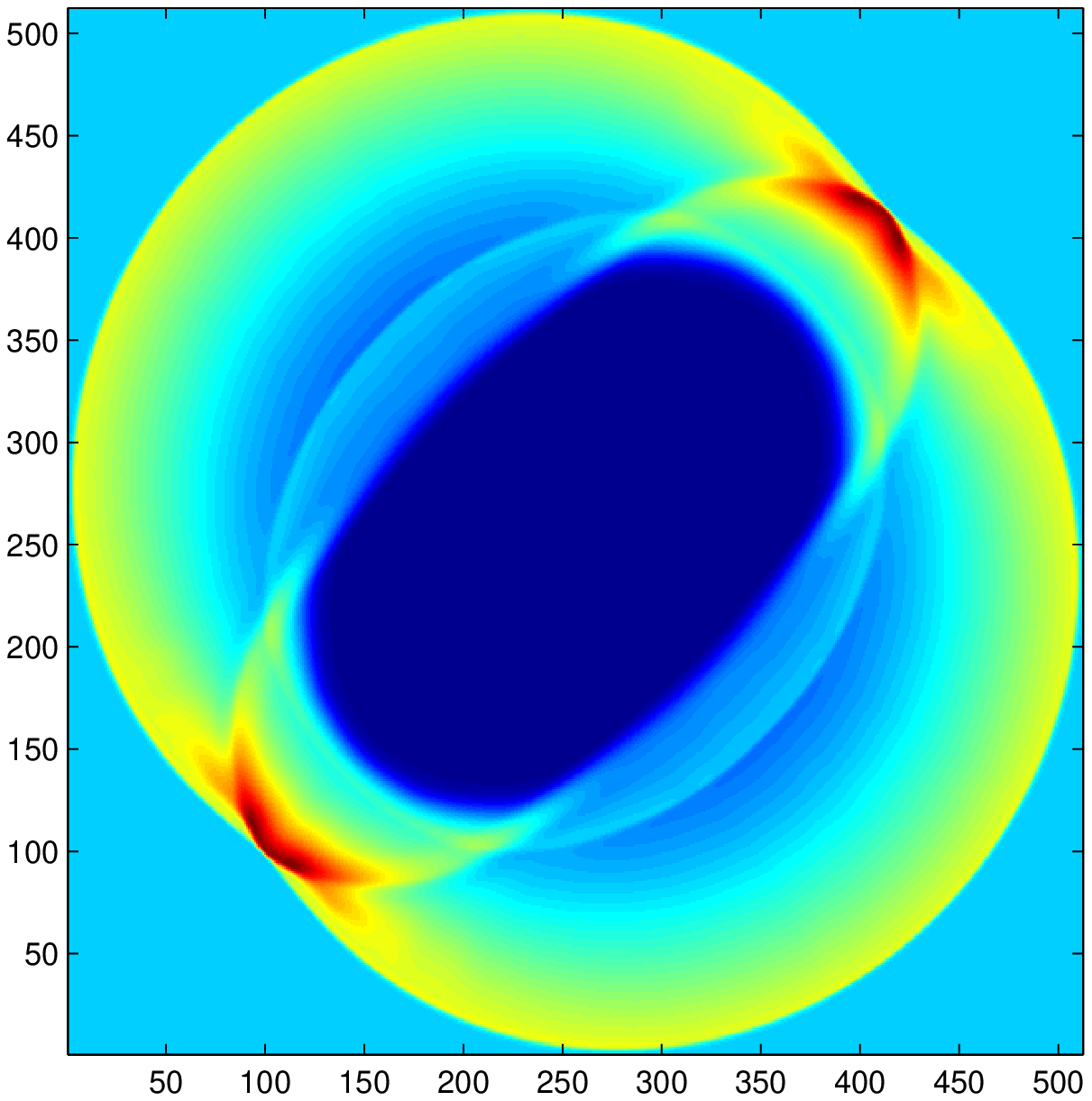}
\includegraphics*[width=2.5in]{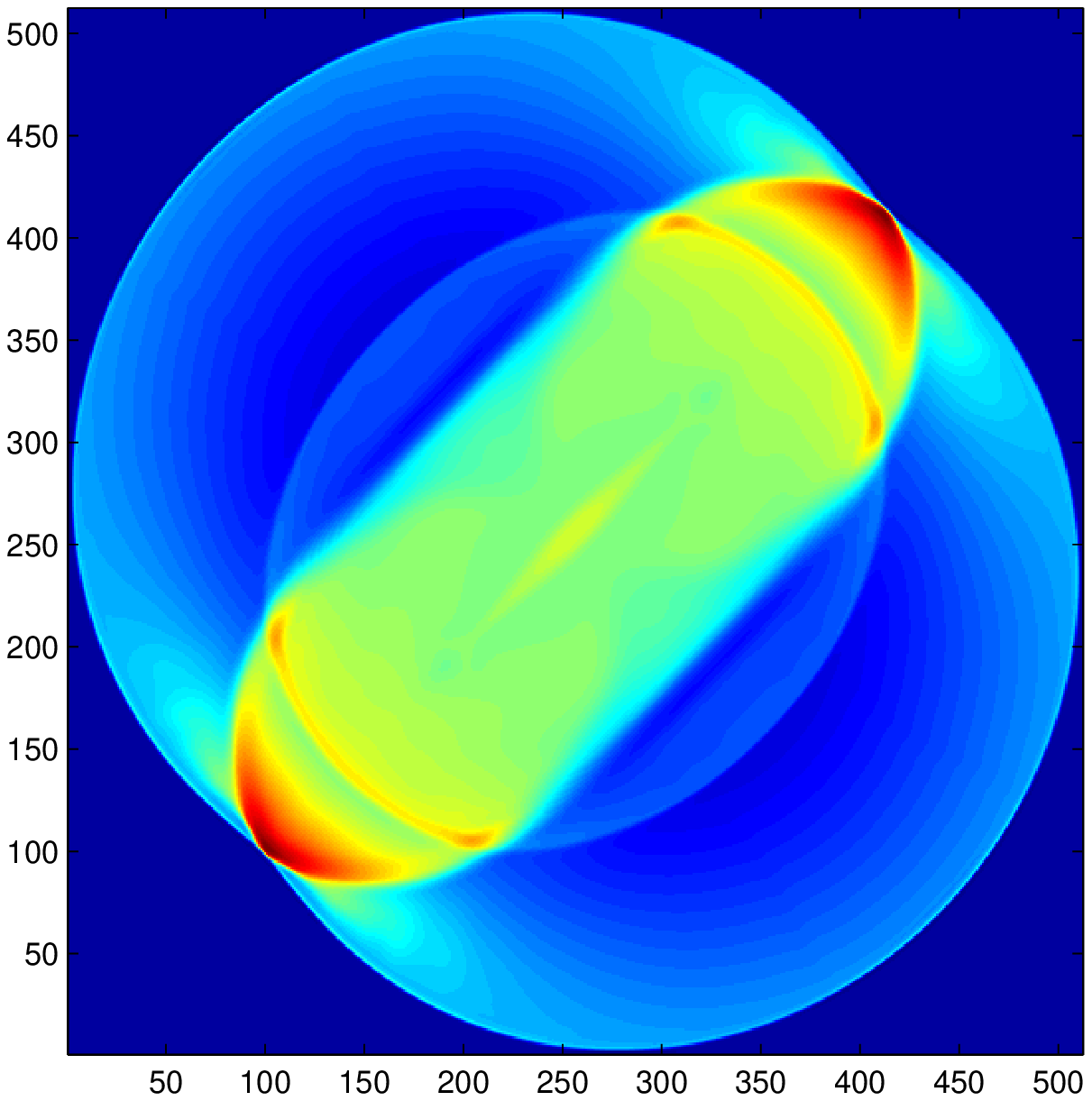}
\hfill \caption{Results of the density (left) and gas pressure (right) of the 2D blast wave test at
$t= 0.2L$, computed with 512 $\times$ 512 grid points.} \label{fig:2DBlast}
\end{center}
\end{figure}

The second 2D test is the MHD blast wave problem. The MHD spherical blast wave problem of Zachary
{\it et al.}~\cite{Zachary1994} is initiated by an over pressured region in the center of the
domain. The result is a strong outward moving spherical shock with rarified fluid inside the
sphere. We followed the test suite~\cite{Spherical} of Athena~\cite{Stone2008}. The condition for
2D MHD blast wave problem is listed as follows~\cite{Spherical}
\begin{equation}
\label{equat_2DBlast_init1} \left\{\begin{array}{c} v_{x} \\
v_{y} \\ v_{z} \end{array} \right\} \mbox{=} \left\{\begin{array}{c} 0 \\
0 \\ 0 \end{array} \right\}
\end{equation}
\begin{equation}
\label{equat_2DBlast_init2} \left\{\begin{array}{c} B_{x} \\
B_{y} \\ B_{z} \end{array} \right\} \mbox{=} \left\{\begin{array}{c} 1/\sqrt{2} \\
1/\sqrt{2} \\ 0 \end{array} \right\}
\end{equation}
\begin{equation}
\label{equat_2DBlast_init3} \mbox{$p = $}\left\{\begin{array}{c} 10 \\
0.1 \end{array} \begin{array}{c} \mbox{inside the spherical region} \\
\,\,\,\mbox{outside the spherical region} \end{array}\right.
\end{equation}
\begin{equation}
\label{equat_2DBlast_init4} \begin{array}{l}\rho = 1, \,\,\,\,\, p =
5/(12\pi), \,\,\,\,\, \gamma = 5/3 \\ \mbox{spherical region center} = (0.5, 0.5),\,\, r = 0.1 \\
(0 \leq x \leq 1)\,\,(0 \leq y \leq 1) \end{array}
\end{equation}
In Fig.~\ref{fig:2DBlast}, we present images of the density and gas pressure at $t=0.2L$ computed
with 512 $\times$ 512 grid points. The results are in excellent agreement with those presented
in~\cite{Spherical}.

\subsubsection{MHD rotor problem}

\begin{figure}[h]
\begin{center}
\includegraphics*[width=2.5in]{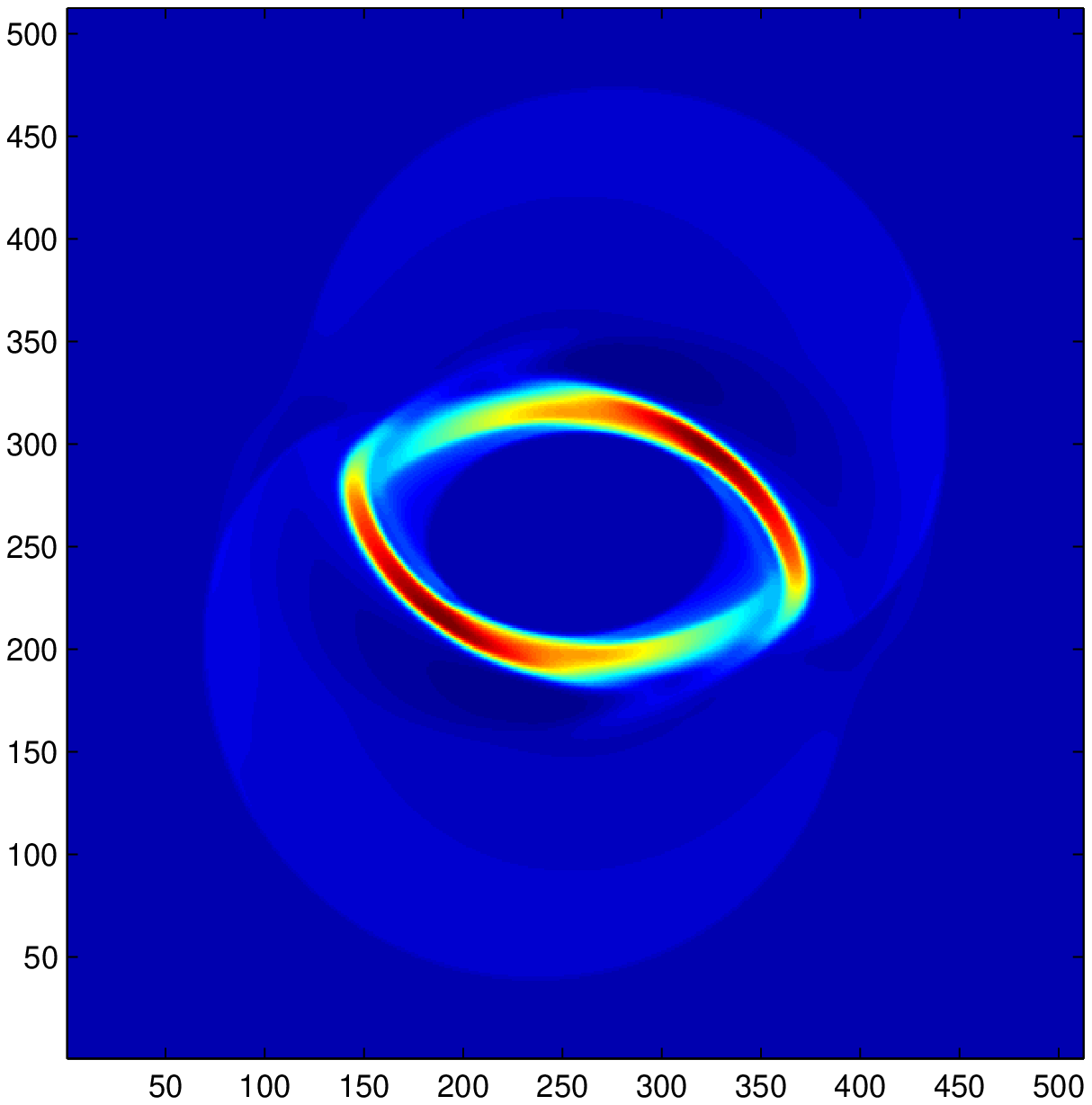}
\includegraphics*[width=2.5in]{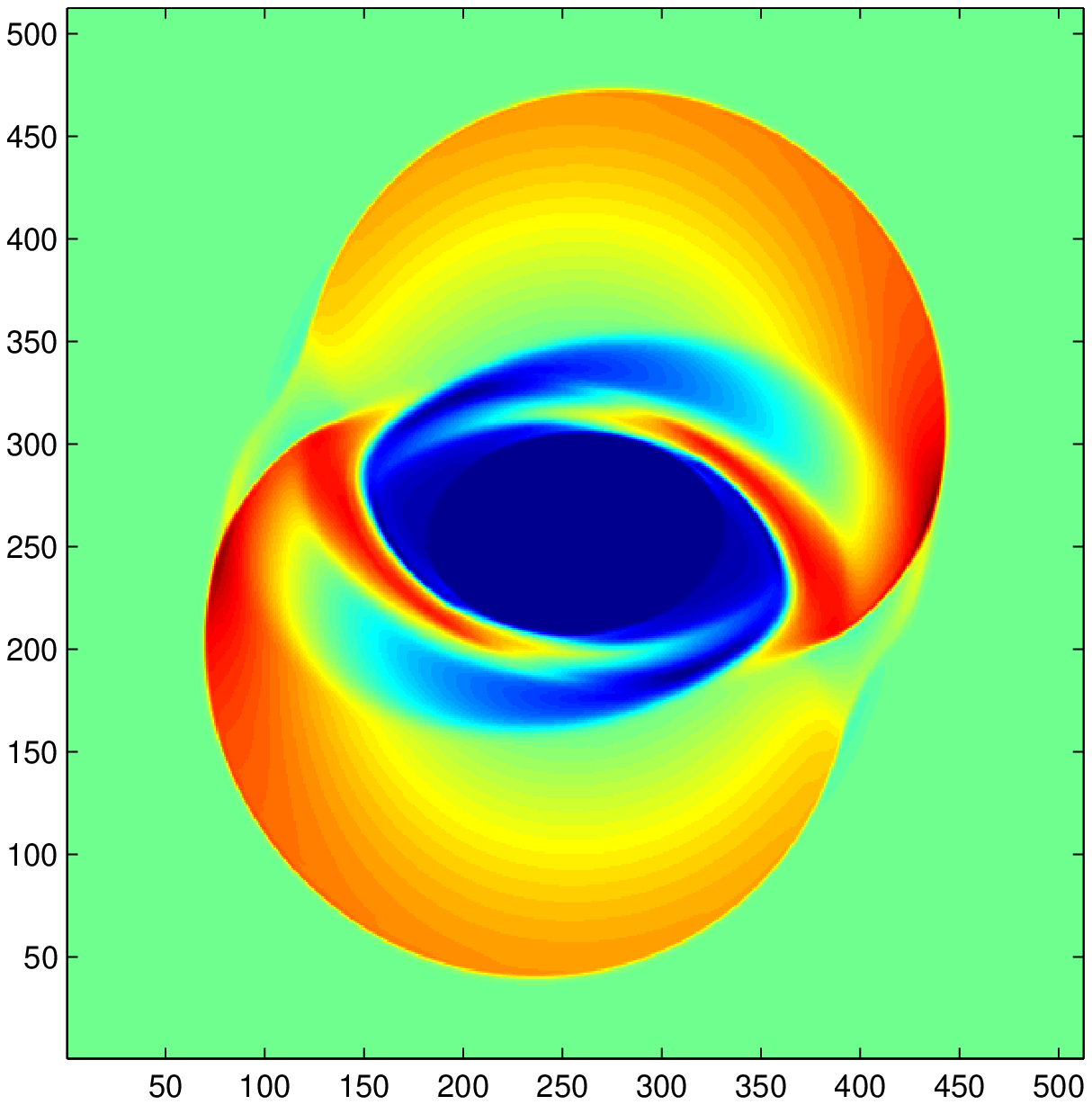}
\includegraphics*[width=2.5in]{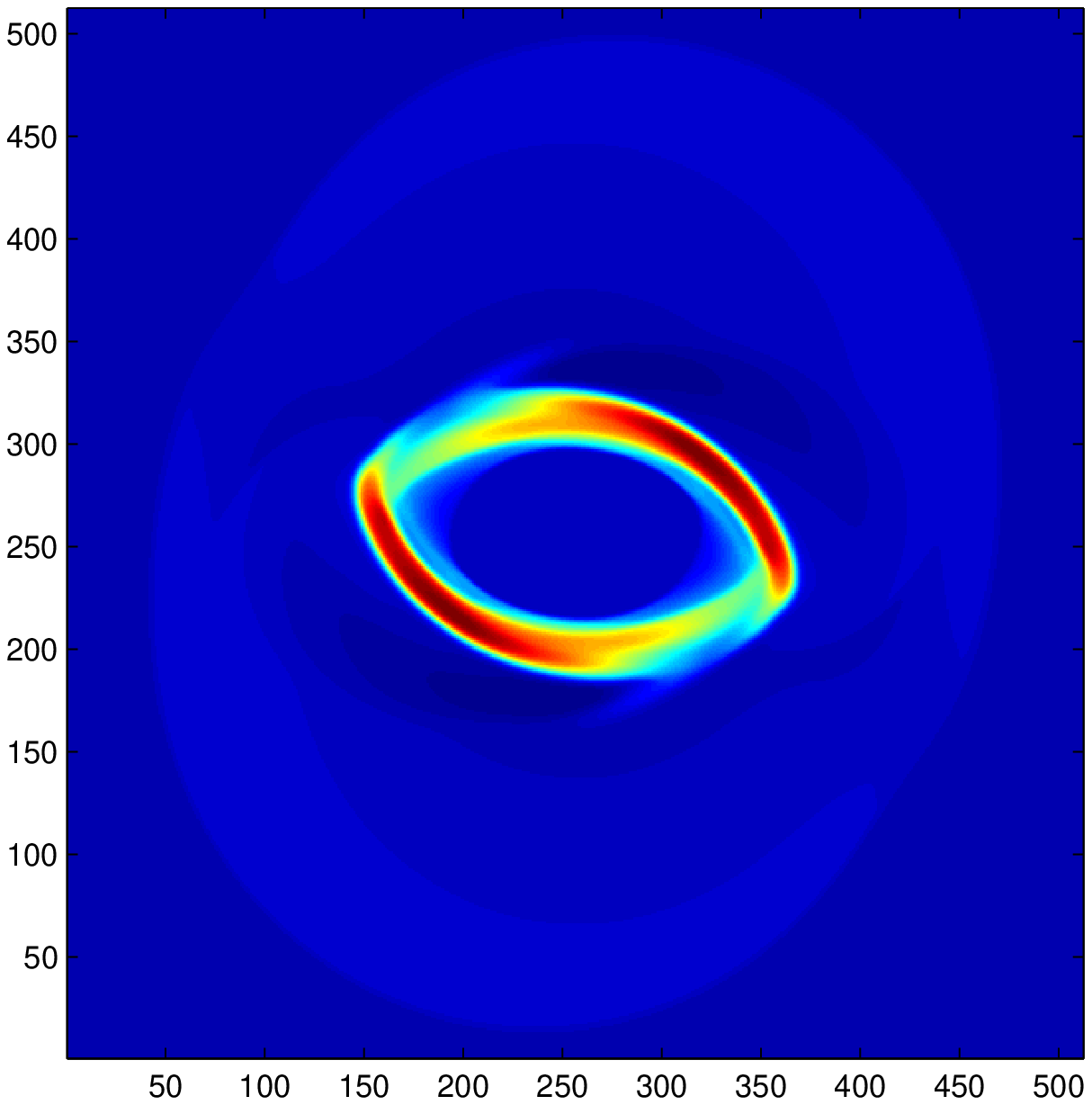}
\includegraphics*[width=2.5in]{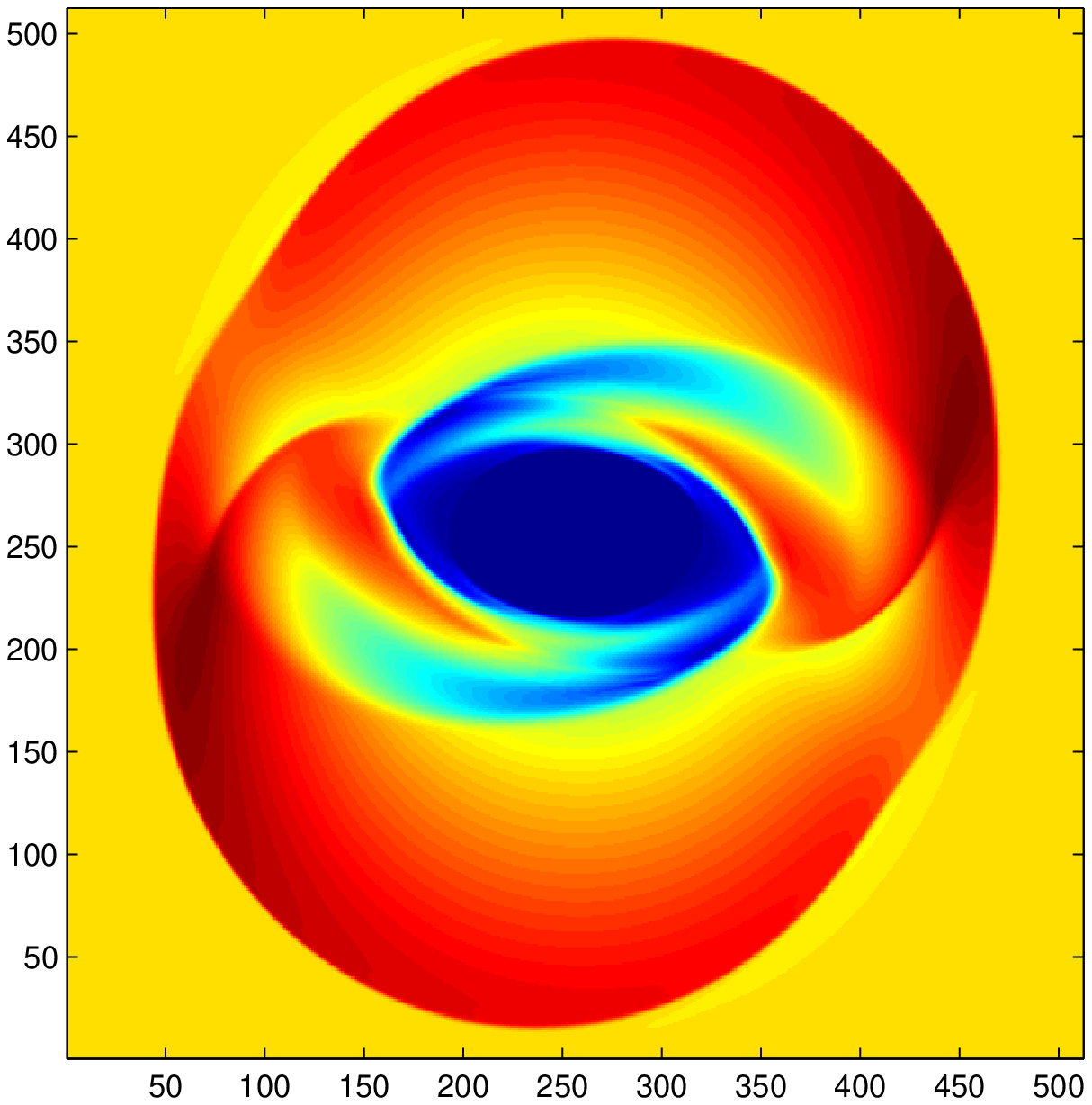}
\hfill \caption{Results of the density (top-left), gas pressure (top-right) of the first MHD rotor
test at $t= 0.15L$, results of the density (bottom-left), gas pressure (bottom-right) of the second
MHD rotor test at $t= 0.295L$, both computed with 512 $\times$ 512 grid points.}
\label{fig:2DRotor}
\end{center}
\end{figure}

The third 2D test is the MHD rotor problem. The problem was taken from~\cite{Balsara1999}. It
initiates a high density rotating disk with radius $r_{0} = 0.1$ of fluid measured from the center
point $(x,y) = (0.5, 0.5)$. The ambient fluid outside of the spherical region of $r_{1} = 0.115)$
has low density and $v_{x} = v_{y} = 0$, and the fluid between the high density disk fluid and
ambient fluid ($r_{1}
> r > r_{0},$ where $r = \sqrt{(x - 0.5)^{2} + (y - 0.5)^{2}}$) has
linear density and angular speed profile with $\rho = 1 + 9f, v_{x} = -fv_{0}(y-0.5)/r$ and $v_{y}
= -fv_{0}(x-0.5)/r$ where $f = (r_{1}-r)/(r_{1}-r_{0})$. Two initial value sets of $v_{0}, p,
B_{x}$ and $\gamma$ provided in~\cite{Balsara1999} and~\cite{Toth2000} were tested. The initial
condition for 2D MHD Rotor problem is listed as follows
\begin{equation}
\label{equat_2DRotor_init0} \begin{array}{l}\mbox{spherical region center} = (0.5, 0.5),\,\, r_{0} = 0.1,\,\, r_{1} = 0.115 \\
f = (r_{1}-r)/(r_{1}-r_{0}), (0 \leq x \leq 1)\,\,(0 \leq y \leq 1)
\end{array}
\end{equation}
$r < r_{0}$
\begin{equation}
\label{equat_2DRotor_init1} \left\{\begin{array}{c} v_{x} \\
v_{y} \\ v_{z} \end{array} \right\} \mbox{=} \left\{\begin{array}{c} -v_{0}(y-0.5)/r_{0} \\
v_{0}(x-0.5)/r_{0} \\ 0 \end{array} \right\}
\end{equation}
$r_{0} < r < r_{1}$
\begin{equation}
\label{equat_2DRotor_init2} \left\{\begin{array}{c} v_{x} \\
v_{y} \\ v_{z} \end{array} \right\} \mbox{=} \left\{\begin{array}{c} -fv_{0}(y-0.5)/r \\
-fv_{0}(x-0.5)/r \\ 0 \end{array} \right\}
\end{equation}
$r > r_{1}$
\begin{equation}
\label{equat_2DRotor_init3} \left\{\begin{array}{c} v_{x} \\
v_{y} \\ v_{z} \end{array} \right\} \mbox{=} \left\{\begin{array}{c} 0 \\
0 \\ 0 \end{array} \right\}
\end{equation}
\begin{equation}
\label{equat_2DRotor_init4} \mbox{$\rho = $}\left\{\begin{array}{c} 10 \\
1+9f \\ 1 \end{array} \begin{array}{c} \mbox{$r < r_{0}$} \\
\,\,\,\mbox{$r_{0} < r < r_{1}$} \\ \,\,\,\mbox{$r > r_{1}$}
\end{array}\right.
\end{equation}
First rotor problem:
\begin{equation}
\label{equat_2DRotor_init5} v_{0}=2, p=1, \gamma=1.4, t_{max}=0.15, \left\{\begin{array}{c} B_{x} \\
B_{y} \\ B_{z} \end{array} \right\} \mbox{=} \left\{\begin{array}{c} 5/\sqrt{4\pi} \\
0 \\ 0 \end{array} \right\}
\end{equation}
Second rotor problem:
\begin{equation}
\label{equat_2DRotor_init6} v_{0}=1, p=0.5, \gamma=5/3, t_{max}=0.295 \left\{\begin{array}{c} B_{x} \\
B_{y} \\ B_{z} \end{array} \right\} \mbox{=} \left\{\begin{array}{c} 2.5/\sqrt{4\pi} \\
0 \\ 0 \end{array} \right\}
\end{equation}
In Fig.~\ref{fig:2DRotor}, we present images of the density, gas pressure of the two rotor problems
computed with 512 $\times$ 512 grid points. The results are in excellent agreement with those
presented in~\cite{Balsara1999} and~\cite{Toth2000}.

\subsection{Three-dimensional blast wave problem}\label{sub:3DBlast}

The 3D version of MHD spherical blast wave problem was also tested. The condition is listed as
follows~\cite{Spherical}
\begin{equation}
\label{equat_3DBlast_init1} \left\{\begin{array}{c} v_{x} \\
v_{y} \\ v_{z} \end{array} \right\} \mbox{=} \left\{\begin{array}{c} 0 \\
0 \\ 0 \end{array} \right\}
\end{equation}
\begin{equation}
\label{equat_3DBlast_init2} \left\{\begin{array}{c} B_{x} \\
B_{y} \\ B_{z} \end{array} \right\} \mbox{=} \left\{\begin{array}{c} 1/\sqrt{3} \\
1/\sqrt{3} \\ 1/\sqrt{3} \end{array} \right\}
\end{equation}
\begin{equation}
\label{equat_3DBlast_init3} \mbox{$p = $}\left\{\begin{array}{c} 10 \\
0.1 \end{array} \begin{array}{c} \mbox{inside the spherical region} \\
\mbox{outside the spherical region} \end{array}\right.
\end{equation}
\begin{equation}
\label{equat_3DBlast_init4} \begin{array}{l}\rho = 1, \,\,\,\,\,
\gamma = 5/3 \\ \mbox{spherical region center} = (0.5, 0.5, 0.5),\,\, r = 0.1 \\
(0 \leq x \leq 1)\,\,(0 \leq y \leq 1)\,\, (0 \leq z \leq 1)
\end{array}
\end{equation}

\begin{figure}[h]
\begin{center}
\includegraphics*[width=2.5in]{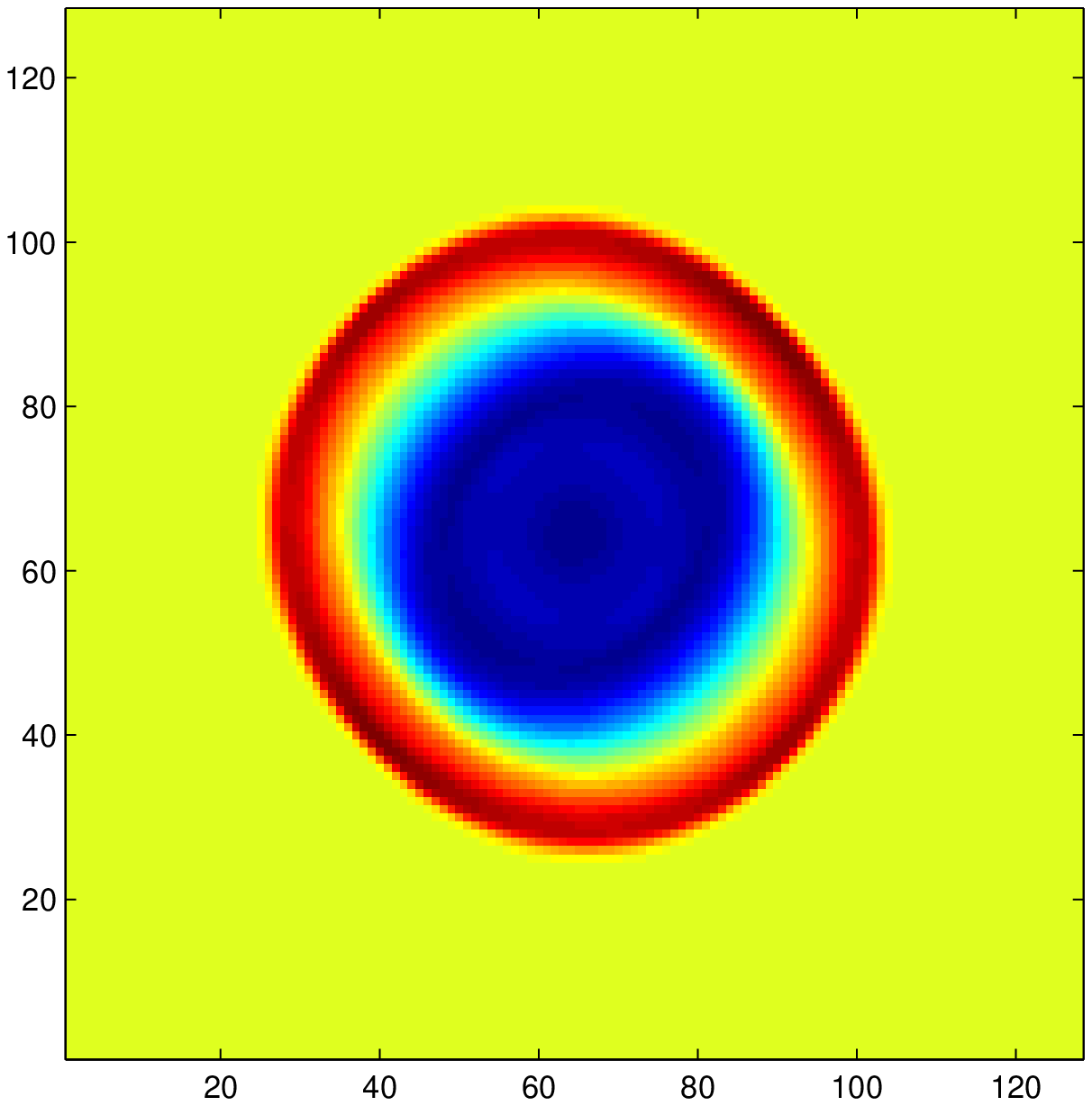}
\includegraphics*[width=2.5in]{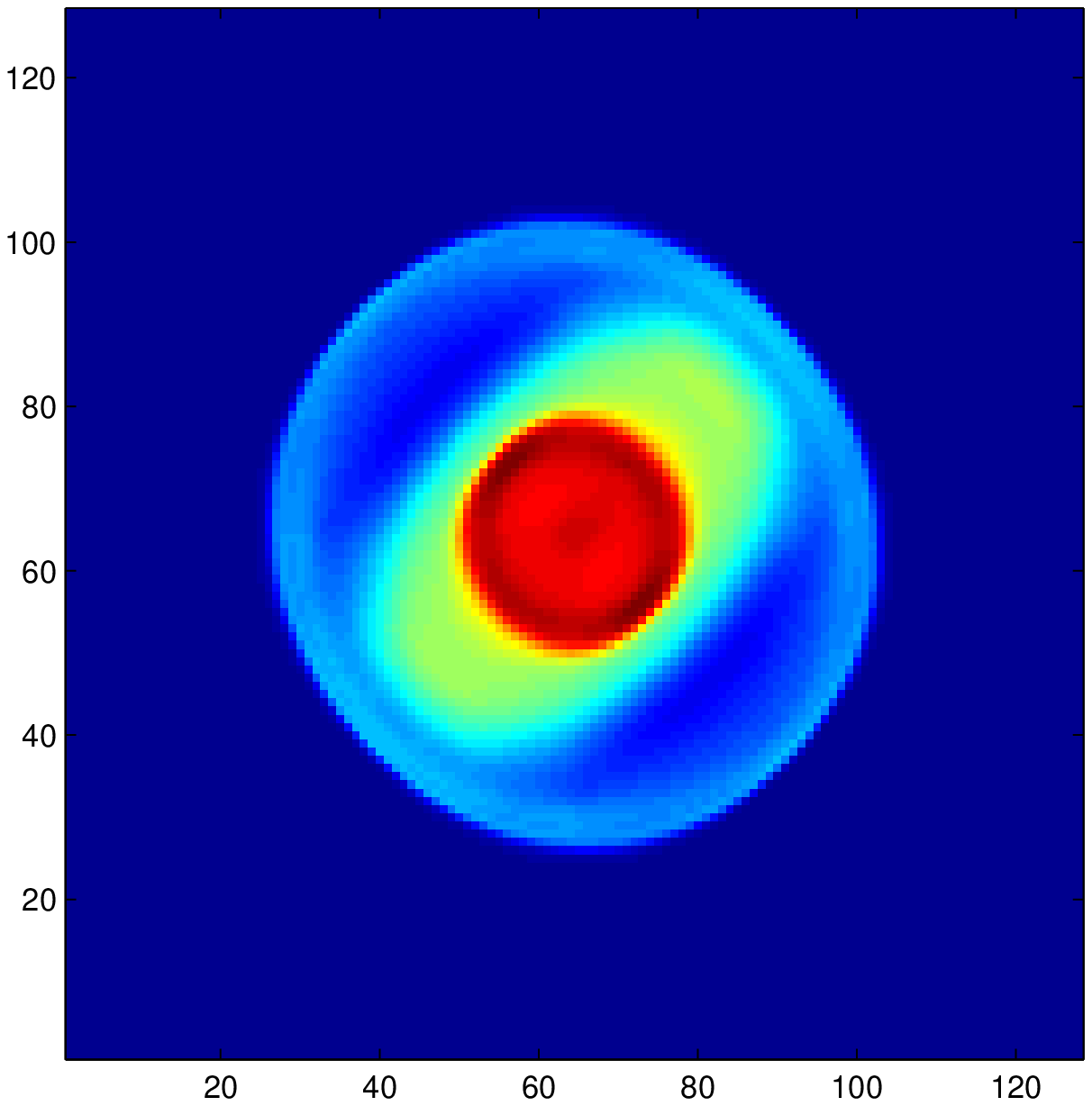}
\includegraphics*[width=2.5in]{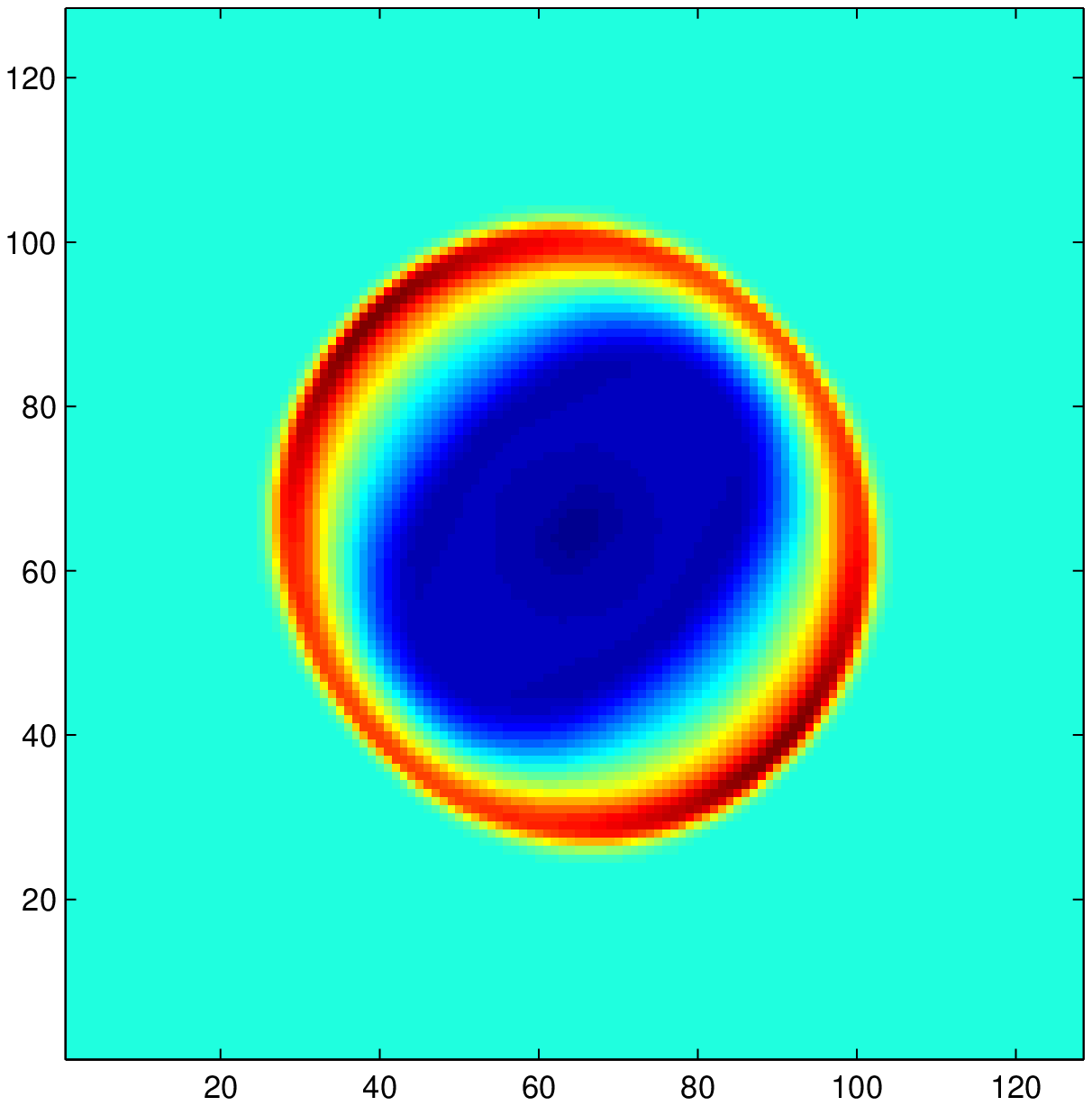}
\hfill \caption{Results of the density (top-left), gas pressure (top-right) and magnetic pressure
(bottom) of 3D blast wave test at $t = 0.1L$ sliced along $x$-$y$ plane at $z$ = 0.5 and computed
with 128 $\times$ 128 $\times$ 128 grid points.}\label{fig:3DBlast01}
\end{center}
\end{figure}

\begin{figure}[h]
\begin{center}
\includegraphics*[width=2.5in]{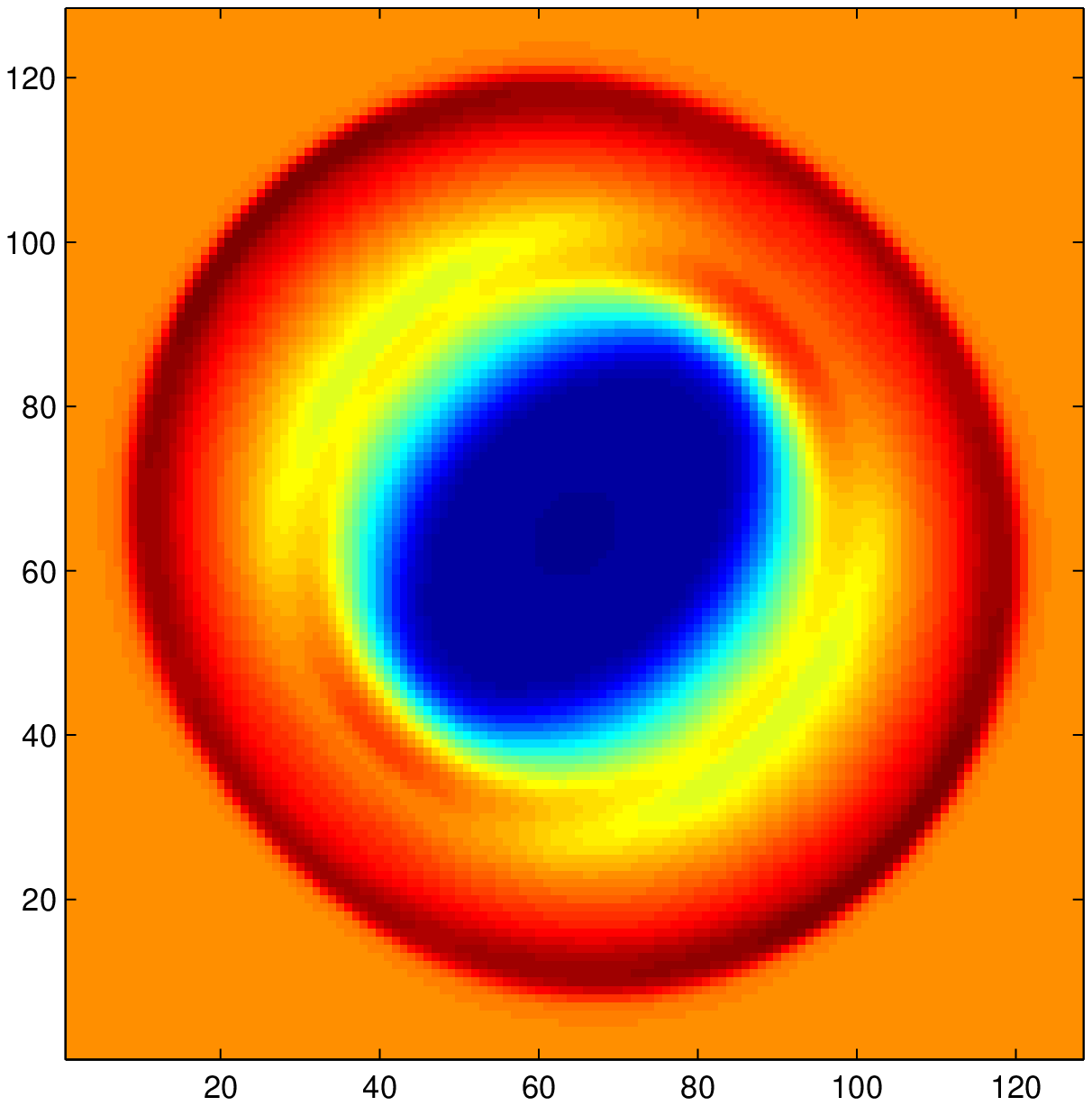}
\includegraphics*[width=2.5in]{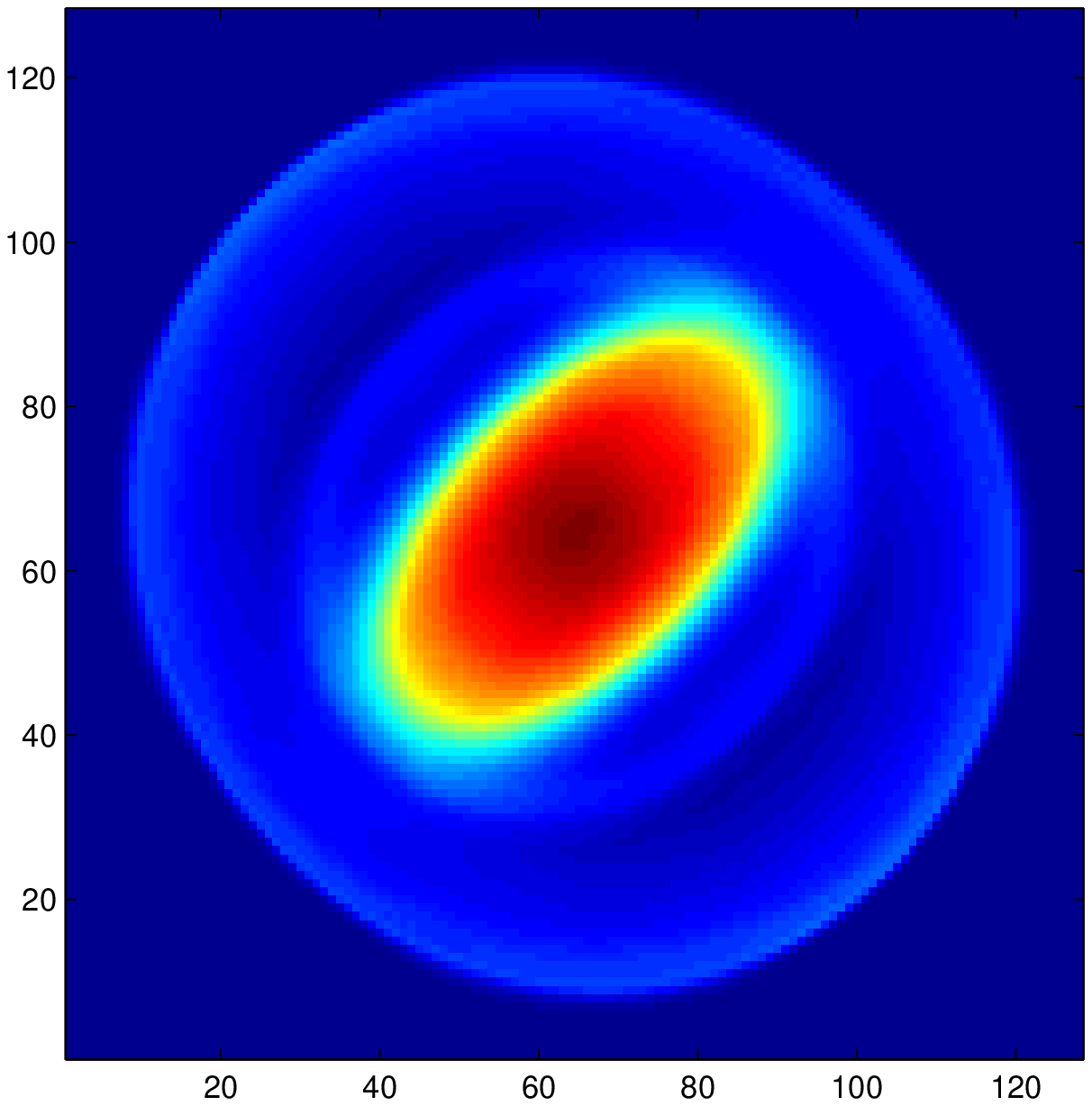}
\includegraphics*[width=2.5in]{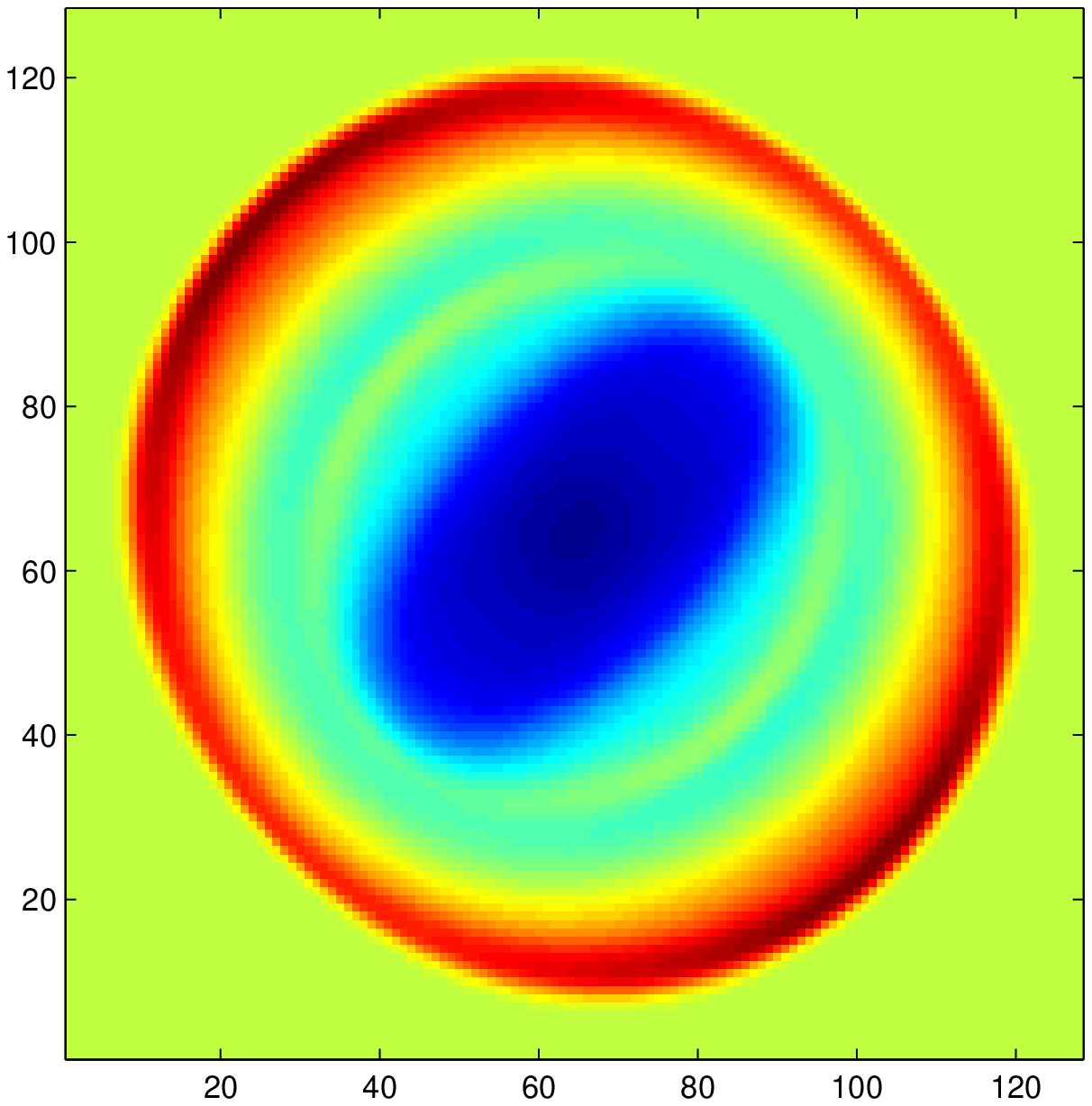}
\hfill \caption{Results of the density (top-left), gas pressure (top-right) and magnetic pressure
(bottom) of 3D blast wave test at $t = 0.2L$ sliced along $x$-$y$ plane at $z$ = 0.5 and computed
with 128 $\times$ 128 $\times$ 128 grid points.}\label{fig:3DBlast02}
\end{center}
\end{figure}

Fig.~\ref{fig:3DBlast01} and Fig.~\ref{fig:3DBlast02} show the results of 3D blast wave problem,
which include the density, gas pressure, and magnetic pressure at $t = 0.1L$ and $t = 0.2L$ sliced
along $x$-$y$ plane at $z$ = 0.5. The test was computed with 128 $\times$ 128 $\times$ 128 grid
points. Due to the scarcity of published 3D test results, we do not make direct contact with
results presented in the literature here. Considering only the ${\bu}$ and ${\bB}$, the memory
requirement of $256^3$ MHD problem is about 512MB GRAM for single precision and 1024MB GRAM for
double precision, respectively. If the storage of intermediate results such as ${\bB_{temp}}$,
${\bu_{temp}}$, ${\bf flux_{temp}}$ and $F$ etc. (See Section~\ref{sec:sweeping}) are considered,
the amount of memory requirement will be about 2.25GB (single precision). As we mentioned in
Section~\ref{sec:sweeping}, not all the capacity of GRAM can be used to store data arbitrarily. As
we said in the beginning of this section, there are actually two GPUs inside the GTX 295 and the
1.75 GB GRAM is the total amount of the GRAM shared by two GPUs, so that only less than $1.75 / 2 =
0.875$ GB GRAM can be used. As a result, the test of 3D problem with $256^3$ resolution are not
able to be provided on a graphics card.

\section{Accuracy evaluation}\label{sec:accuracy}
\par
In MHD simulations, accuracy is always to be considered since the error may increase fast and crash
the simulation if low precision is used for computation. Scientific computations such as MHD
simulation mostly use double precision to reduce errors. In this section, the results generated by
{\it GPU-MHD} using single precision and double precision modes are shown and compared.

The difference between the results of double precision and single precision computation of the $512
\times 1 \times 1$ one-dimensional Brio-Wu shock tube problem is shown in
Fig.~\ref{fig:BrioWu_diff}. Two curves are almost the same but there are actually some differences
with the amount of $error \leq \pm10^{-6}$:

\begin{figure}[h]
\begin{center}
\includegraphics*[width=2.5in]{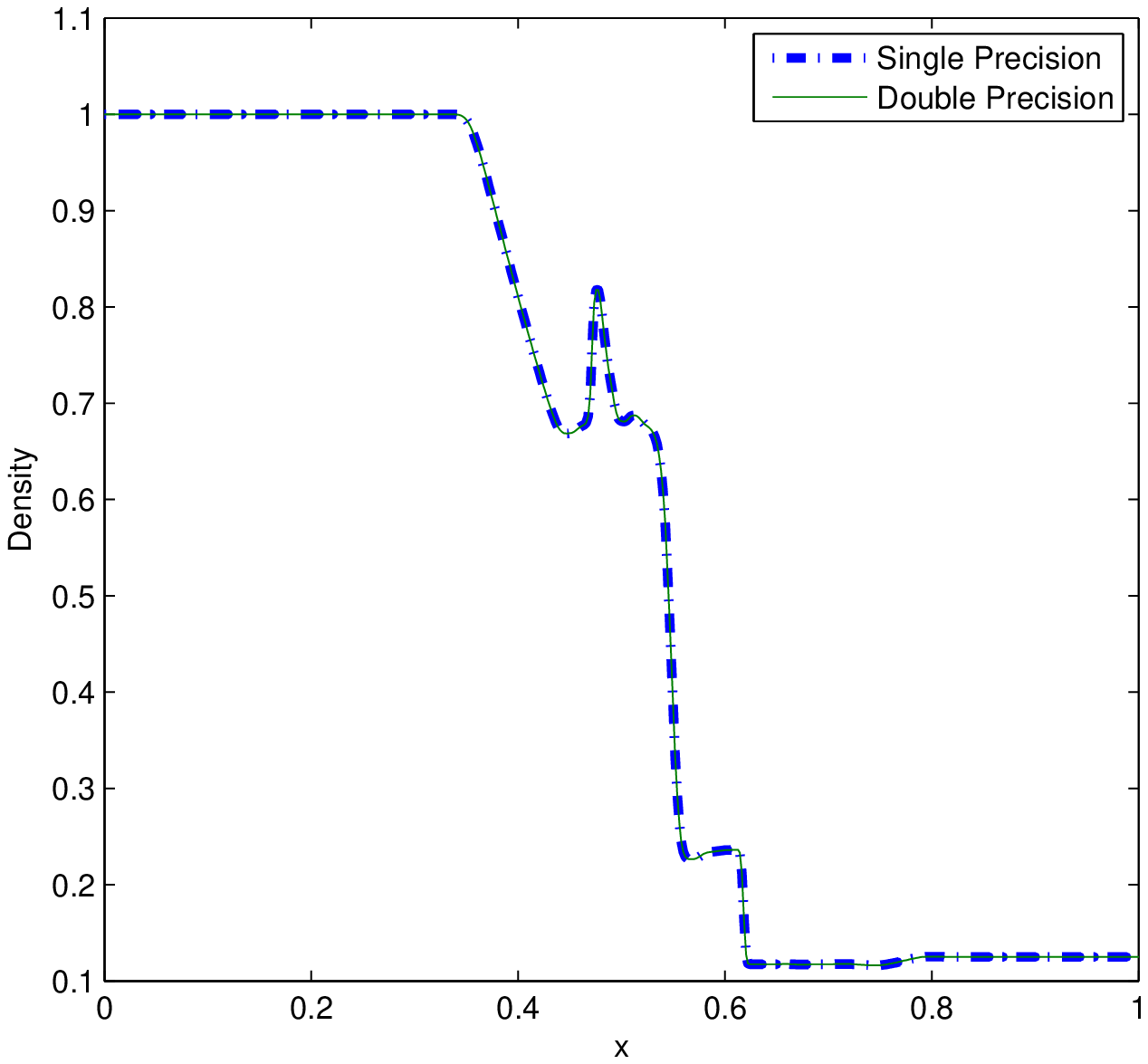}
\includegraphics*[width=2.5in]{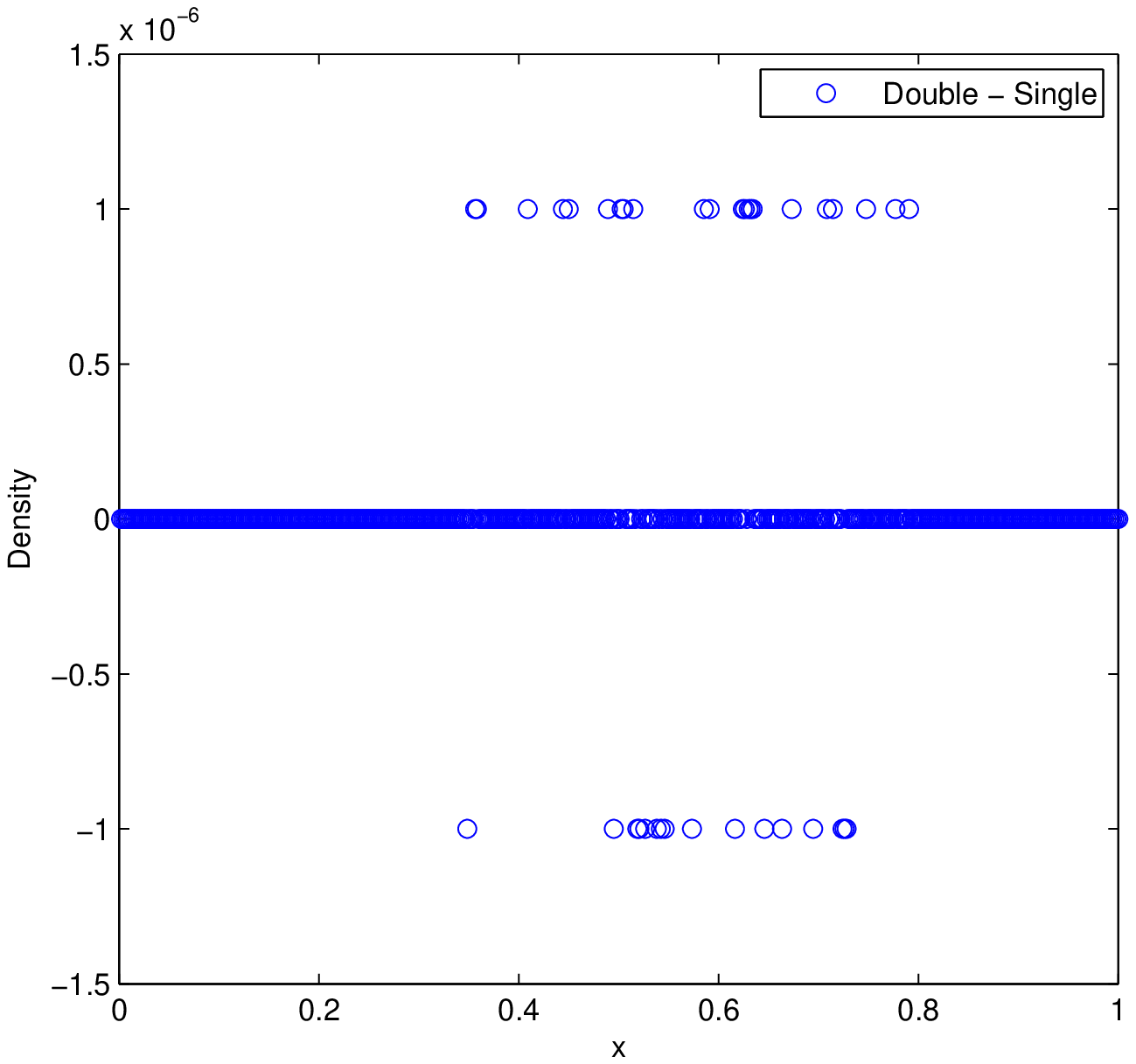}
\includegraphics*[width=2.5in]{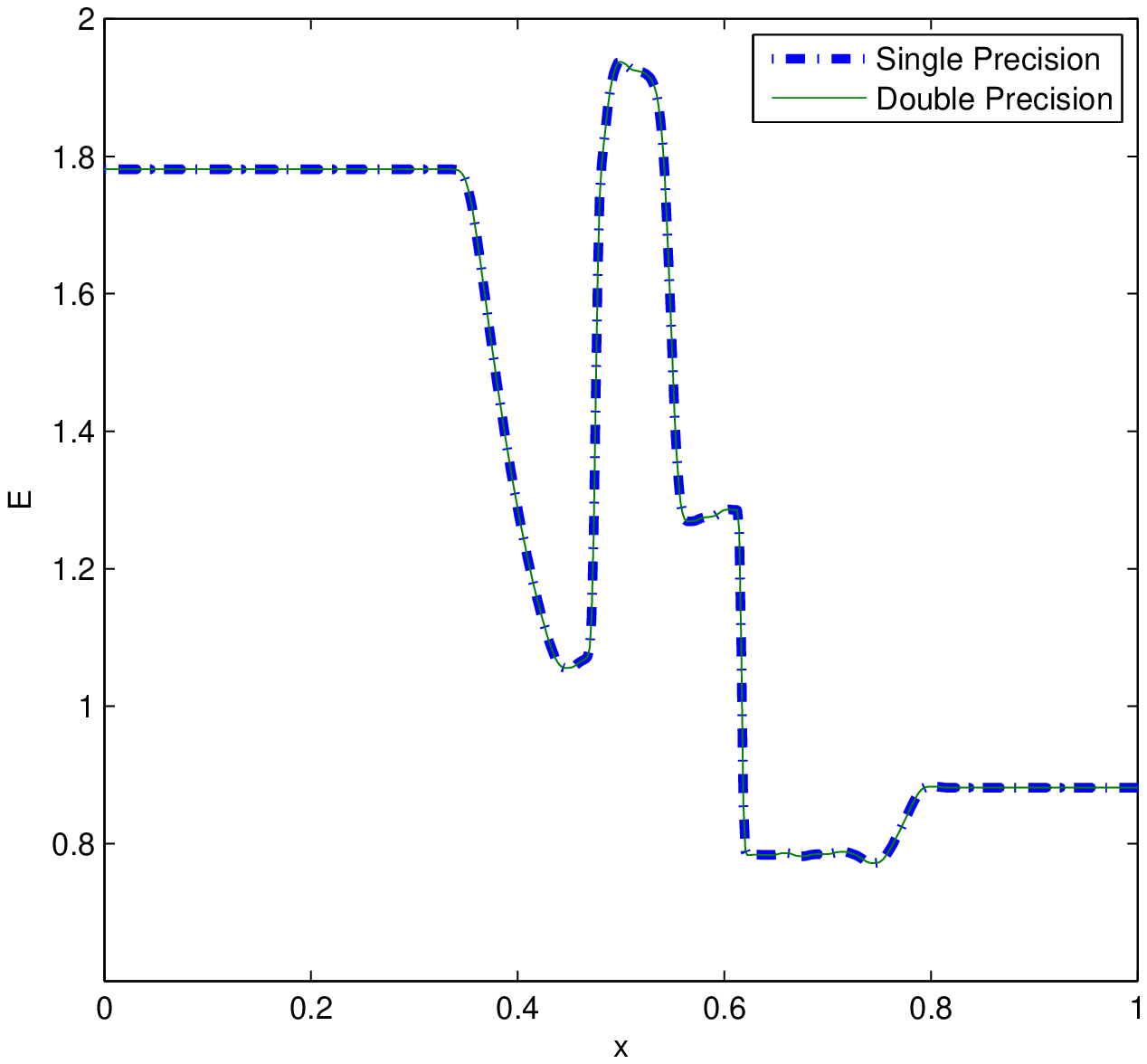}
\includegraphics*[width=2.5in]{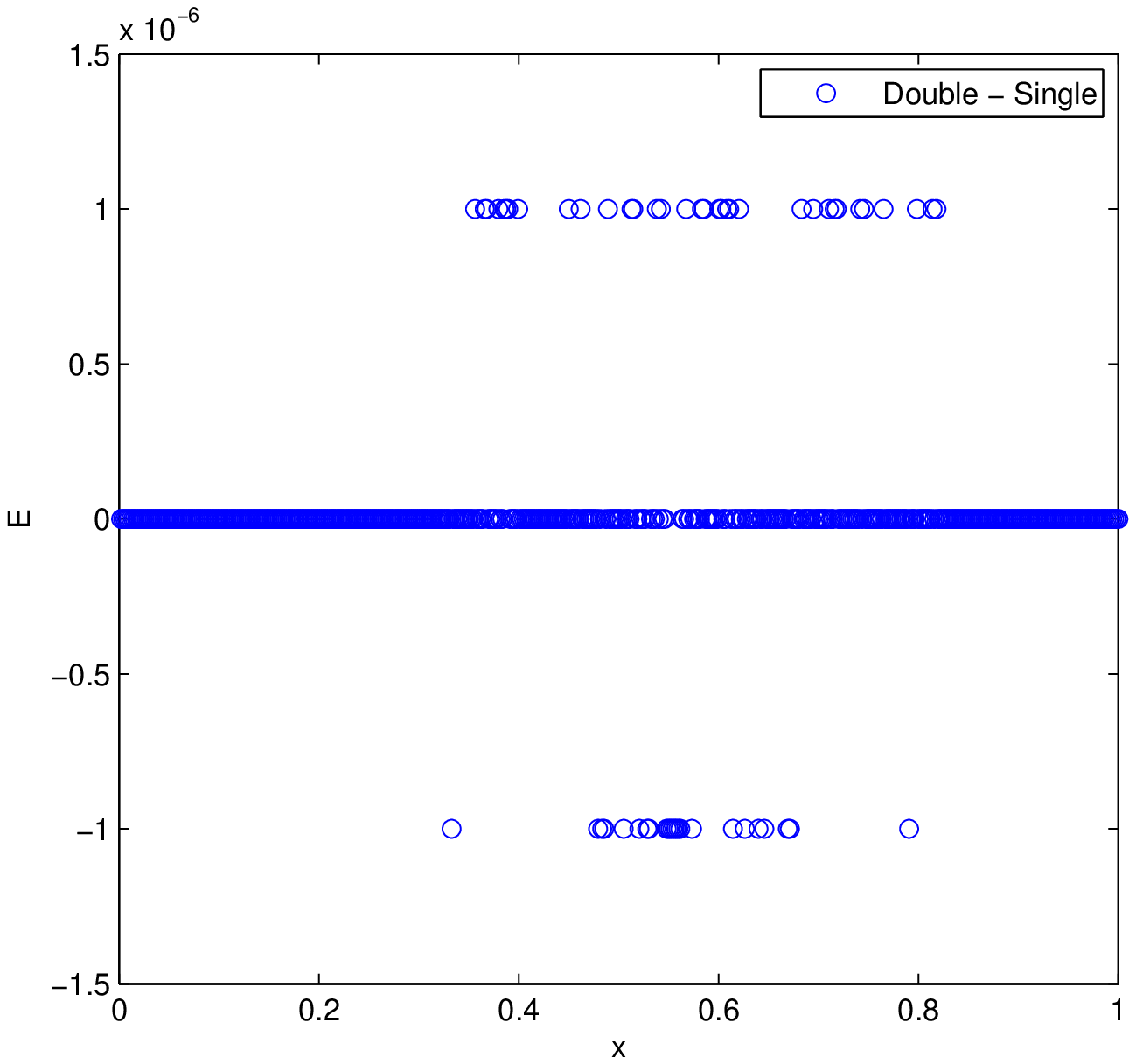}
\includegraphics*[width=2.5in]{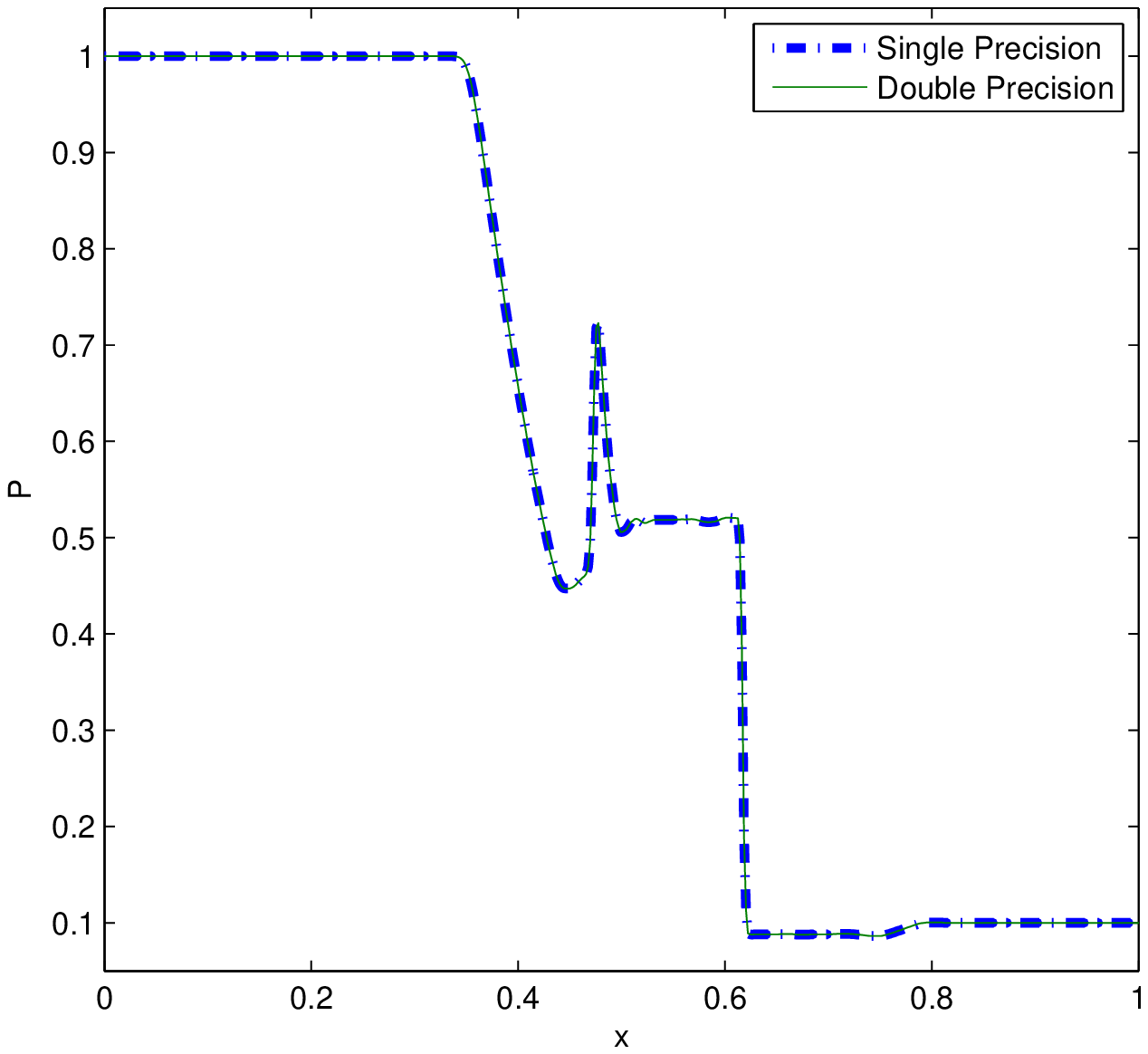}
\includegraphics*[width=2.5in]{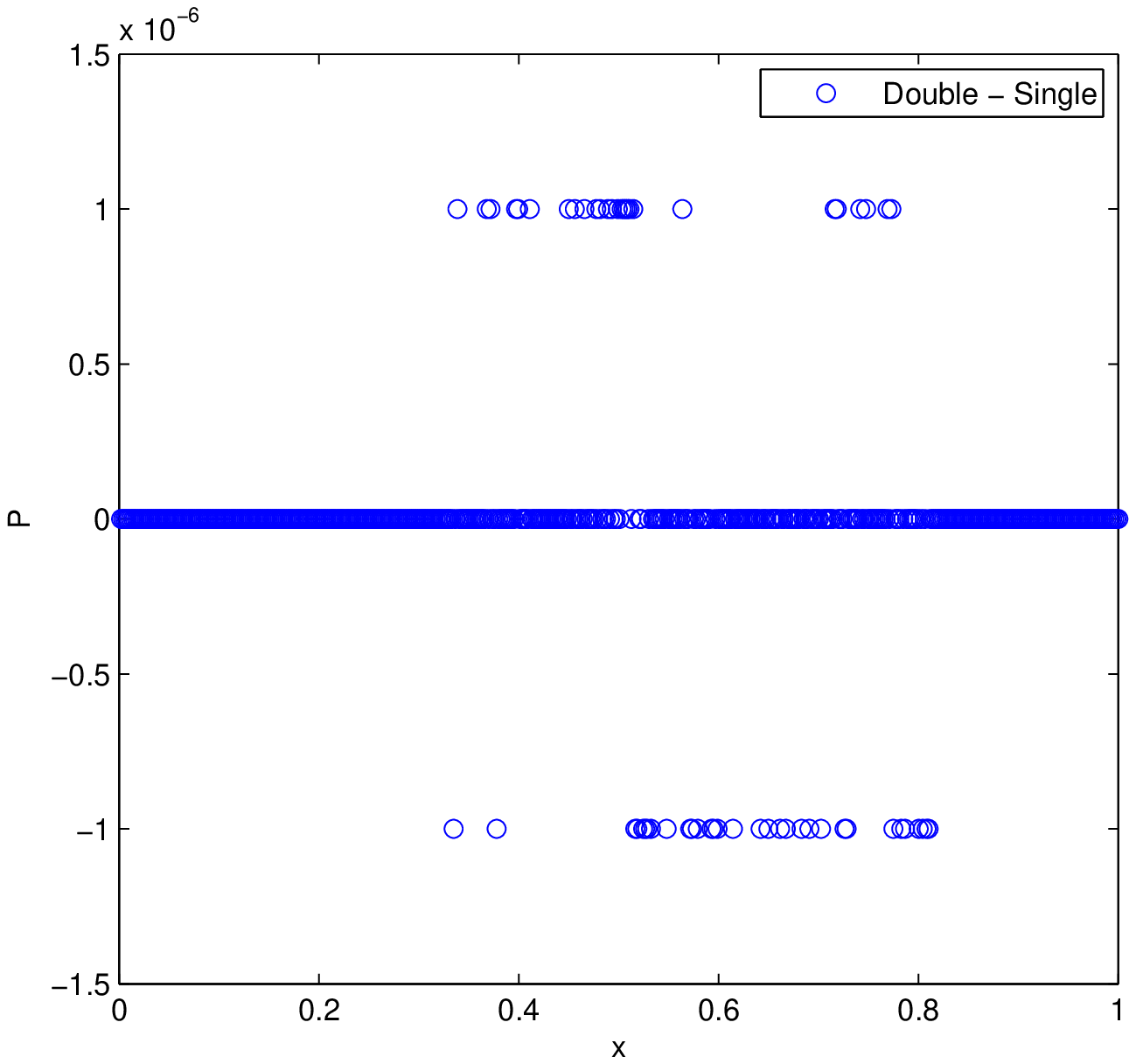}
\hfill \caption{Result of $\rho_{double}-\rho_{single}$ (top), $E_{double}-E_{single}$ (middle) and
$p_{double}-p_{single}$ (bottom) of 1D Brio-Wu shock tube problem at $t = 0.08L$ with 512 grid
points} \label{fig:BrioWu_diff}
\end{center}
\end{figure}

\begin{figure}[h]
\begin{center}
\includegraphics*[width=2.5in]{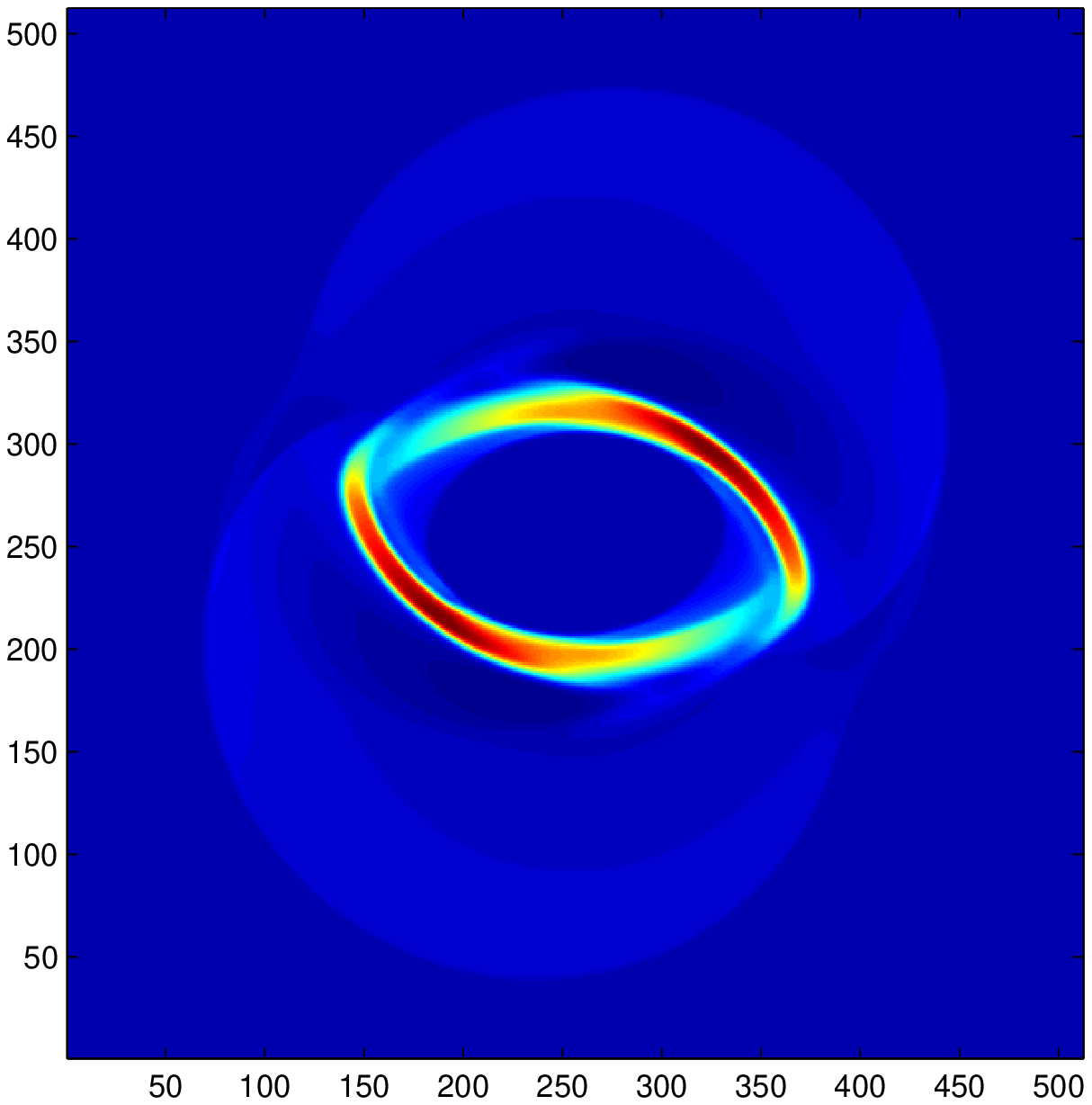}
\includegraphics*[width=2.5in]{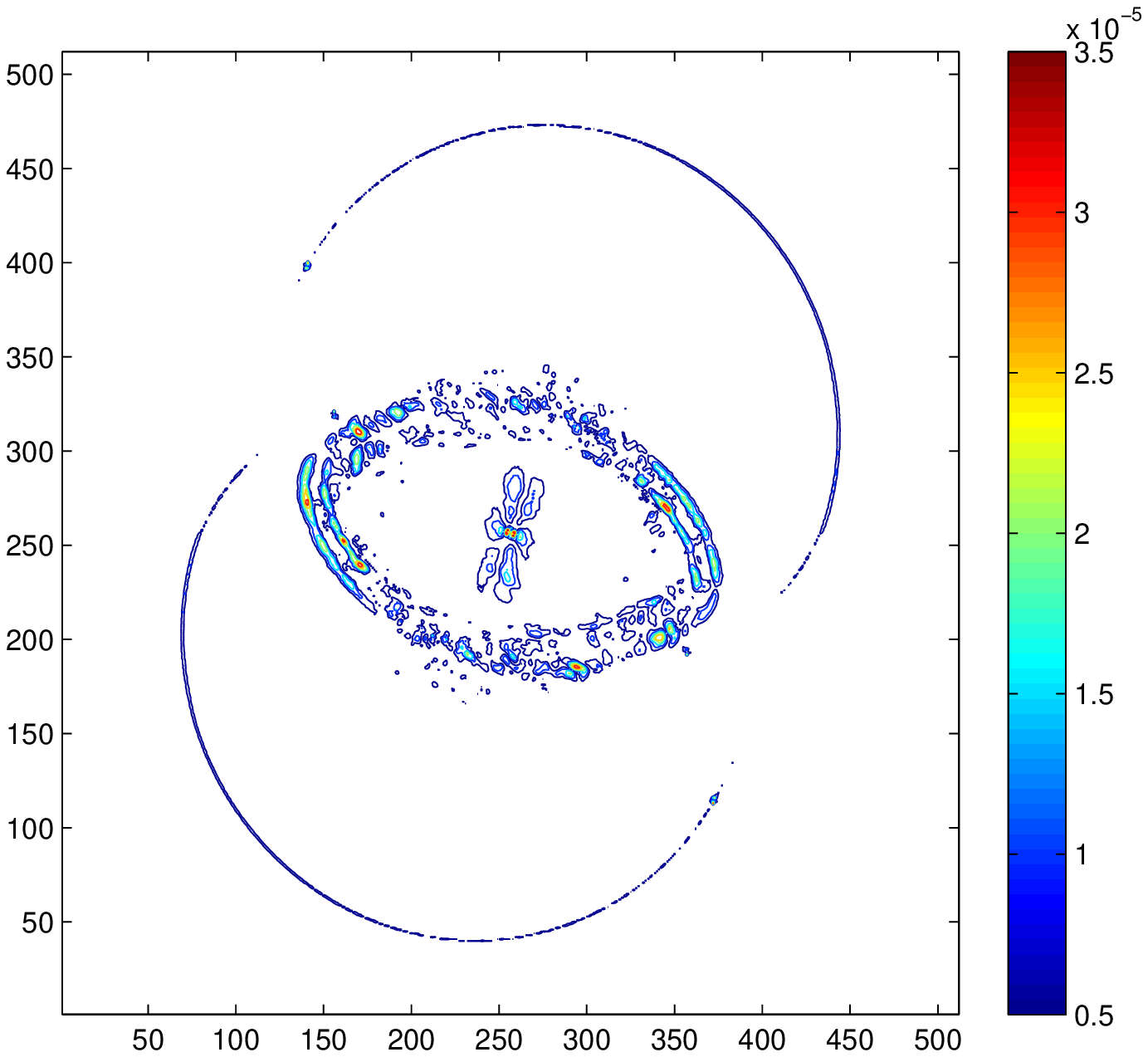}
\hfill \caption{Results of $\rho_{double}$ (left) and $|\rho_{double}-\rho_{single}|$ (right) of
MHD rotor problem at $t = 0.15L$ with $512^2$ grid points} \label{fig:Rotor_diff}
\end{center}
\end{figure}

\begin{figure}[h]
\begin{center}
\includegraphics*[width=2.5in]{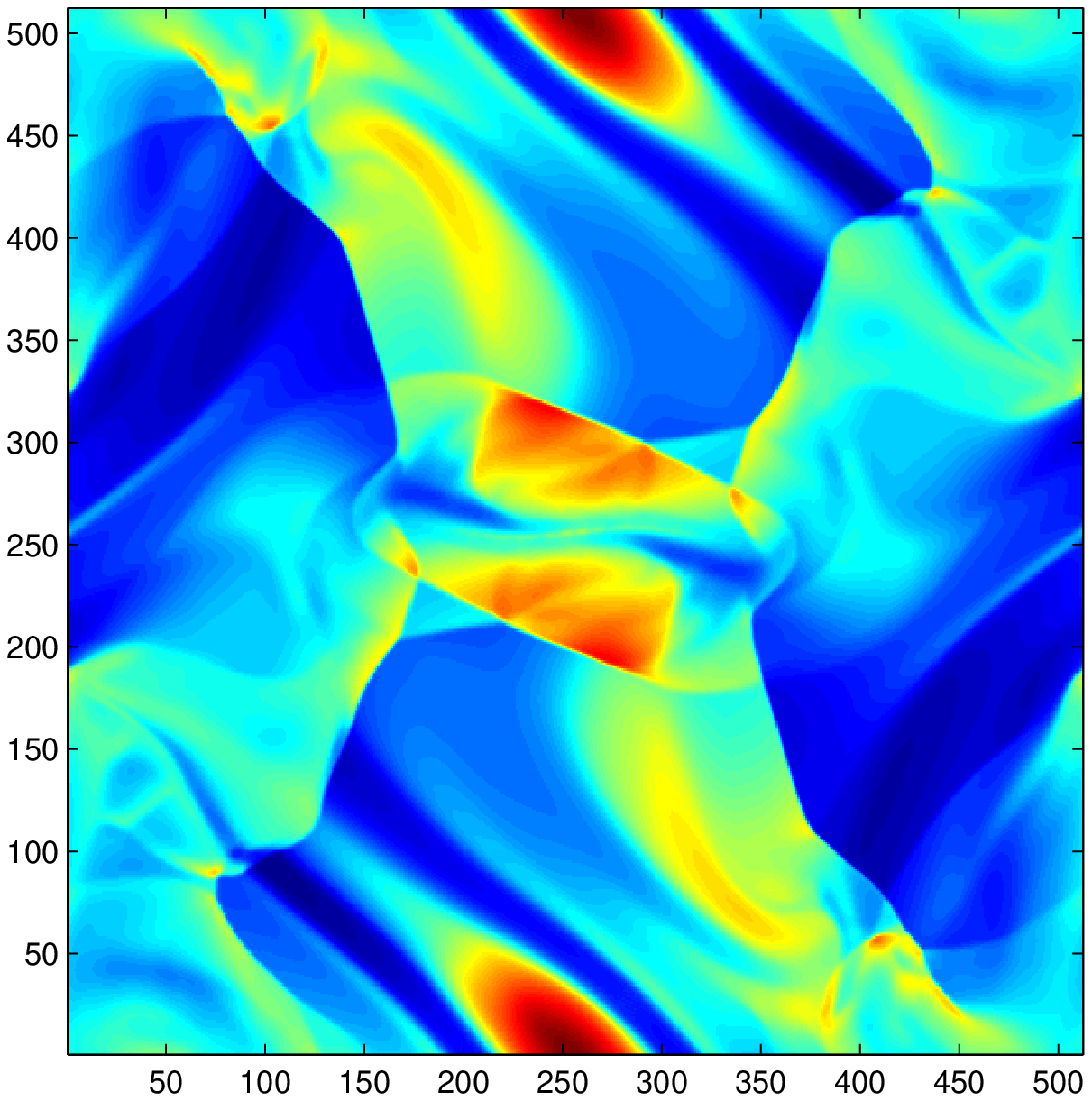}
\includegraphics*[width=2.5in]{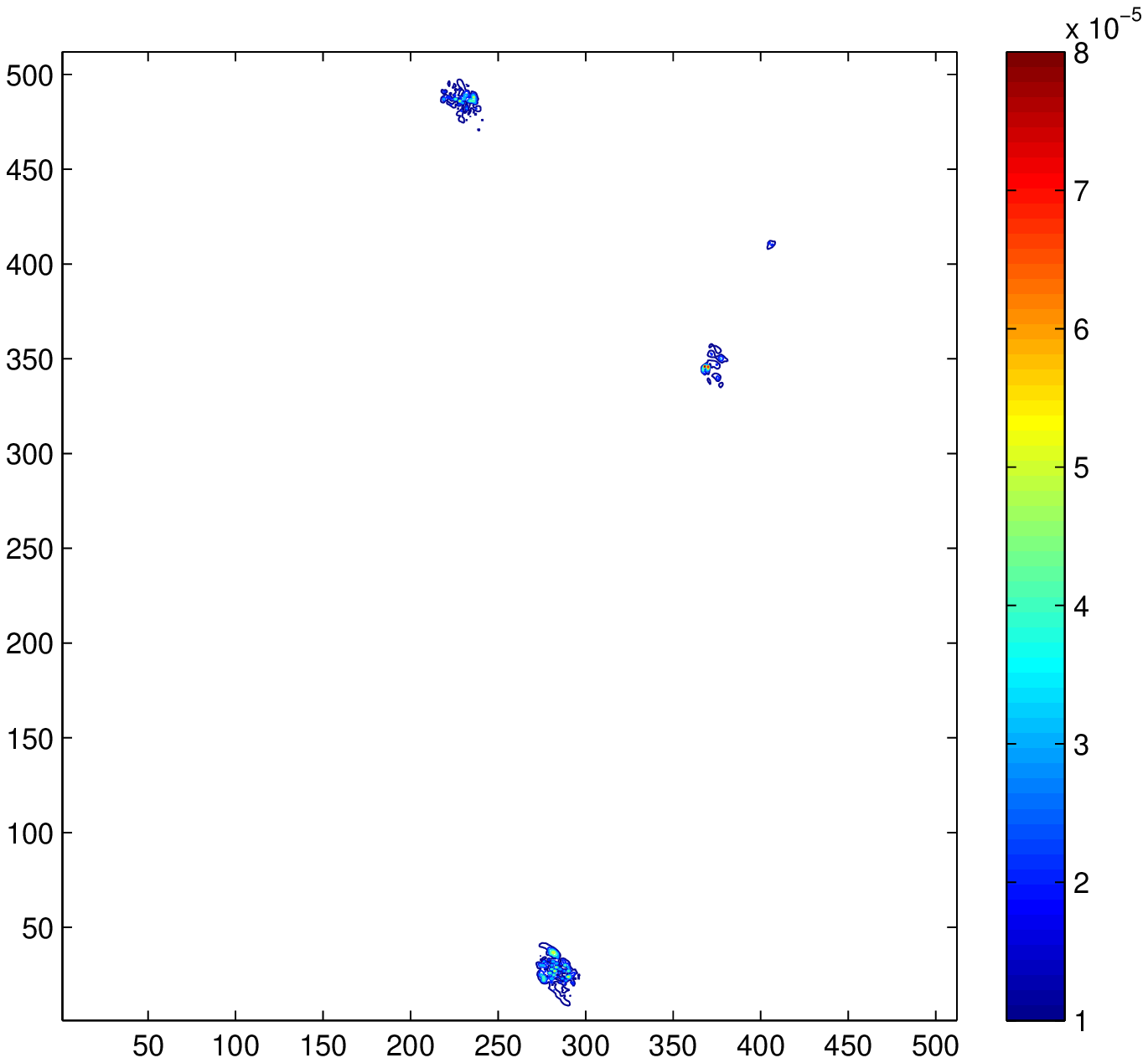}
\includegraphics*[width=2.5in]{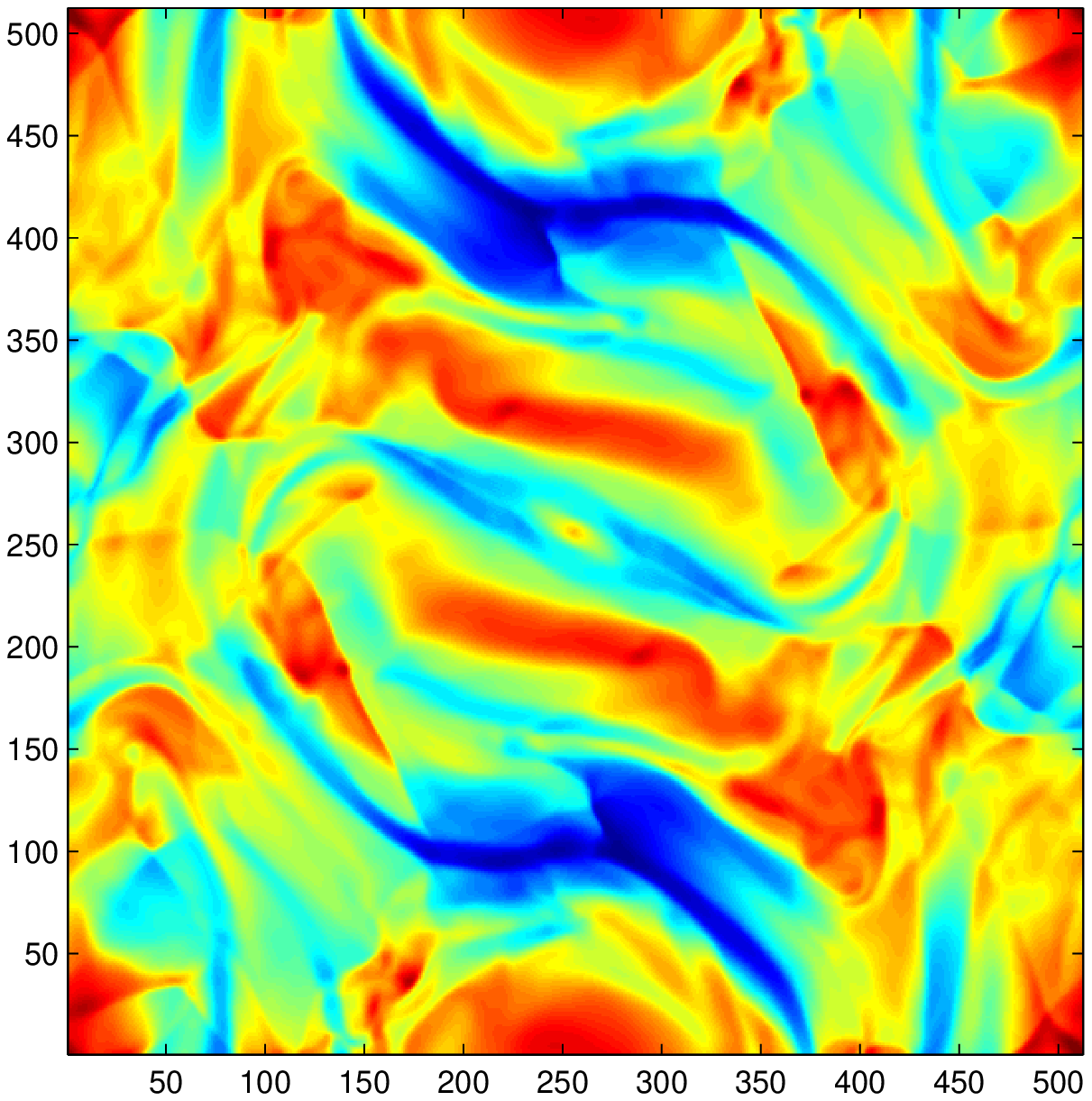}
\includegraphics*[width=2.5in]{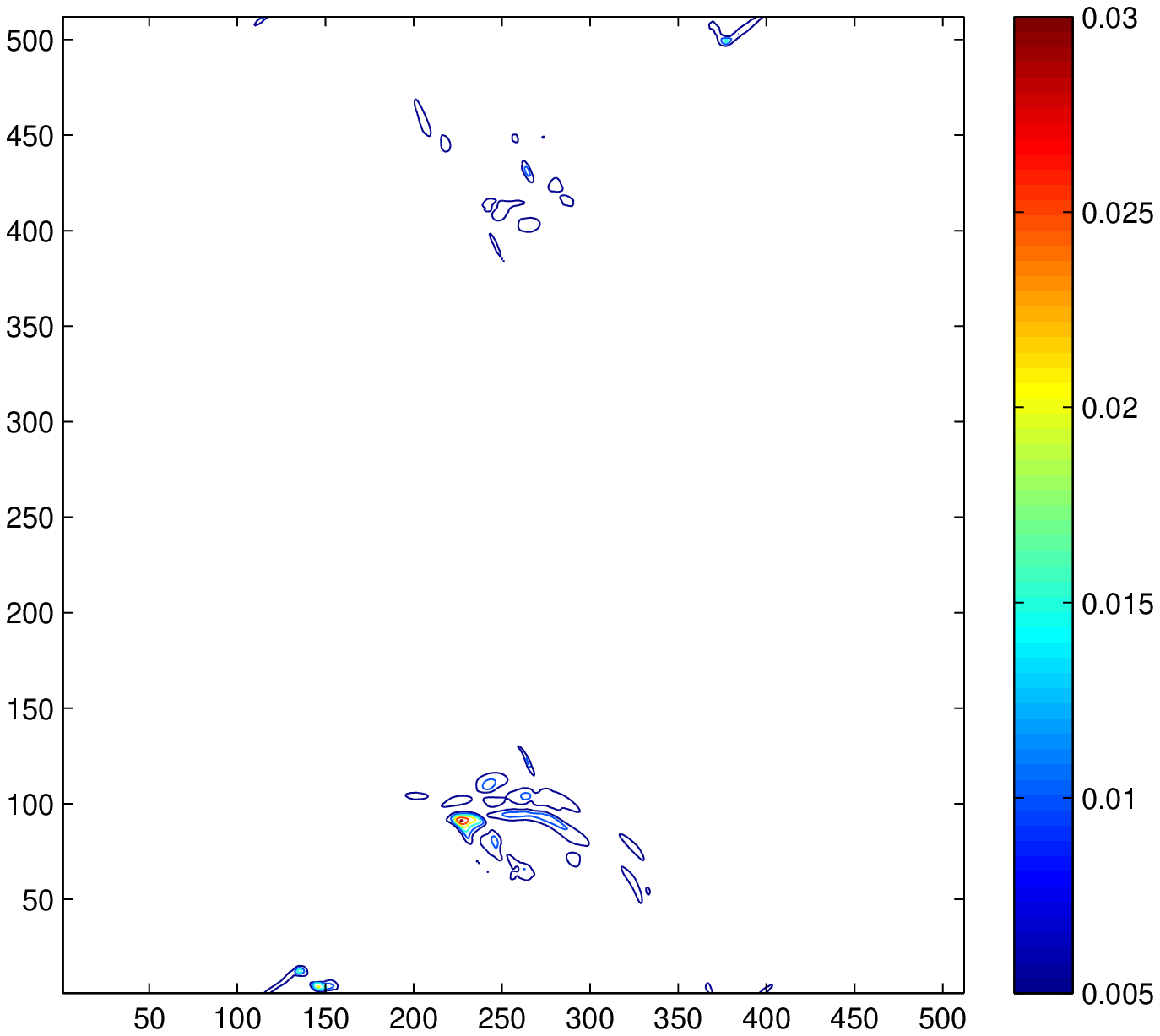}
\hfill \caption{Results of $\rho_{double}$ (left) and $|\rho_{double}-\rho_{single}|$ (right) of
Orszag-Tang problem at $t = 0.5L$ (top) and $t = 1.0L$ (bottom) with $512^2$ grid points}
\label{fig:OT_diff}
\end{center}
\end{figure}

In 2D cases, the absolute difference between the results of double precision and single precision
computation of MHD Rotor test ($t = 0.15L$) and Orszag-Tang vortex test ($t = 0.5L$ and $t = 1.0L$)
are shown in Fig.~\ref{fig:Rotor_diff} and Fig~\ref{fig:OT_diff}, respectively. The double
precision computation results of both tests are also shown in the left-hand side of these figures.

For the MHD Rotor test, even the resulting image (left in Fig.~\ref{fig:Rotor_diff}) looks similar
to the single precision resulting image (top-left of Fig.~\ref{fig:2DRotor}), the high differences
at the dense region can be found. Experimental result shows that the maximum $error$ is larger than
$\pm 3.5 \times 10^{-4}$.

Fig.~\ref{fig:OT_diff} shows the absolute difference between the results of double precision and
single precision computation of Orszag-Tang test at $t = 0.5L$ and $t = 1.0L$. As the simulation
time increases, the maximum $error$ increases from about $\pm 8\times 10^{-5}$ to $\pm 0.03$.

\begin{figure}[h]
\begin{center}
\includegraphics*[width=2.5in]{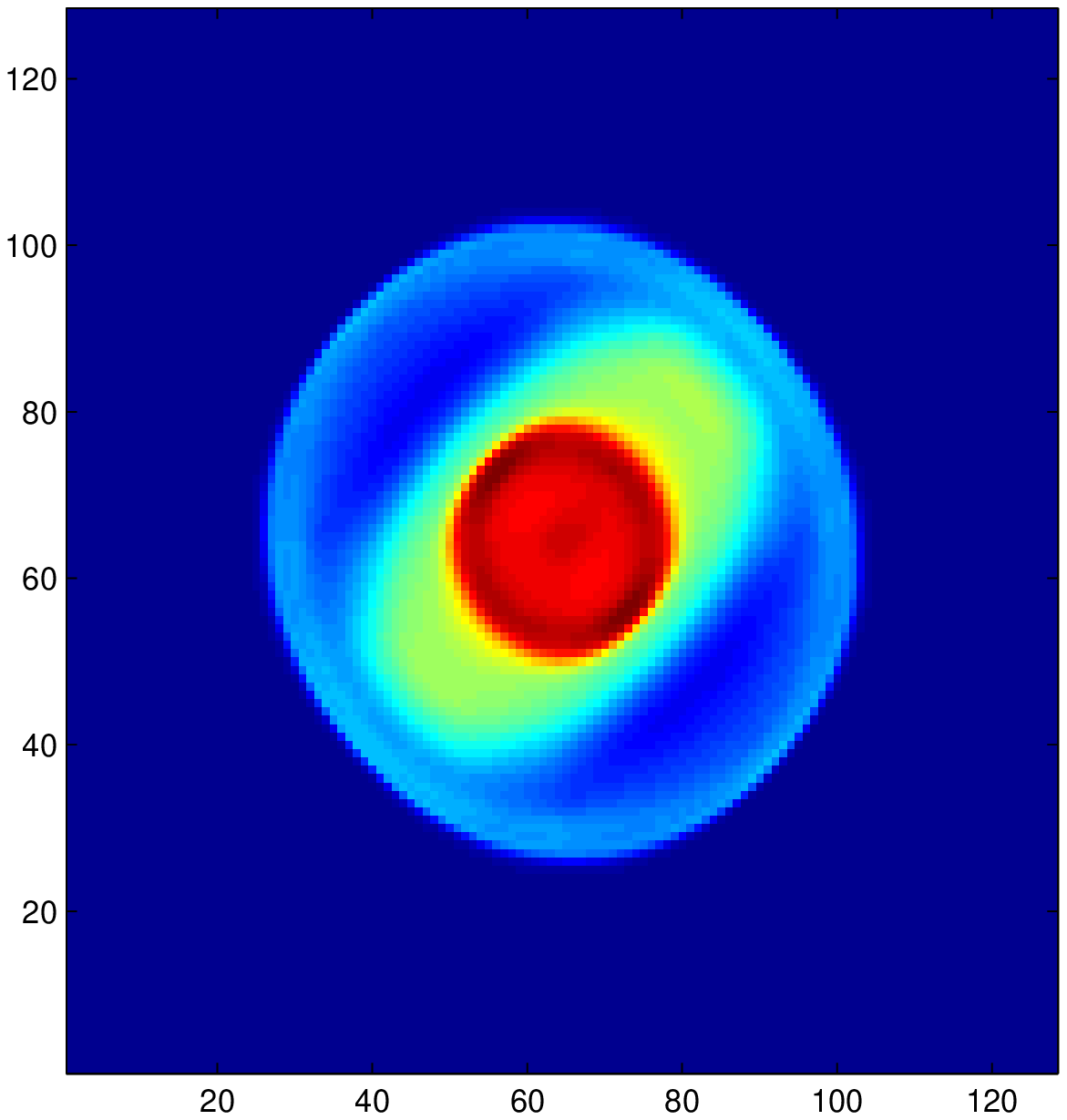}
\includegraphics*[width=2.5in]{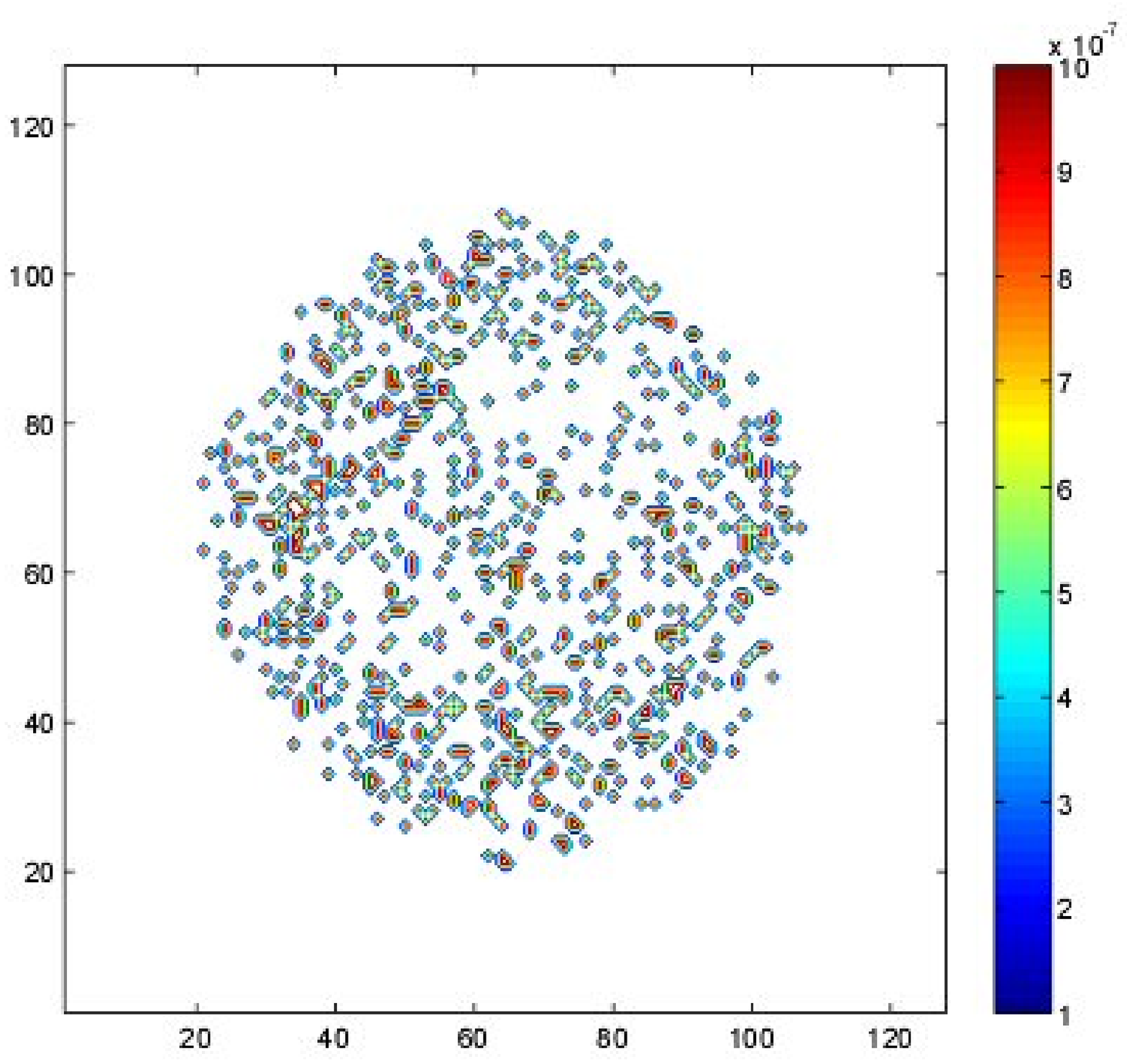}
\includegraphics*[width=2.5in]{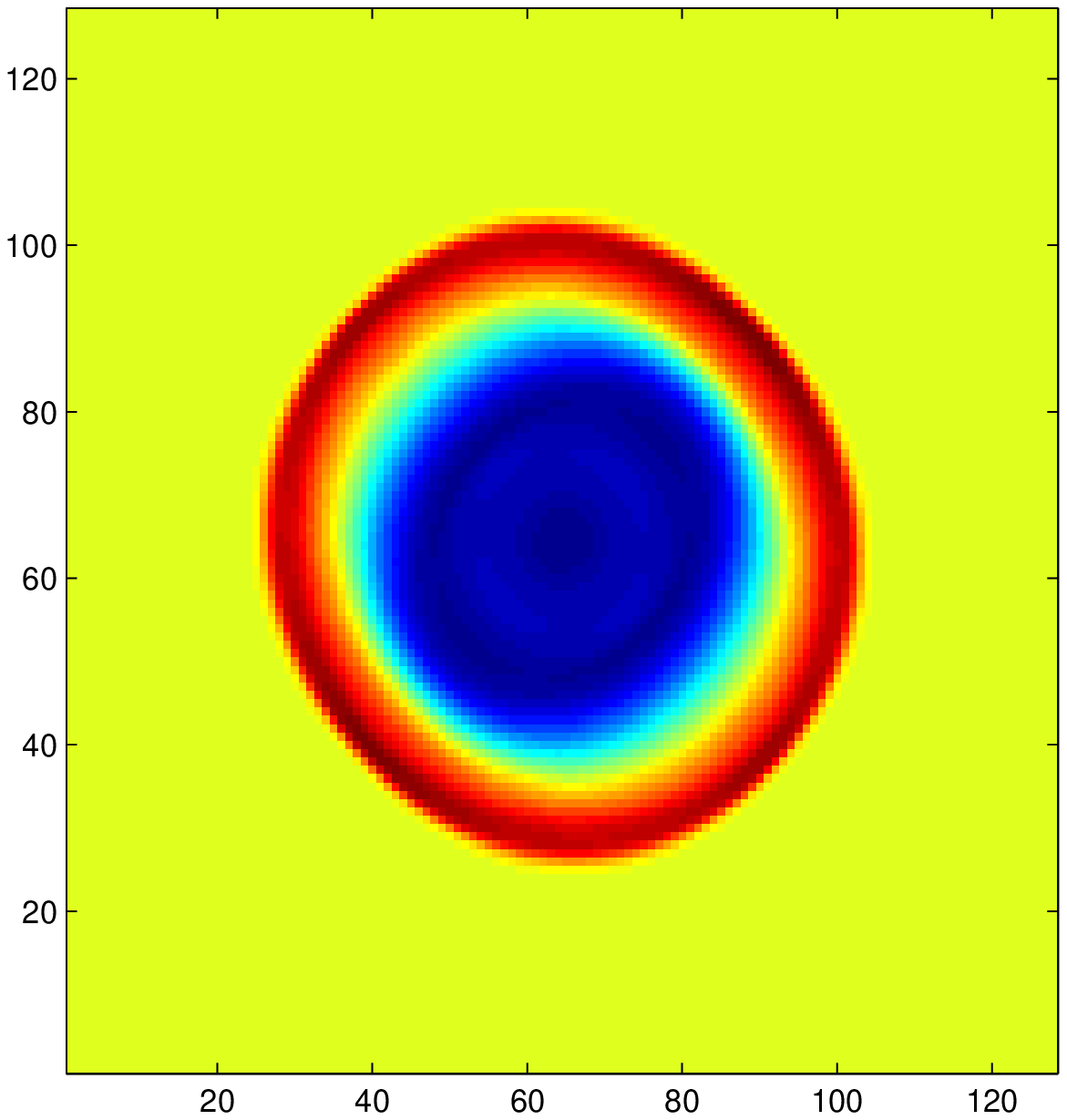}
\includegraphics*[width=2.5in]{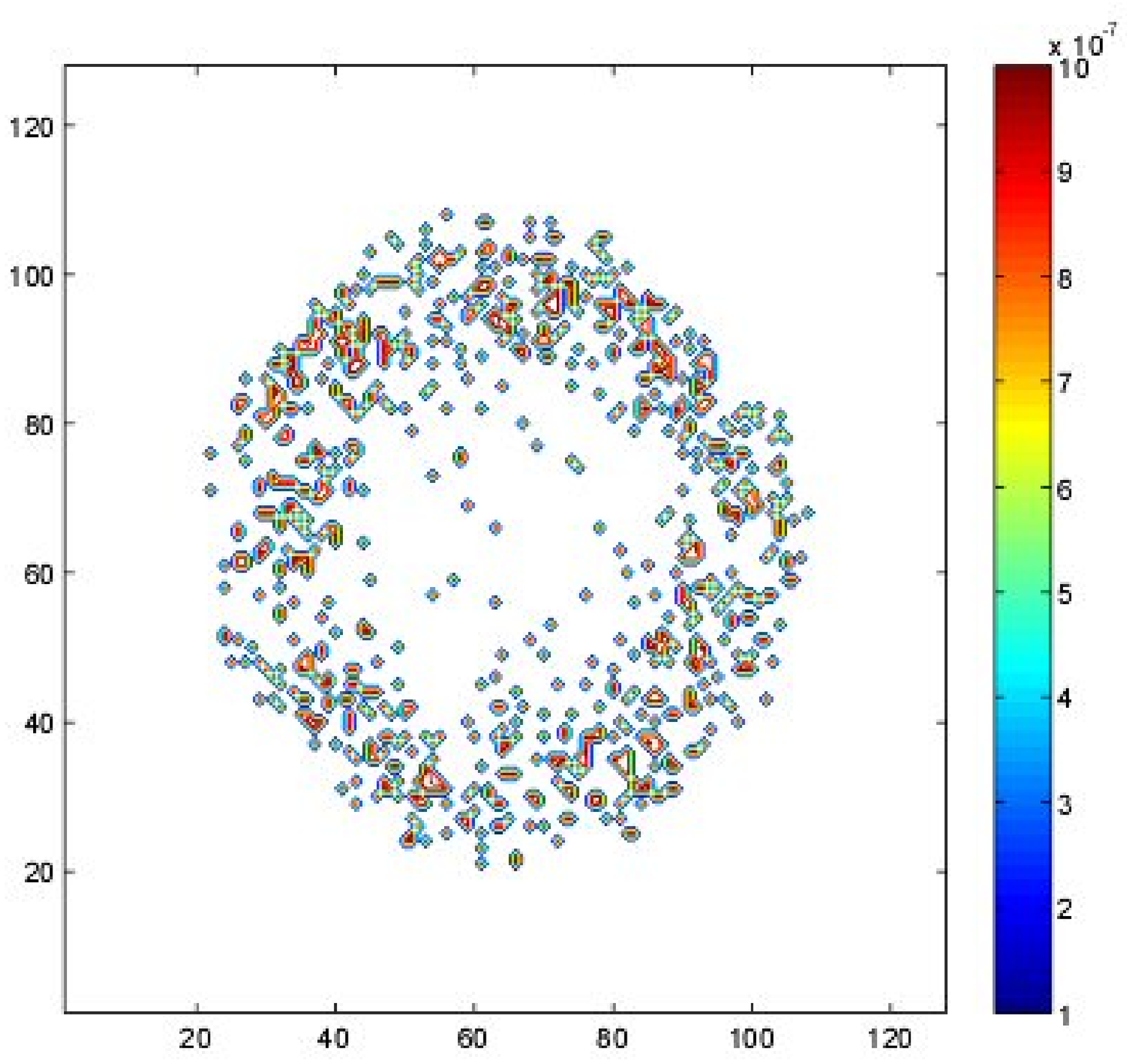}
\hfill \caption{Results of $p_{double}$ (top-left), $\rho_{double}$ (bottom-left) and
$|p_{double}-p_{single}|$ (top-right), $|\rho_{double}-\rho_{single}|$ (bottom-right) of 3D blast
wave problem at $t = 0.1L$ with $128^3$ grid points}\label{fig:3DBlast1_diff}
\end{center}
\end{figure}

\begin{figure}[h]
\begin{center}
\includegraphics*[width=2.5in]{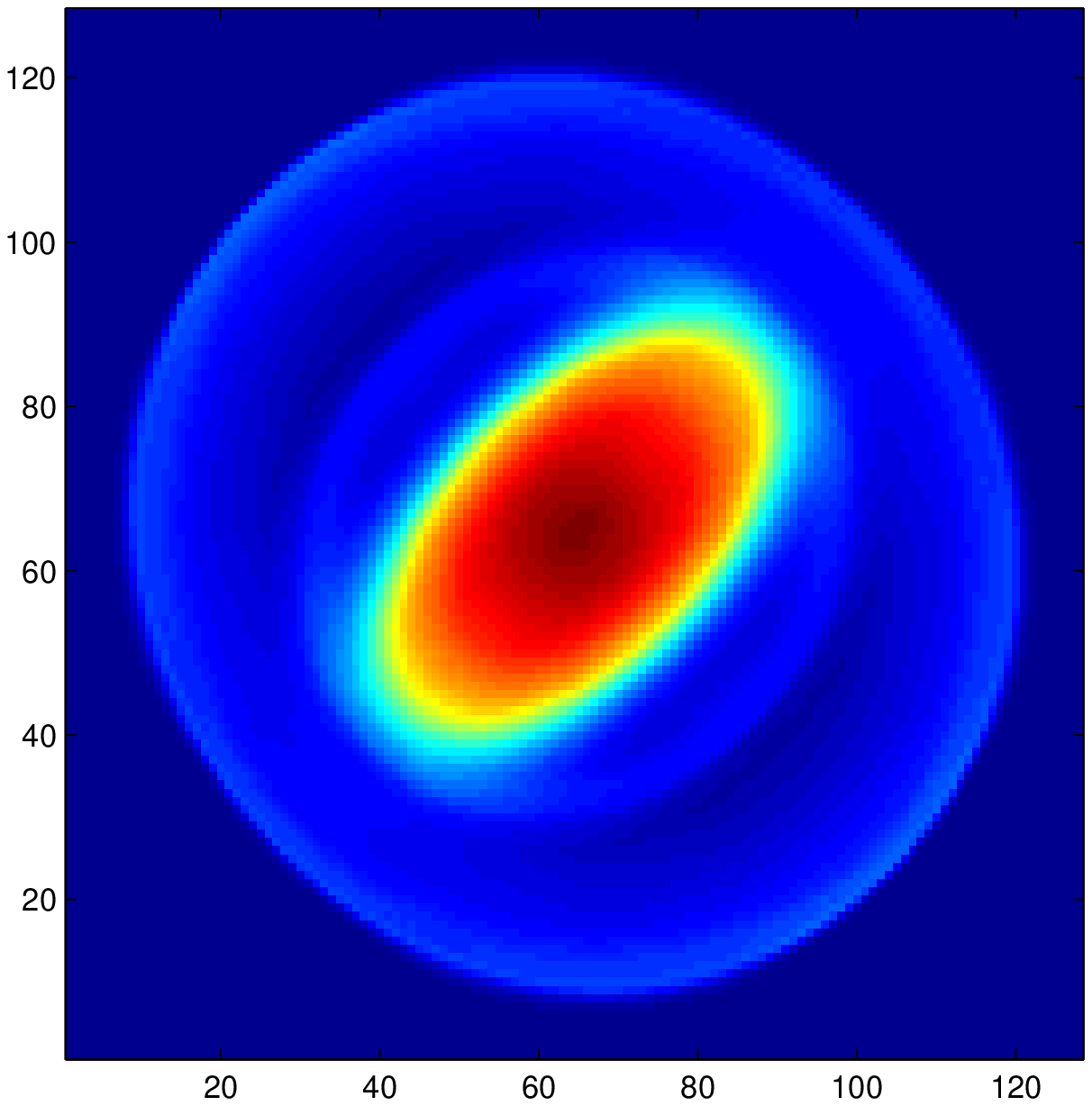}
\includegraphics*[width=2.7in]{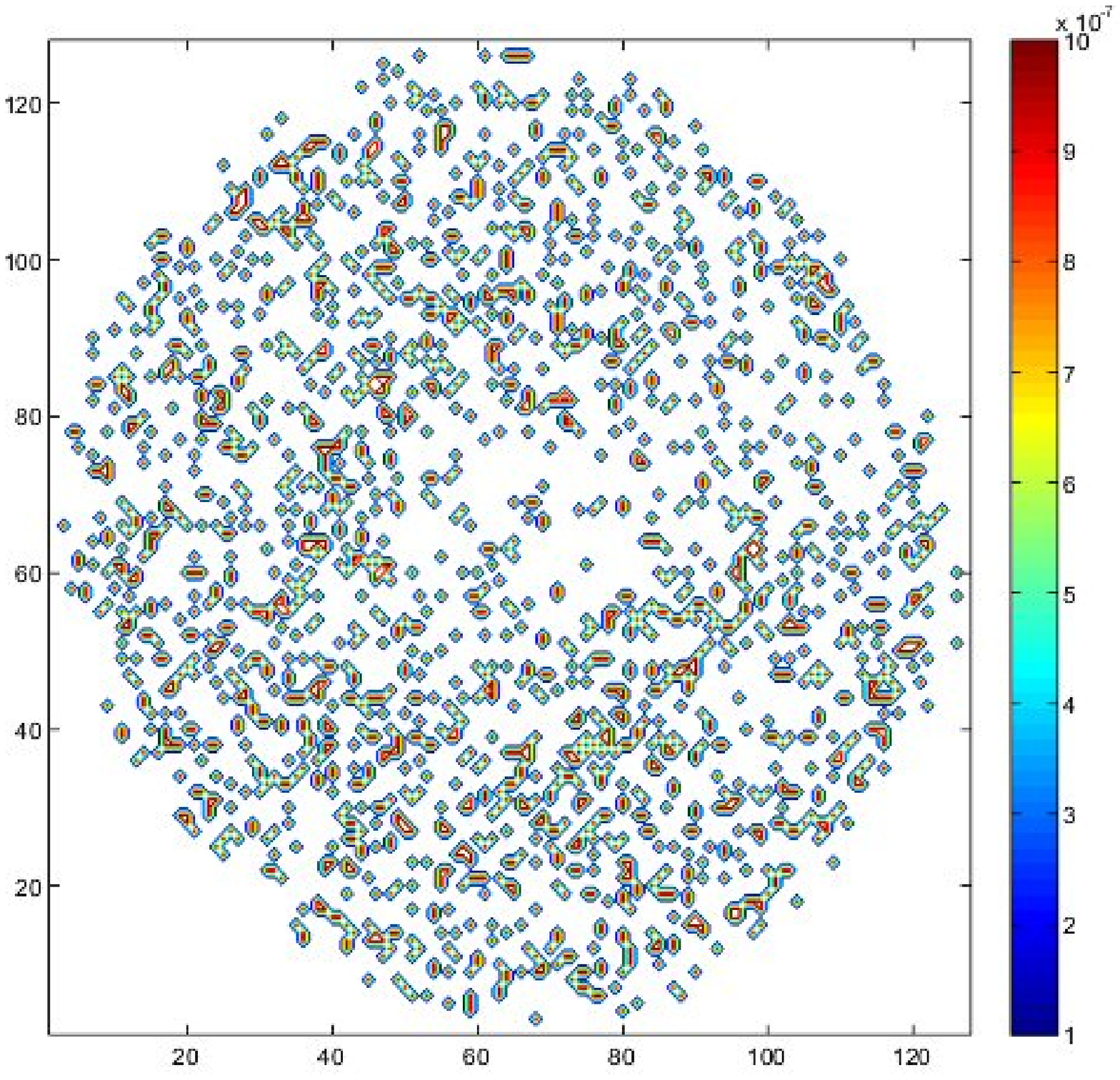}
\includegraphics*[width=2.5in]{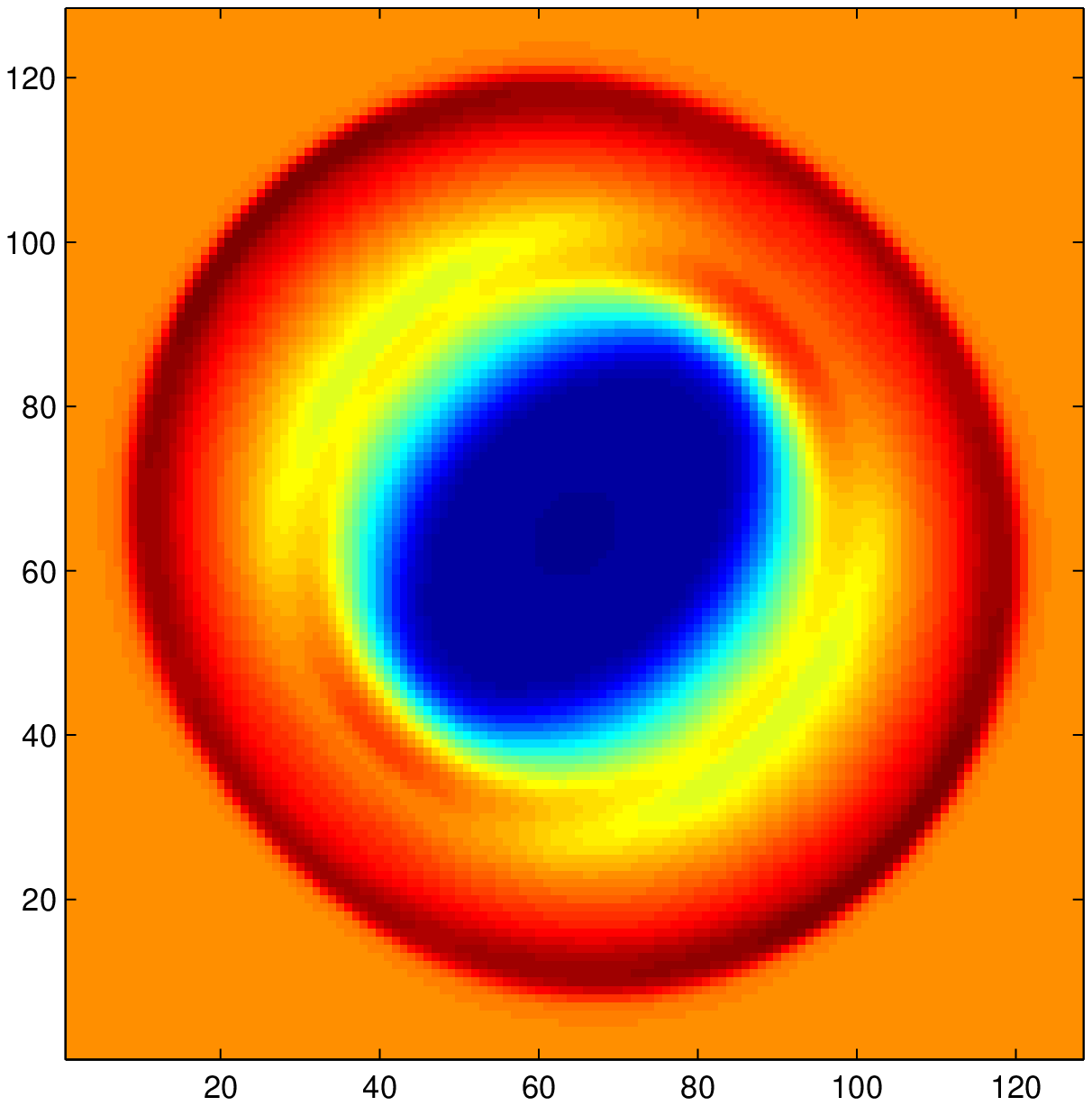}
\includegraphics*[width=2.7in]{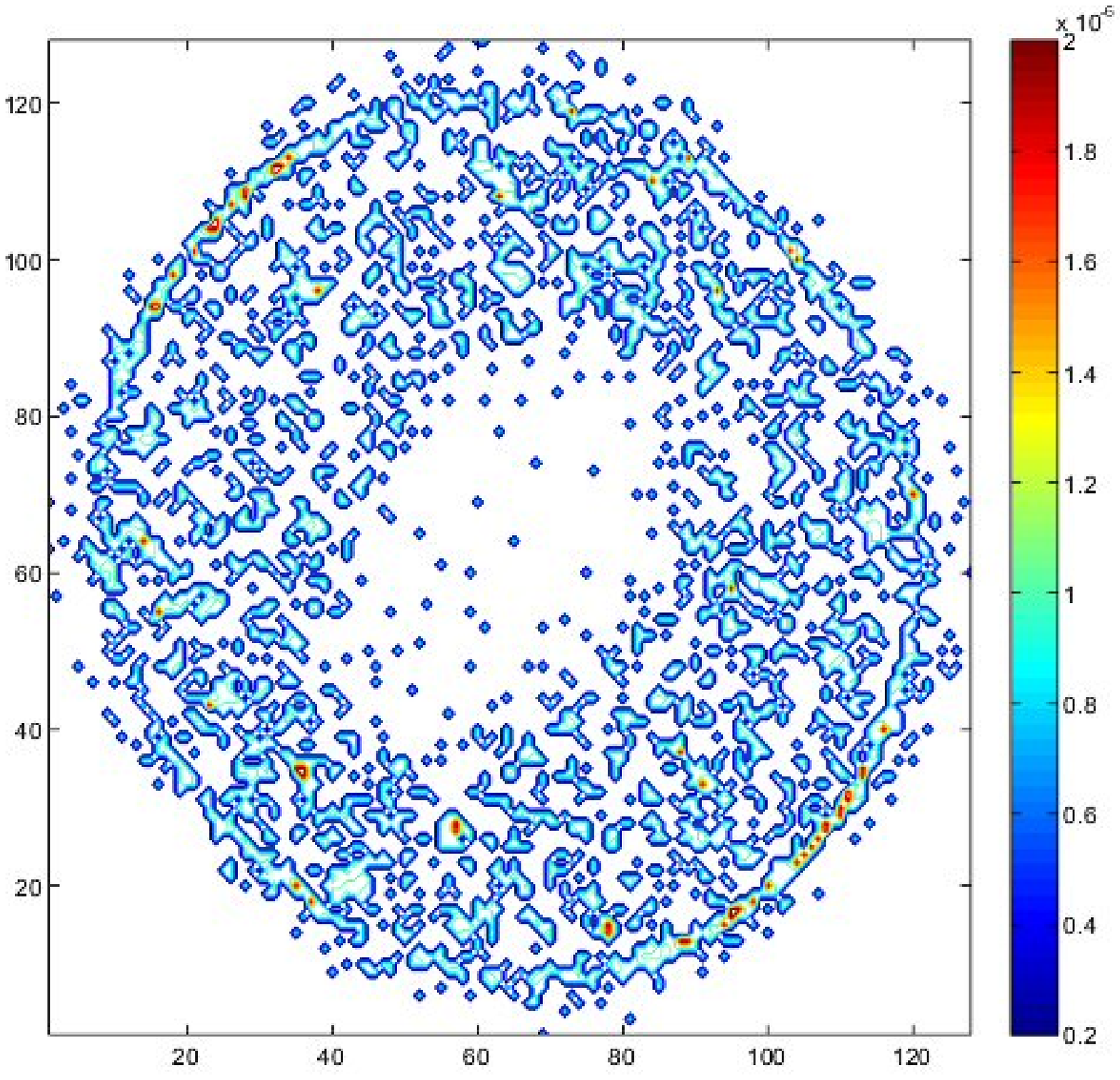}
\hfill \caption{Result of $p_{double}$ (top-left), $\rho_{double}$ (bottom-left) and
$|p_{double}-p_{single}|$ (top-right), $|\rho_{double}-\rho_{single}|$ (bottom-right) of 3D blast
wave problem at $t = 0.2L$ with $128^3$ grid points} \label{fig:3DBlast2_diff}
\end{center}
\end{figure}

Fig.~\ref{fig:3DBlast1_diff} and Fig.~\ref{fig:3DBlast2_diff} show the resulting images of the
simulation using double precision and the contours of the absolute differences between the results
of double precision and single precision computation of 3D blast wave test with $128^3$ grid points
at $t = 0.1L$ and $t = 0.2L$. As it is a high dimension computation in low resolution, the
differences between them are clear. The number of grid points having higher difference value
increases, and the $error$ is still less than $10^{-6}$. Small difference value makes the double
precision resulting images (Fig.~\ref{fig:3DBlast1_diff} and Fig.~\ref{fig:3DBlast2_diff}) looked
similar to the single precision resulting images (Fig.~\ref{fig:3DBlast01} and
Fig.~\ref{fig:3DBlast02}).

An important point can be realized that not only the grid points at the high density region has
high difference value, but also the number of grid points having high difference value and the
amount of the difference values are increasing along with the increase of the simulation time.
Higher dimension is another factor to introduce noticeable differences between the computation
results with different precisions because higher dimension means a grid point has more neighbors
and more neighbors need more computation steps in one time step. As a result the differences become
more obvious. Therefore, for a long-term simulation, double precision computation is a must.

The original Fortran code~\cite{Pen2003} is a second-order accurate high-resolution TVD MHD code.
Theoretically, we consider that {\em GPU-MHD} is sufficient to capture forward and reverse shocks
as well as any other discontinuities such as contact discontinuities which are important in space
physics. As {\em GPU-MHD} is a dimensional-splitting based code, there are two drawbacks: (i) the
code is unable to evolve the normal (to the sweep direction) magnetic field during each sweep
direction~\cite{Gardiner2005}, and (ii) splitting errors will generally be introduced due to the
linearized Jacobian flux matrices do not commute in most of the nonlinear multidimensional
problems~\cite{LeVeque1992}.

\section{Performance measurements}
\par

The performance measurements of the GPU and CPU implementations as well as the computation using
double precision and single precision are carried out in this section. Different numbers of grid
points and different dimensions were used in the performance tests. We run both {\it GPU-MHD} and
Pen {\it et al.}'s FORTRAN/CPU MHD code~\cite{Pen2009} to perform the simulations on a PC with
Intel Core i7 965 3.20 GHz CPU, 6G main memory, running Microsoft Windows XP 64-bit Professional.
Two graphics cards were tested: NVIDIA GeForce GTX 295 with 1.75G video memory and GTX 480 (Fermi)
with 1.5G video memory. The Fortran compiler and GPU development toolkit we used are G95 Stable
Version 0.92 and NVIDIA CUDA 3.2, respectively. {\it GPU-MHD} was designed for three-dimensional
problems, thus the dimensions are expressed in three-dimensional form in all figures. For 1D test,
1D Brio-Wu shock tube problem (see Section~\ref{sub:Brio-Wu}) was used. For 2D test, 2D Orszag-Tang
problem (see Section~\ref{sub:O-T}) was used  was used. For 3D test, 3D blast wave problem (see
Section~\ref{sub:3DBlast}) was used.

\begin{figure}[h]
\begin{center}
\includegraphics*[width=2.5in]{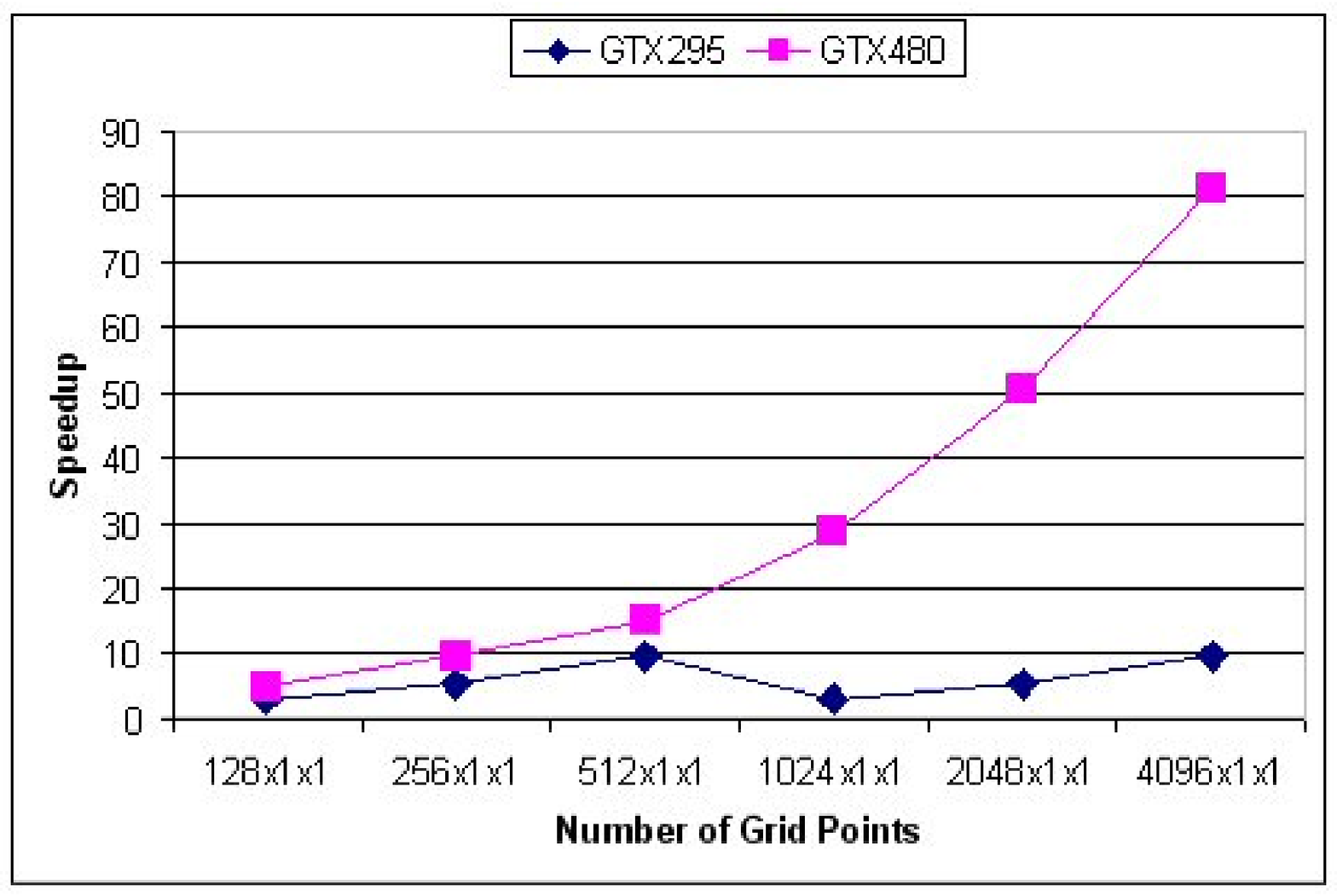}
\includegraphics*[width=2.5in]{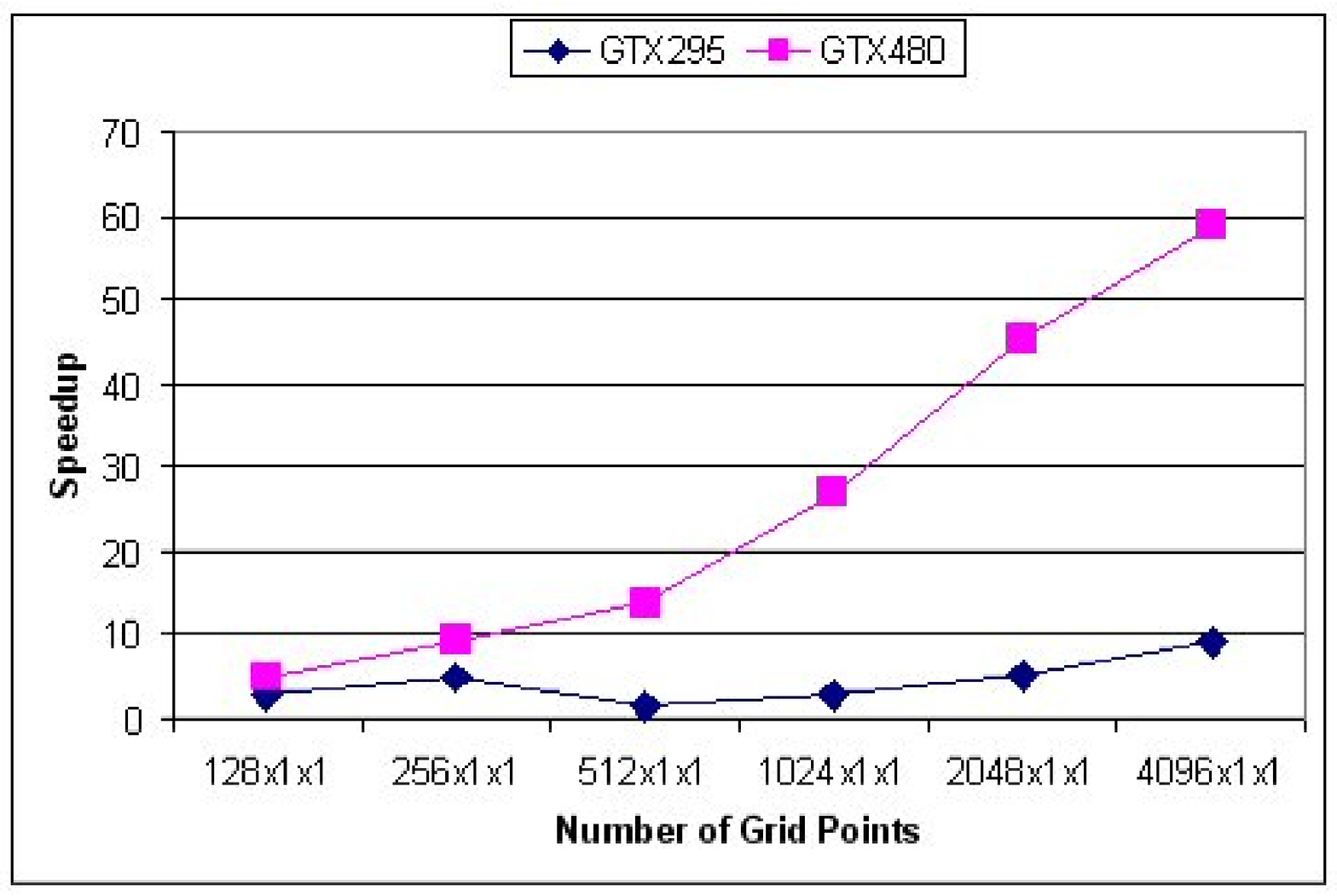}
\hfill \caption{The speedup of 1D test using {\it GPU-MHD} (Left: Single precision; Right: Double
precision), compared with the FORTRAN/CPU code at different resolutions} \label{fig:speedup1D}
\end{center}
\end{figure}

Fig.~\ref{fig:speedup1D} reports the comparison of {\it GPU-MHD} and the FORTRAN/CPU code of 1D
test with different numbers of grid points in single and double precisions. In single precision
mode, basically there is only about 10 times speedup ($4096\times 1\times 1$ case) since the number
of grid points is small. And it should be realized that the amount of speedup is increased as long
as the resolution is increased but dropped when the resolution reaches $512$. It is because the
``max threads per block'' of GTX 295 is 512, all the computations are handled within one block and
a very high processing speed can be archived. On GTX 480, there is about 80 times speedup
($4096\times 1\times 1$ case) and the amount of speedup is increased linearly due to the higher
floating point capability (512 MAD ops/clock). In double precision mode, around 10 times and 60
times speedup ($4096\times 1\times 1$ case) is achieved on GTX 295 and GTX 480, respectively.

Table~\ref{Table_1D_precision} gives the comparison of {\it GPU-MHD} using single precision and
double precision of 1D test with different numbers of grid points. On GTX 295, a similar speed drop
happened both in single and double precision modes, but it occurred in different resolutions: $512$
in single precision and $256$ in double precision. This is not strange and not difficult to
understand since the double precision has double size of data to be handled by the processors.
Except the special case of $512$ resolution, the processing speed in both modes are very closed. On
GTX 480, the performance between single and double precision is quite close.

\begin{table}[h]
\caption{The performance results of 1D test between single precision and double precision of {\it
GPU-MHD} at different resolutions}\label{Table_1D_precision}
\begin{tabular}{|c|c|c|c|c|}
  \hline
  Number of grid points & GPU & Double precision & Single precision & Ratio \\
  & & (ms/step) & (ms/step) & \\
  \hline
  128 $\times$ 1 $\times$ 1 & GTX295 & 3.9 & 3.6 & 1.0833 \\
                            & GTX480 & 2.26 & 2.24 & 1.0089 \\
                            \hline
  256 $\times$ 1 $\times$ 1 & GTX295 & 4.5 & 4.0 & 1.1250 \\
                            & GTX480 & 2.35 & 2.29 & 1.0262 \\
                            \hline
  512 $\times$ 1 $\times$ 1 & GTX295 & 29.5 & 4.5 & 6.5555 \\
                            & GTX480 & 3.21 & 2.95 & 1.0881 \\
                            \hline
  1024 $\times$ 1 $\times$ 1 & GTX295 & 31.0 & 30.0 & 1.0333 \\
                            & GTX480 & 3.25 & 3.03 & 1.0726 \\
                            \hline
  2048 $\times$ 1 $\times$ 1 & GTX295 & 33.0 & 32.2 & 1.0248 \\
                            & GTX480 & 3.82 & 3.43 & 1.1137 \\
                            \hline
  4096 $\times$ 1 $\times$ 1 & GTX295 & 39.1 & 36.4 & 1.0742 \\
                            & GTX480 & 5.95 & 4.31 & 1.3805 \\
  \hline
\end{tabular}
\end{table}

\begin{figure}[h]
\begin{center}
\includegraphics*[width=2.5in]{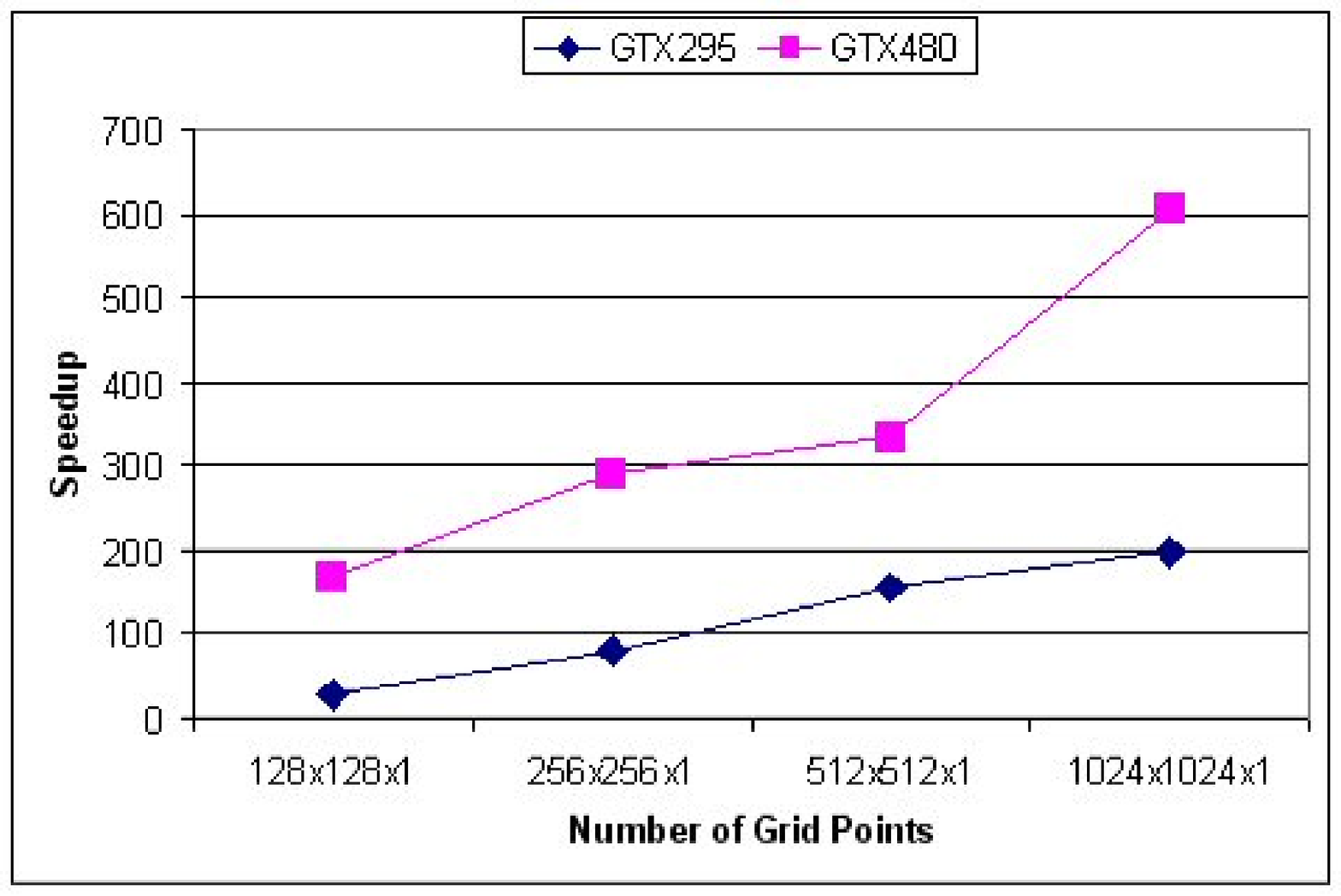}
\includegraphics*[width=2.5in]{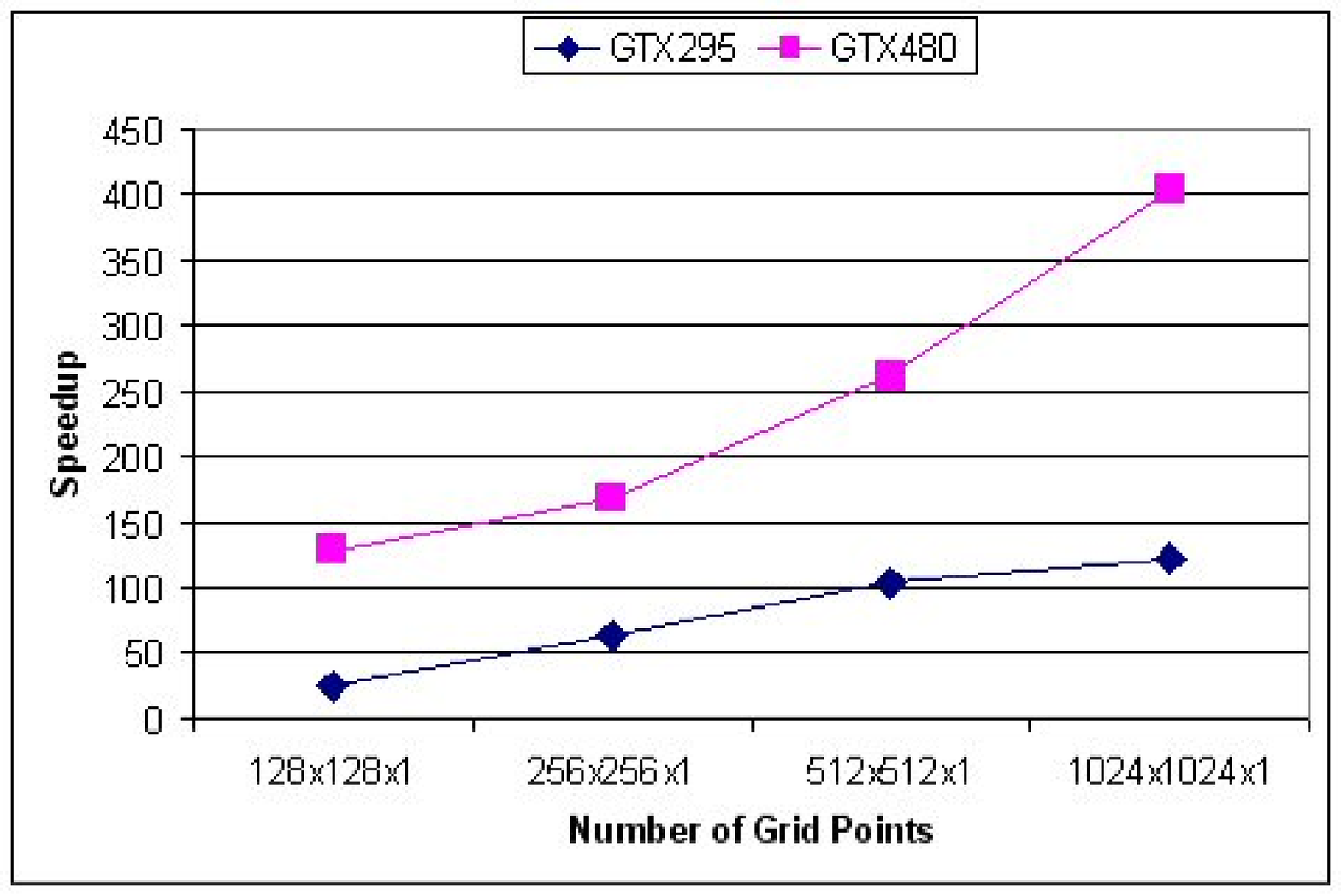}
\hfill \caption{The speedup of 2D test using {\it GPU-MHD} (Left: Single precision; Right: Double
precision), compared with the FORTRAN/CPU code at different resolutions} \label{fig:speedup2D}
\end{center}
\end{figure}

The comparison of {\it GPU-MHD} and the FORTRAN/CPU code of 2D test with different numbers of grid
points in single and double precisions is presented in Fig.~\ref{fig:speedup2D}. In 2D case, a
significant performance improvement is observed, especially when the numbers of grid points are
$512^2$ are $1024^2$, a speedup of around 150 and around 200 is achieved on GTX 295, respectively.
On GTX 480, a speedup of around 320 and around 600 is achieved, respectively. This is due to the
the significant improvement of the performance in double precision on GTX 480 (with double
precision floating point capability of 256 FMA ops/clock) is better than that of GTX 295 (30 FMA
ops/clock).

Table~\ref{Table_2D_precision} presents the comparison of {\it GPU-MHD} using single precision and
double precision of 2D test with different numbers of grid points. The significant performance
difference is noticeable when the number of grid points is increased. However, it still keeps a
ratio increasing slowly from $1.118$ to $1.6218$ while the resolution increases from $128^2$ to
$1024^2$. The double/single precision ratios on GTX 480 vary between $1.2804$ and $1.7300$.

\begin{table}[h]
\caption{The performance results of 2D test between single precision and double precision of {\it
GPU-MHD} at different resolutions}\label{Table_2D_precision}
\begin{tabular}{|c|c|c|c|c|}
  \hline
  Number of grid points & GPU & Double precision & Single precision & Ratio \\
  & & (ms/step) & (ms/step) & \\
 \hline
  128 $\times$ 128 $\times$ 1 & GTX295 & 35.8 & 32.2 & 1.1118 \\
                            & GTX480 & 7.01 & 5.36 & 1.3078 \\
                            \hline
  256 $\times$ 256 $\times$ 1 & GTX295 & 57.3 & 44.8 & 1.2790 \\
                            & GTX480 & 21.59 & 12.48 & 1.7300 \\
                            \hline
  512 $\times$ 512 $\times$ 1 & GTX295 & 142.3 & 94.0 & 1.5138 \\
                            & GTX480 & 56.26 & 43.94 & 1.2804 \\
                            \hline
  1024 $\times$ 1024 $\times$ 1 & GTX 295 & 478.6 & 295.1 & 1.6218 \\
                            & GTX480 & 144.48 & 96.43 & 1.4983 \\
  \hline
\end{tabular}
\end{table}

\begin{figure}[h]
\begin{center}
\includegraphics*[width=2.5in]{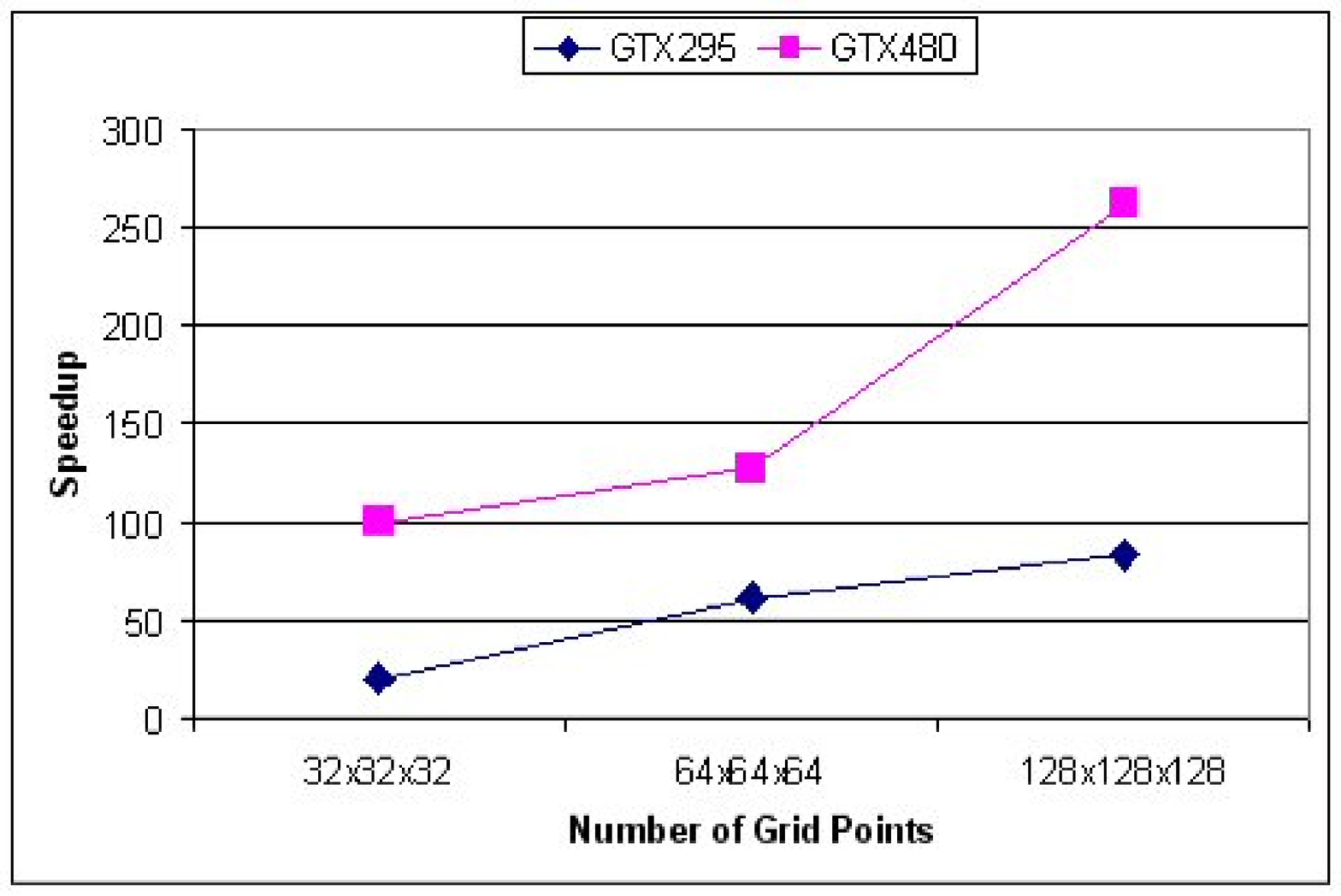}
\includegraphics*[width=2.5in]{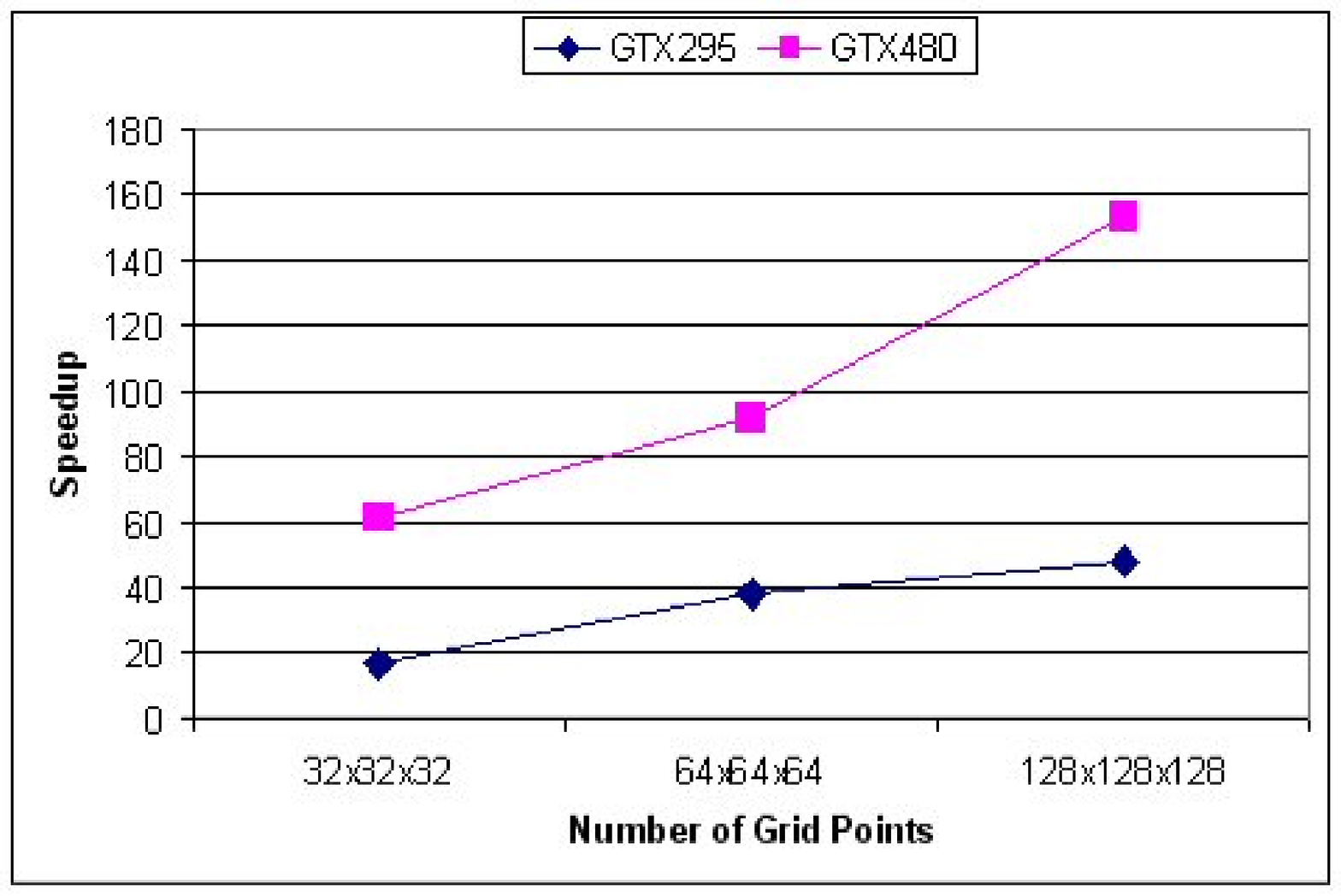}
\hfill \caption{The speedup of 3D test using {\it GPU-MHD} (Left: Single precision; Right: Double
precision), compared with the FORTRAN/CPU code at different resolutions} \label{fig:speedup3D}
\end{center}
\end{figure}

Fig.~\ref{fig:speedup3D} shows the comparison of {\it GPU-MHD} and the FORTRAN/CPU code of 3D test
with different numbers of grid points in single and double precisions. The performance of {\it
GPU-MHD} is faster than the FORTRAN/CPU code about 60 times and 84 times when the numbers of grid
points are $64^3$ and $128^3$ on GTX 295, respectively. The corresponding speedups on GTX 480 are
about 260 and 155, respectively.

Table~\ref{Table_3D_precision} shows the comparison of {\it GPU-MHD} using single precision and
double precision of 3D test with different numbers of grid points. The ratio is $1.6020$ when the
number of grid points is $64^3$, and is $1.7389$ when the number of grid points is $128^3$ on GTX
295. The corresponding ratio on GTX 480 are $1.3973$ and $1.7133$, respectively.

\begin{table}[h]
\caption{The performance results of 3D test between single precision and double precision of {\it
GPU-MHD} at different resolutions}\label{Table_3D_precision}
\begin{tabular}{|c|c|c|c|c|}
  \hline
  Number of grid points & GPU & Double precision & Single precision & Ratio \\
  & & (ms/step) & (ms/step) & \\
 \hline
  32 $\times$ 32 $\times$ 32 & GTX295 & 44.6 & 36.6 & 1.2186 \\
                            & GTX480 & 12.03 & 7.41 & 1.6235 \\
                            \hline
  64 $\times$ 64 $\times$ 64 & GTX295 & 145.3 & 90.7 & 1.6020 \\
                            & GTX480 & 60.18 & 43.07 & 1.3973 \\
                            \hline
  128 $\times$ 128 $\times$ 128 & GTX295 & 880.6 & 506.4 & 1.7389 \\
                            & GTX480 & 276.40 & 161.33 & 1.7133 \\
  \hline
\end{tabular}
\end{table}

The performance tests show that when the number of grid points of the test problems is small, such
as those in 1D case, {\it GPU-MHD} can give a significant performance improvement. When the number
of grid points increases, an obvious disparity of performance becomes clear, especially for
multidimensional cases (see Fig.~\ref{fig:speedup2D} and Fig.~\ref{fig:speedup3D}). Computation
using double precision on GPUs prior to Fermi is known to be very low performance compared to
single precision. However, in the performance comparison between single precision and double
precision modes in {\it GPU-MHD}, the ratios of the processing speed between two modes show that
{\it GPU-MHD} is efficient enough in double precision computation. The performance results show
that CUDA is an attractive parallel computing environment for MHD simulations.

Tables~\ref{Table_1D_part},~\ref{Table_2D_part}, and~\ref{Table_3D_part} show the performance
comparisons between FORTRAN/CPU and {\it GPU-MHD} of major calculations in single precision at
different resolutions of the 1D, 2D and 3D tests, respectively. These tables provide the
information of how the computational effort spent on which step in 1D, 2D, and 3D problems. From
these tables we can find out that the calculations of fluids ($fluid_x$, $fluid_y$, and $fluid_z$)
spend most of the computational times in all kinds of problems. And even though the data accesses
of $L_x$ and $L_y$ are not coalesced (see Section 4.2), the computational times spent on each part
of $L_x$, $L_y$, and $L_z$ in the same problem are very close. It shows that GT200 and Fermi
architectures handle the data accesses that are not perfectly coalesced quite well in our study.
\begin{table}[h]
\caption{Performance comparison (ms/step) between FORTRAN/CPU and {\it GPU-MHD} of major
calculations in single precision at different resolutions of the 1D test.}\label{Table_1D_part}
\begin{tabular}{|c|c|c|c|c|c|c|}
  \hline
  Number of grid points & Operations & FORTRAN/CPU & GTX295 & GTX480 \\
  \hline
  512 $\times$ 1 $\times$ 1
        & CFL & 0.18 & 0.14 & 0.28 \\
        & ${fluid}_x$ & 1.07 & 0.93 & 0.61 \\
        & $B_{x \rightarrow y}$ and $B_{x \rightarrow z}$ & 0.36 & 0.18 & 0.11 \\
        & ${fluid}_y$ & 5.85 & 0.40 & 0.27 \\
        & $B_{y \rightarrow x}$ and $B_{y \rightarrow z}$ & 3.80 & 0.19 & 0.10 \\
        & ${fluid}_z$ & 5.88 & 0.40 & 0.27 \\
        & $B_{z \rightarrow x}$ and $B_{z \rightarrow y}$ & 3.75 & 0.18 & 0.10 \\
        & Transposition & 1.96 & - & - \\
  \hline
  1024 $\times$ 1 $\times$ 1
        & CFL & 0.44 & 0.61 & 0.24 \\
        & ${fluid}_x$ & 2.16 & 4.90 & 0.58 \\
        & $B_{x \rightarrow y}$ and $B_{x \rightarrow z}$ & 0.53 & 0.69 & 0.10 \\
        & ${fluid}_y$ & 11.81 & 4.09 & 0.26 \\
        & $B_{y \rightarrow x}$ and $B_{y \rightarrow z}$ & 7.57 & 0.66 & 0.10 \\
        & ${fluid}_z$ & 11.70 & 4.03 & 0.26 \\
        & $B_{z \rightarrow x}$ and $B_{z \rightarrow y}$ & 7.55 & 0.80 & 0.10 \\
        & Transposition & 4.10 & - & - \\
  \hline
  2048 $\times$ 1 $\times$ 1
        & CFL & 0.90 & 0.62 & 0.24 \\
        & ${fluid}_x$ & 4.31 & 6.09 & 0.77 \\
        & $B_{x \rightarrow y}$ and $B_{x \rightarrow z}$ & 1.07 & 0.67 & 0.10 \\
        & ${fluid}_y$ & 23.31 & 4.10 & 0.26 \\
        & $B_{y \rightarrow x}$ and $B_{y \rightarrow z}$ & 15.09 & 0.67 & 0.10 \\
        & ${fluid}_z$ & 23.26 & 4.04 & 0.26 \\
        & $B_{z \rightarrow x}$ and $B_{z \rightarrow y}$ & 14.96 & 0.67 & 0.10 \\
        & Transposition & 8.22 & - & - \\
  \hline
  4096 $\times$ 1 $\times$ 1
        & CFL & 1.24 & 0.63 & 0.28 \\
        & ${fluid}_x$ & 8.92 & 7.93 & 1.19 \\
        & $B_{x \rightarrow y}$ and $B_{x \rightarrow z}$ & 1.59 & 0.66 & 0.10 \\
        & ${fluid}_y$ & 46.83 & 4.01 & 0.26 \\
        & $B_{y \rightarrow x}$ and $B_{y \rightarrow z}$ & 30.41 & 0.67 & 0.10 \\
        & ${fluid}_z$ & 46.74 & 3.98 & 0.26 \\
        & $B_{z \rightarrow x}$ and $B_{z \rightarrow y}$ & 30.15 & 0.67 & 0.10 \\
        & Transposition & 16.50 & - & - \\
  \hline
\end{tabular}
\end{table}

\begin{table}[h]
\caption{Performance comparison (ms/step) between FORTRAN/CPU and {\it GPU-MHD} of major
calculations in single precision at different resolutions of the 2D test}\label{Table_2D_part}
\begin{tabular}{|c|c|c|c|c|c|c|}
  \hline
  Number of grid points & Operations & FORTRAN/CPU  & GTX295  & GTX480  \\
  \hline
  128 $\times$ 128 $\times$ 1
        & CFL & 9.44 & 0.69 & 0.30 \\
        & ${fluid}_x$ & 33.30 & 4.43 & 0.79 \\
        & $B_{x \rightarrow y}$ and $B_{x \rightarrow z}$ & 8.54 & 0.77 & 0.11 \\
        & ${fluid}_y$ & 33.28 & 4.48 & 0.83 \\
        & $B_{y \rightarrow x}$ and $B_{y \rightarrow z}$ & 7.15 & 0.83 & 0.12 \\
        & ${fluid}_z$ & 179.00 & 4.37 & 0.57 \\
        & $B_{z \rightarrow x}$ and $B_{z \rightarrow y}$ & 116.17 & 0.78 & 0.11 \\
        & Transposition & 105.22 & - & - \\
  \hline
  256 $\times$ 256 $\times$ 1
        & CFL & 28.26 & 1.01 & 0.56 \\
        & ${fluid}_x$ & 136.87 & 6.05 & 1.73 \\
        & $B_{x \rightarrow y}$ and $B_{x \rightarrow z}$ & 33.41 & 1.42 & 0.37 \\
        & ${fluid}_y$ & 133.76 & 6.32 & 1.77 \\
        & $B_{y \rightarrow x}$ and $B_{y \rightarrow z}$ & 32.61 & 1.41 & 0.41 \\
        & ${fluid}_z$ & 713.28 & 5.94 & 1.33 \\
        & $B_{z \rightarrow x}$ and $B_{z \rightarrow y}$ & 467.13 & 1.24 & 0.36 \\
        & Transposition & 406.36 & - & - \\
  \hline
  512 $\times$ 512 $\times$ 1
        & CFL & 108.00 & 1.90 & 1.46 \\
        & ${fluid}_x$ & 513.75 & 12.21 & 6.11 \\
        & $B_{x \rightarrow y}$ and $B_{x \rightarrow z}$ & 121.67 & 2.86 & 1.18 \\
        & ${fluid}_y$ & 503.29 & 12.73 & 6.12 \\
        & $B_{y \rightarrow x}$ and $B_{y \rightarrow z}$ & 120.46 & 3.53 & 1.33 \\
        & ${fluid}_z$ & 2747.38 & 11.78 & 5.33 \\
        & $B_{z \rightarrow x}$ and $B_{z \rightarrow y}$ & 1796.88 & 2.94 & 1.16 \\
        & Transposition & 1554.66 & - & - \\
  \hline
  1024 $\times$ 1024 $\times$ 1
        & CFL & 459.22 & 5.28 & 3.18 \\
        & ${fluid}_x$ & 2113.20 & 36.24 & 12.80 \\
        & $B_{x \rightarrow y}$ and $B_{x \rightarrow z}$ & 532.89 & 9.37 & 3.08 \\
        & ${fluid}_y$ & 2078.98 & 40.44 & 13.27 \\
        & $B_{y \rightarrow x}$ and $B_{y \rightarrow z}$ & 534.93 & 11.89 & 3.70 \\
        & ${fluid}_z$ & 11611.11 & 37.14 & 10.72 \\
        & $B_{z \rightarrow x}$ and $B_{z \rightarrow y}$ & 7538.00 & 9.70 & 3.05 \\
        & Transposition & 6537.16 & - & - \\
  \hline
\end{tabular}
\end{table}

\begin{table}[h]
\caption{Performance comparison (ms/step) between FORTRAN/CPU and {\it GPU-MHD} of major
calculations in single precision at different resolutions of the 3D test}\label{Table_3D_part}
\begin{tabular}{|c|c|c|c|c|c|c|}
  \hline
  Number of grid points & Operations & FORTRAN/CPU  & GTX295  & GTX480\\
  \hline
  32 $\times$ 32 $\times$ 32
        & CFL & 14.30 & 0.78 & 0.35 \\
        & ${fluid}_x$ & 76.74 & 4.90 & 1.00 \\
        & $B_{x \rightarrow y}$ and $B_{x \rightarrow z}$ & 23.09 & 0.97 & 0.16 \\
        & ${fluid}_y$ & 74.35 & 4.95 & 0.99 \\
        & $B_{y \rightarrow x}$ and $B_{y \rightarrow z}$ & 22.09 & 1.04 & 0.18 \\
        & ${fluid}_z$ & 74.02 & 4.97 & 1.04 \\
        & $B_{z \rightarrow x}$ and $B_{z \rightarrow y}$ & 22.44 & 1.00 & 0.17 \\
        & Transposition & 125.12 & - & - \\
  \hline
  64 $\times$ 64 $\times$ 64
        & CFL & 111.92 & 1.90 & 1.43 \\
        & ${fluid}_x$ & 545.58 & 11.82 & 5.57 \\
        & $B_{x \rightarrow y}$ and $B_{x \rightarrow z}$ & 144.58 & 2.94 & 1.18 \\
        & ${fluid}_y$ & 533.83 & 12.11 & 5.59 \\
        & $B_{y \rightarrow x}$ and $B_{y \rightarrow z}$ & 143.21 & 3.61 & 1.31 \\
        & ${fluid}_z$ & 537.75 & 12.37 & 5.90 \\
        & $B_{z \rightarrow x}$ and $B_{z \rightarrow y}$ & 142.58 & 3.25 & 1.27 \\
        & Transposition & 963.6 & - & - \\
  \hline
  128 $\times$ 128 $\times$ 128
        & CFL & 871.00 & 10.98 & 5.50 \\
        & ${fluid}_x$ & 4231.83 & 63.83 & 19.64 \\
        & $B_{x \rightarrow y}$ and $B_{x \rightarrow z}$ & 1047.42 & 17.30 & 5.47 \\
        & ${fluid}_y$ & 4110.67 & 65.29 & 18.63 \\
        & $B_{y \rightarrow x}$ and $B_{y \rightarrow z}$ & 1029.96 & 22.42 & 6.45 \\
        & ${fluid}_z$ & 4096.88 & 70.65 & 21.56 \\
        & $B_{z \rightarrow x}$ and $B_{z \rightarrow y}$ & 1031.92 & 19.52 & 6.18 \\
        & Transposition & 7686.58 & - & - \\
  \hline
\end{tabular}
\end{table}


\section{Visualization of the simulation results}
\par
There is a need to visualize the MHD simulation data, for examples, Daum~\cite{Daum2007} developed
a toolbox called {\it VisAn MHD} in MATLAB for MHD simulation data visualization and analysis. With
the help of GPUs, Stantchev {\em et al.}~\cite{Stantchev2009} used GPUs for computation and
visualization of plasma turbulence. In {\it GPU-MHD}, using the parallel computation power of GPUs
and CUDA, the simulation results of one time step can be computed in dozens or hundreds
milliseconds. According to the efficiency of {\it GPU-MHD}, near real-time visualization is able to
be provided for 1D and 2D problems. The motion or attributes of the magnetic fluid can be computed
and rendered on the fly. So the changes of the magnetic fluid during the simulation can be observed
in real-time.

By adding the real-time visualization, the flow of {\it GPU-MHD}, Fig.~\ref{fig:FlowChart} is
extended as Fig.~\ref{fig:FlowChart_v}:
\begin{figure}[hbt]
\begin{center}
\includegraphics*[width=4.0in]{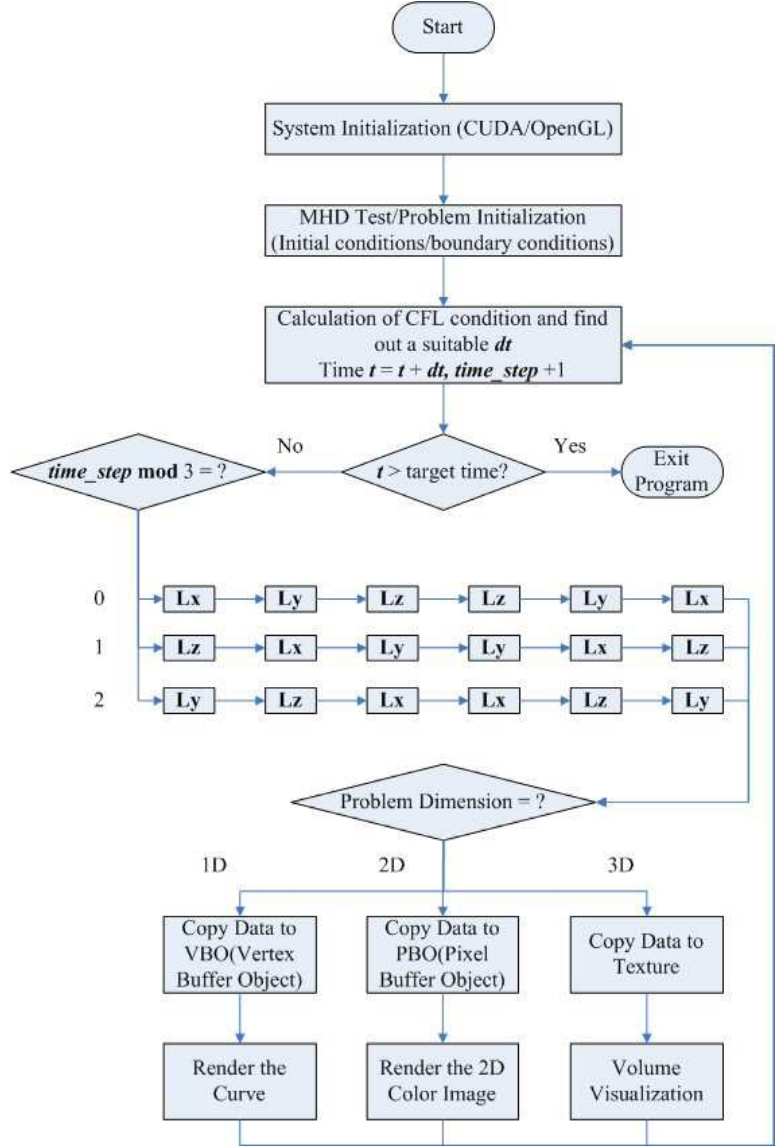}
\hfill \caption{The flow chat of {\it GPU-MHD}.} \label{fig:FlowChart_v}
\end{center}
\end{figure}

{\it GPU-MHD} provides different visualization methods for one-dimensional, two-dimensional and
three-dimensional problems.

\begin{figure}[hbt]
\begin{center}
\includegraphics*[width=2.5in]{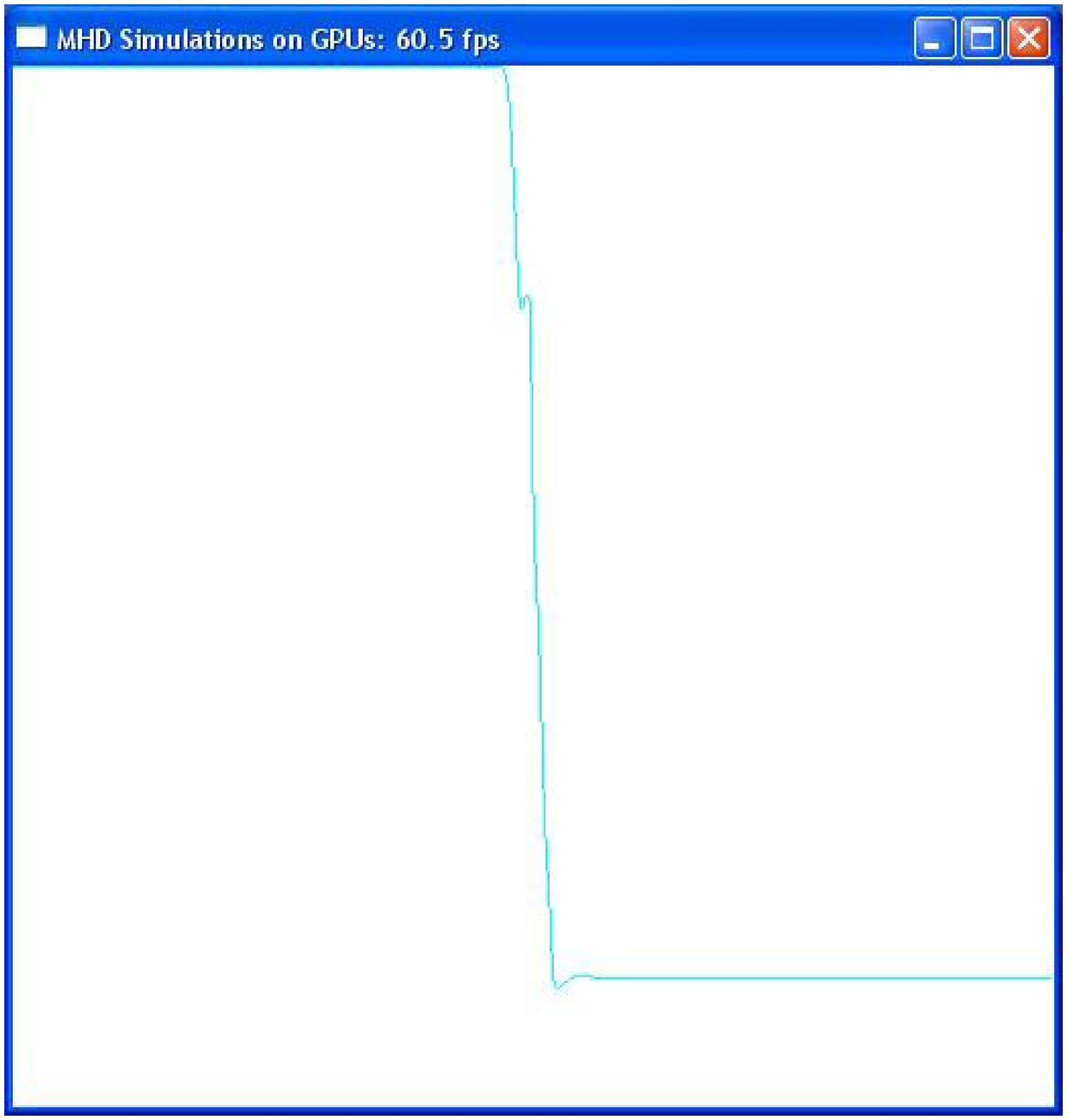}
\includegraphics*[width=2.5in]{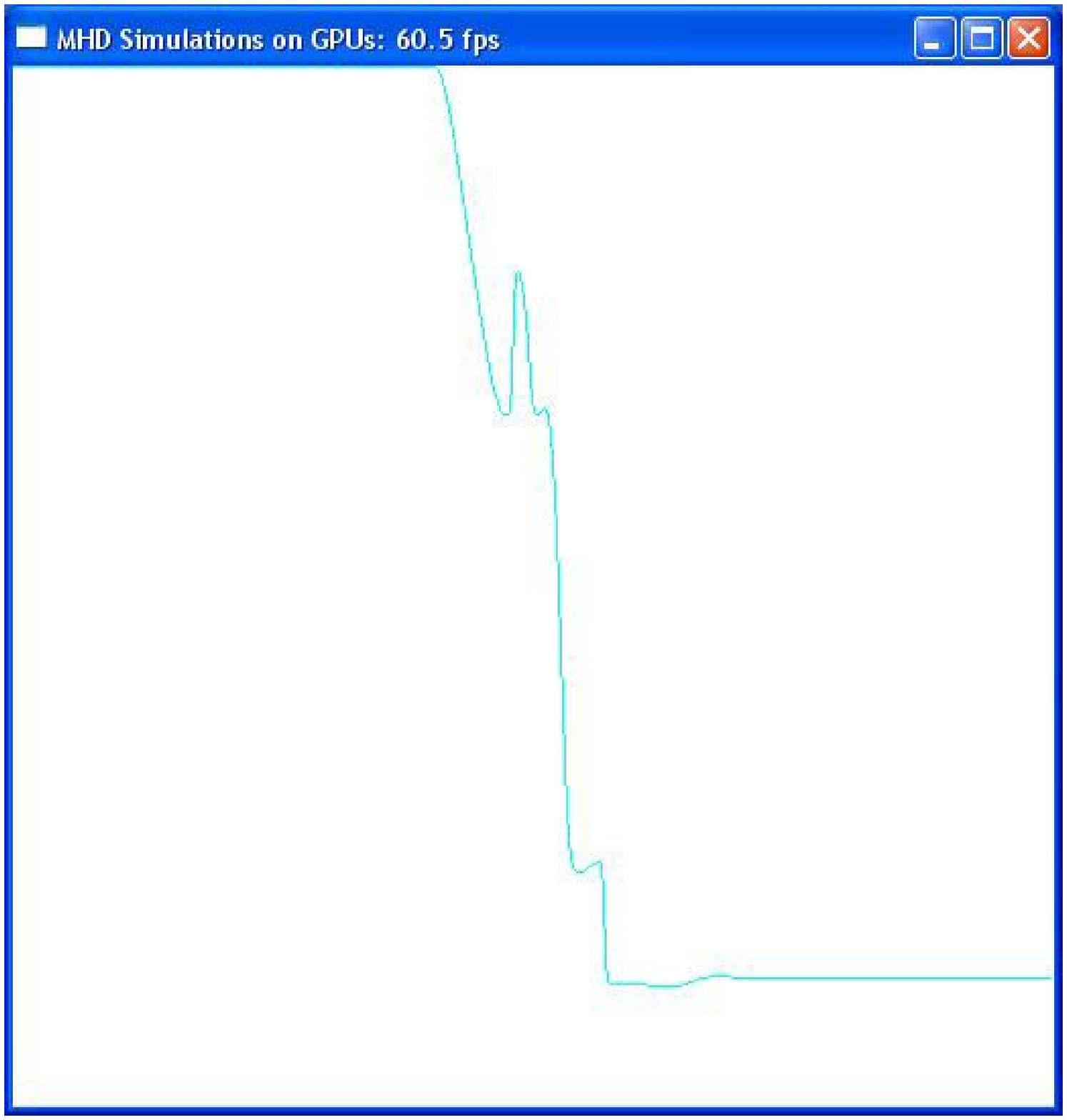}
\hfill \caption{1D real-time visualizations of the density ($\rho$) of Brio-Wu shock tube problem
with $512$ grid points using {\it GPU-MHD}.}\label{fig:1DVis}
\end{center}
\end{figure}

To visualize one-dimensional problems for each time step, the simulation results are copied to the
CUDA global memory that mapped to the Vertex Buffer Object (VBO)~\cite{Wright2007}. For all grid
points, one grid point is mapped to one vertex. The position of each grid point is mapped as the
$x$-position of the vertex and the selected physical value ($\rho$, $p$, etc.) is mapped as the
$y$-position of the vertex. Then a curve of these vertices is drawn. Since the VBO is mapped to
CUDA global memory and simulation results are stored in GRAM, the copying and mapping operations
are fast. Experimental result shows that {\it GPU-MHD} with real-time visualization can achieve 60
frame per second (FPS) in single precision mode and 30 FPS in double precision mode on GTX 295. On
GTX 480, around 60 FPS in both single and double precisions is achieved. Fig.~\ref{fig:1DVis} shows
two example images of 1D visualizations using {\it GPU-MHD}.

\begin{figure}[hbt]
\begin{center}
\includegraphics*[width=2.5in]{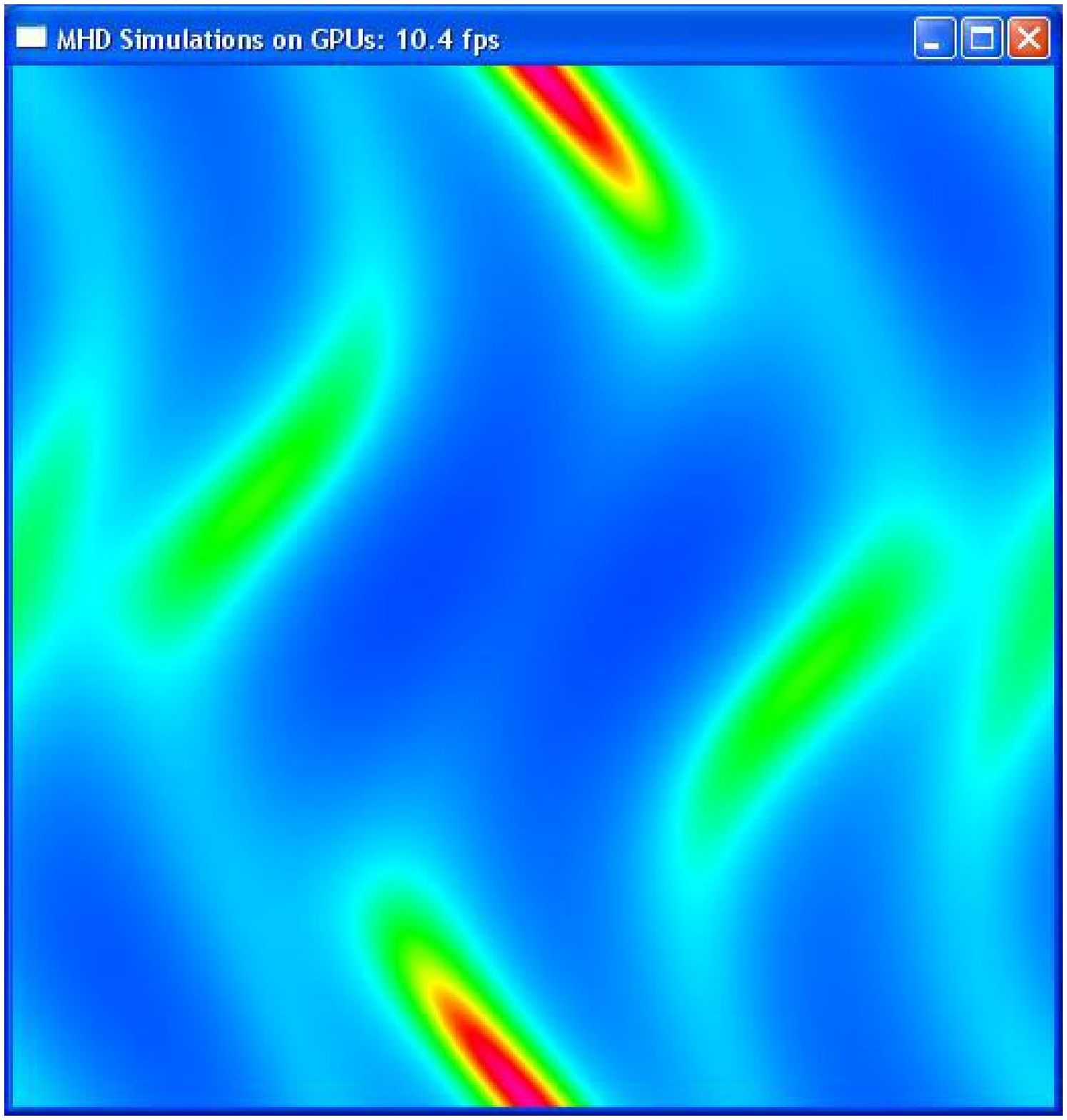}
\includegraphics*[width=2.5in]{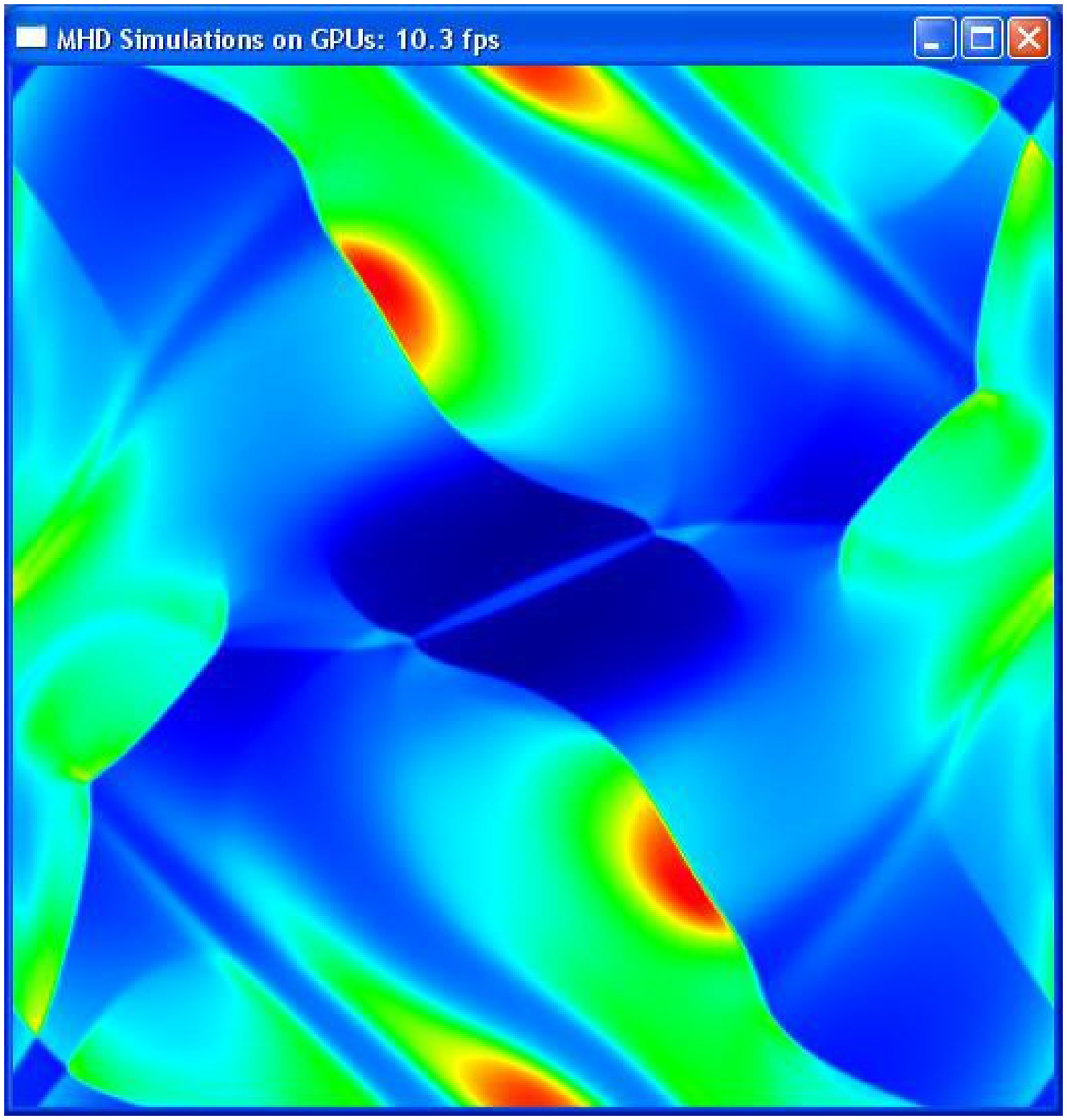}
\hfill \caption{2D visualizations of the density ($\rho$) of Orszag-Tang vortex problem with
$512^2$ grid points using  {\it GPU-MHD}.}\label{fig:2DVis}
\end{center}
\end{figure}

The operational flow of visualization of 2D problems is similar to that in 1D visualization.
However, instead of Vertex Buffer Object (VBO), Pixel Buffer Object(PBO)~\cite{Wright2007} is used.
For each time step, the simulation results are copied to the CUDA global memory that are then
mapped to the PBO. For all grid points, one grid point is mapped to one pixel. The $x$ and $y$
position of each grid point are mapped as the corresponding $x$-position and the $y$-position of
the vertex and the selected physical value ($\rho$, $p$, etc.) is mapped as the color of the pixel
to form a color image. To render this color image, a transfer function is set to map the physical
value to the color of the pixel and then the resulting image is drawn. Similar to VBO, PBO is also
mapped to CUDA global memory and the simulation results are stored in GRAM, so the copying and
mapping operations are also fast and do not affect too much to the performance. Although the number
of grid points in 2D problem is much larger than those in the one-dimension problem, the FPS still
reaches 10 in single precision mode and 6 in double precision mode on GTX 295 when the number of
grid points is $512^2$, still giving acceptable performance to the user. On GTX 480, 22 FPS in
single precision and 17 FPS in double precision are achieved and thus interactive rates are
available. Fig.~\ref{fig:2DVis} shows two example images of 2D visualizations using {\it GPU-MHD}.

\begin{figure}[hbt]
\begin{center}
\includegraphics*[width=2.5in]{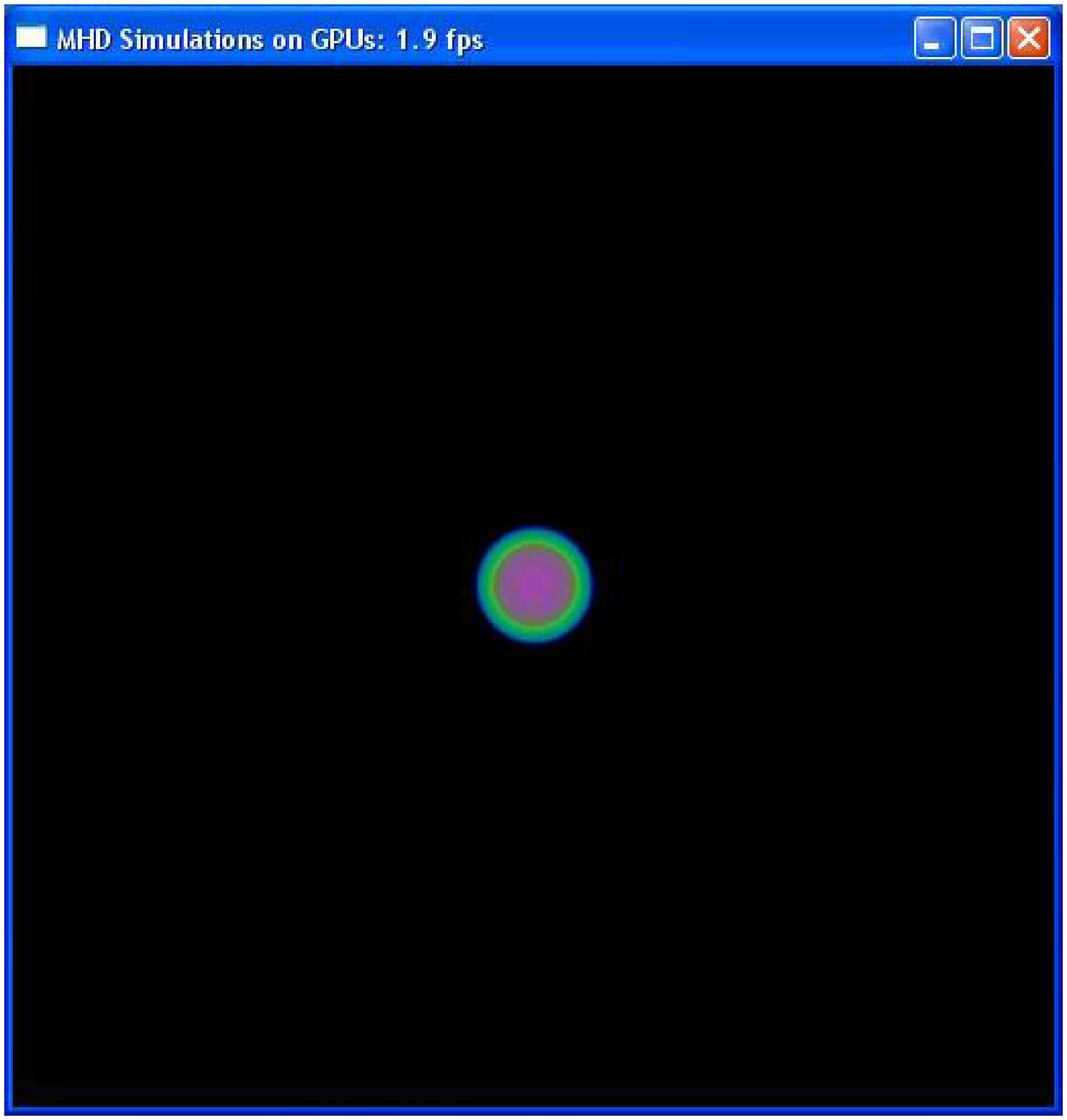}
\includegraphics*[width=2.5in]{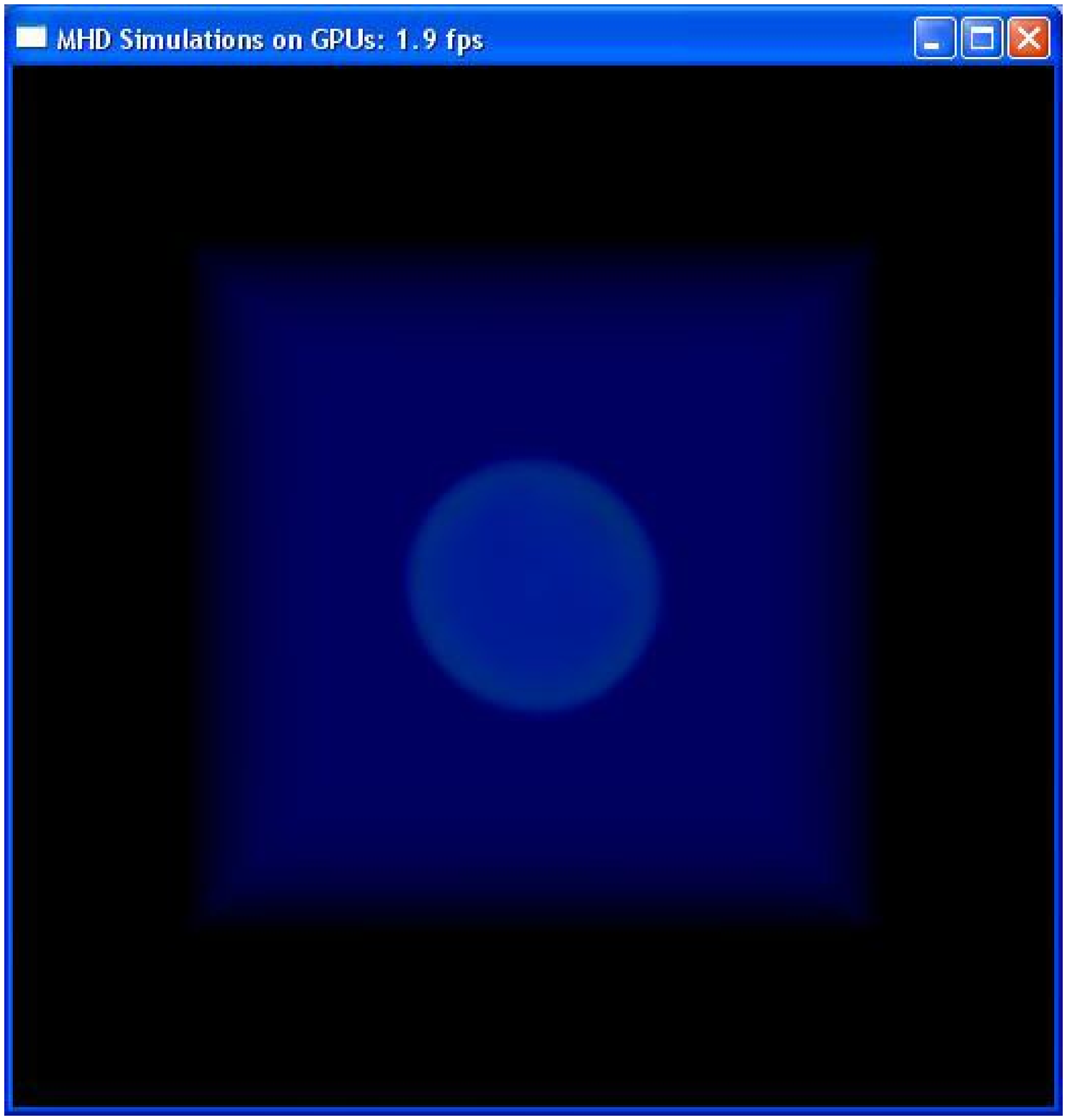}
\hfill \caption{3D visualizations of the density ($E$) of 3D Blast wave problem with $128^3$ grid
points using {\it GPU-MHD}.}\label{fig:3DVis}
\end{center}
\end{figure}

However, visualization of 3D problem is different to 1D and 2D problems. GPU-based volume
visualization method~\cite{Hadwiger2006} and texture memory  (or video memory) are used.
Unfortunately, the current version (Version 2.3) of CUDA does not provide the feature to copy the
data from the CUDA global memory to texture memory directly, even both of them are in GRAM. On the
other hand, texture memory is readable but is not rewritable in CUDA. So the simulation results
have to be copied to the main memory first, and then be copied to texture memory. In addition, the
number of grid points is usually large compared to 2D problems and volume visualization techniques
are somewhat time-consuming. As a result, on GTX 295, {\it GPU-MHD} only gets 2 FPS in single
precision mode and 1 FPS in double precision mode when the number of grid points is $128^3$, and it
is far from real-time. Nevertheless, we still get 10 FPS (single precision mode) and 6 FPS (double
precision mode) for performing the simulation of problems with resolution of $64^3$ and about 20
FPS (single and double precision modes) for problems with resolution of $32^3$. On GTX 480, we can
get 60 FPS for both single and double precision for $32^3$ grid points, 20 FPS (single) and 16 FPS
(double) for $64^3$ grid points, and 6.1 FPS (single) and 3.6 FPS (double) for $128^3$ grid points.
Fig.~\ref{fig:3DVis} shows two example images of 2D visualizations using {\it GPU-MHD}.

\section{Conclusion and future work}
\par

In this paper we present, to the author's knowledge, the first implementation of MHD simulations
entirely on GPUs with CUDA, named {\it GPU-MHD}, to accelerate the simulation process. The aim of
this paper is to present a GPU implementation in detail, demonstrating how a TVD based MHD
simulations can be implemented efficiently for NVIDIA GPUs with CUDA. A series of numerical tests
have been performed to validate the correctness of our code. Accuracy evaluation by comparing
single and double precision computation results is also given, indicating that double precision
support on GPUs is a must for long-term MHD simulation. Performance measurements of both single and
double precision modes of {\it GPU-MHD} are conducted on GT200 architecture (GTX 295) and Fermi
architecture (GTX 480). These measurements show that our GPU-based implementation achieves between
one and two orders of magnitude depending on the used graphics card, problem size, and precision
when comparing to the original serial CPU MHD implementation. In order to provide the user better
understanding of the problems being investigated during the simulation process, we have extended
{\it GPU-MHD} to support visualization of the simulation results. With {\it GPU-MHD}, the whole MHD
simulation and visualization process can be performed entirely on GPUs.

There are two directions in our future work, firstly, we are going to extend {\it GPU-MHD} for
multiple GPUs and GPU cluster~\cite{Schive2008} to fully exploit the power of GPUs. Secondly, we
will investigate implementing other recent high-order Godunov MHD algorithms such as~\cite{Lee2009}
and~\cite{Stone2009} on GPUs. These GPU-based algorithms will be served as the base of our GPU
framework for simulating large-scale MHD problems in space weather modeling.

\section{Acknowledgments}

This work has been supported by the Science and Technology Development Fund of Macao SAR
(03/2008/A1) and the National High-Technology Research and Development Program of China
(2009AA122205). Xueshang Feng is supported by the National Natural Science Foundation of China
(40874091 and 40890162). The authors would like to thank Dr. Ue-Li Pen and Bijia Pang at the
Canadian Institute for Theoretical Astrophysics, University of Toronto for providing the FORTRAN
MHD code. Thanks to Dr. Yuet Ming Lam for his suggestions on the revision of the paper. Special
thanks to anonymous reviewers for their constructive comments on the paper.

\end{document}